\documentclass[10pt,prd,preprintnumbers,floatfix,aps,nofootinbib,showpacs,twocolumn,superscriptaddress,noshowpacs,aas_macros]{revtex4-2} 

\usepackage{epsfig}
\usepackage{url}
\usepackage{hyperref}
\usepackage[normalem]{ulem} 
\usepackage{latexsym}
\usepackage{epsfig}
\usepackage{amsmath}
\usepackage{amssymb}
\usepackage{wasysym}
\usepackage{graphicx}
\usepackage{verbatim}
\usepackage{enumerate,mdwlist}
\usepackage[titletoc]{appendix}
\usepackage{amsfonts}
\usepackage{fancyvrb} 
\usepackage{tikz} 
\usetikzlibrary{calc}
\usepackage{pgfplots}
\usepackage[export]{adjustbox}
\usepackage[normalem]{ulem}

\newcommand{\tr}{\mbox{tr}}

\newcommand{\ba}{\begin{eqnarray}}
\newcommand{\ea}{\end{eqnarray}}
\newcommand{\be}{\begin{equation}}
\newcommand{\ee}{\end{equation}}

\newcommand{\nn}{\nonumber}

\newcommand{\de}{{\rm d}}
\newcommand{\vect}[1]{\boldsymbol{#1}}

\newcommand{\al}{\alpha}

\definecolor{grey}{rgb}{0.4,0.4,0.4}
\definecolor{dullmagenta}{rgb}{0.4,0,0.4}
\definecolor{darkblue}{rgb}{0,0,0.4}
\definecolor{midblue}{rgb}{0,0,0.5}
\definecolor{midred}{rgb}{0.5,0,0}
\definecolor{orange}{rgb}{1,0.5,0}
\definecolor{lightbrown}{rgb}{0.75,0.5,0.25}
\definecolor{tan}{cmyk}{0.14,0.42,0.56,0}
\definecolor{djunglegreen}{cmyk}{0.99,0,0.52,0}
\definecolor{lightgreen}{rgb}{0,1,0}
\definecolor{olivegreen}{cmyk}{0.64,0,0.95,0.40}
\definecolor{midgreen}{rgb}{0.0,0.675,0.0}
\definecolor{darkgreen}{rgb}{0,0.5,0}
\definecolor{ceruleanblue}{rgb}{0.0, 0.2, 0.7}
\definecolor{burgundy}{rgb}{0.5, 0.0, 0.13}

\hypersetup{
    colorlinks=true,
    linkcolor=ceruleanblue,
    filecolor=ceruleanblue,      
    urlcolor=midblue,
    citecolor=burgundy,
}

\newcommand{\mz}[1]{\textcolor{black}{#1}}
\newcommand{\gt}[1]{\textcolor{black}{#1}}
\newcommand{\mzc}[1]{\textcolor{black}{\textbf{[MZ: #1]}}}

\newcommand{\gtc}[1]{\textcolor{black}{\textbf{[GT: #1]}}}
\newcommand{\mcc}[1]{\textcolor{black}{\textbf{[MC: #1]}}}
\newcommand{\mct}[1]{\textcolor{black}{#1}}

\usepackage[all]{xy} 
\usepackage{amsfonts}

\makeatletter
\def\l@subsubsection#1#2{}
\makeatother

\begin{document} 

\title{Gravitational wave lensing as a probe of halo properties and dark matter}

\author{Giovanni Tambalo}
\email{giovanni.tambalo@aei.mpg.de}
\affiliation{Max Planck Institute for Gravitational Physics (Albert Einstein Institute) \\
Am Mühlenberg 1, D-14476 Potsdam-Golm, Germany}

\author{Miguel Zumalac\'arregui}
\email{miguel.zumalacarregui@aei.mpg.de}
\affiliation{Max Planck Institute for Gravitational Physics (Albert Einstein Institute) \\
Am Mühlenberg 1, D-14476 Potsdam-Golm, Germany}

\author{Liang Dai}
\email{liangdai@berkeley.edu}
\affiliation{University of California at Berkeley,
Berkeley, California 94720, USA}

\author{Mark Ho-Yeuk Cheung}
\email{hcheung5@jhu.edu}
\affiliation{William H. Miller III Department of Physics and Astronomy, Johns Hopkins University, 3400 North Charles Street, Baltimore, Maryland, 21218, USA}

\begin{abstract}
Just like light, gravitational waves (GWs) are deflected and magnified by gravitational fields as they propagate through the Universe. However, their low frequency, phase coherence and feeble coupling to matter allow for distinct lensing phenomena, such as diffraction and central images, that are challenging to observe through electromagnetic sources.
Here we explore how these phenomena can be used to probe features of gravitational lenses. We focus on two variants of the singular isothermal sphere, with 1) a variable slope of the matter density and 2) a central core. 
We describe the imprints of these features in the wave- and geometric-optics regimes, including the prospect of detecting central images.
We forecast the capacity of LISA and advanced LIGO to study strongly lensed signals and measure the projected lens mass, impact parameter and slope or core size. A broad range of lens masses allows all parameters to be measured with precision up to $\sim 1/{\rm SNR}$, despite large degeneracies. Thanks to wave-optics corrections, all parameters can be measured, even when no central image forms. 
Although GWs are sensitive to projected quantities, we compute the probability distribution of lens redshift, virial mass and projection scale given a cosmology.
As an application, we consider the prospect of constraining self-interacting and ultra-light dark matter, showing the regions of parameter space accessible to strongly-lensed GWs.
The distinct GW signatures will enable novel probes of fundamental physics and astrophysics, including the properties of dark matter and the central regions of galactic halos.
\end{abstract}

\date{\today}

\maketitle

{
  \hypersetup{hidelinks}
  \tableofcontents
}

\section{Introduction}

Gravitational lensing is a consolidated astronomical probe across vastly different scales. From weak lensing of the cosmic microwave background to microlensing by exoplanets orbiting Milky Way stars, its rich phenomenology offers valuable insights about the matter distribution in the Universe \cite{Bartelmann:2010fz}. Often, gravitational lensing data translates into powerful tests of fundamental physics, such as the properties of dark matter (DM). Given the breadth of DM theories still viable \cite{Bertone:2018krk}, increasing the pool of available observations is of paramount importance.

Applications of gravitational lensing have relied mainly on sources observed via electromagnetic (EM) radiation. Rapid progress of gravitational wave (GW) astronomy may soon open a new window into the Universe with the advent of GW lensing.\footnote{GWs can only be deflected and lensed by gravitational fields. We therefore use ``GW lensing" instead of ``Gravitational lensing of GWs", as repeating ``gravitational" is both redundant and wordy.}
Lensing of GWs is highly complementary to observations of lensed EM sources due to several key properties, such as low emission frequency, phase coherence, negligible absorption and accurate source models.

The lower frequencies of GWs enable the observation of wave-optics (WO) lensing phenomena caused by diffraction. 
The shortest observed GWs have wavelengths ($\sim 1000\,{\rm km}$) orders of magnitude larger than the longest EM waves able to penetrate the atmosphere ($\sim 10\,{\rm m}$).
The rich WO phenomenology includes frequency-dependent patterns in the GW signal, which carry information about the lens' mass and its density distribution. WO amplification contains a wealth of information that is lost in Geometric Optics (GO), the high-frequency limit, characterised by only three parameters per image (magnification, time delay and Morse phase), which depend on the local properties of the lensing potential at the location of the images. In contrast, WO effects probe a large portion of the lens, potentially allowing for a more detailed lens reconstruction \cite{Ulmer:1994ij,Takahashi:2003ix,Tambalo:2022plm,Savastano:2023spl}.
 
The phase coherence of GWs allows one to measure the interference between different GO images in the strong-lensing regime and more easily observe fainter images. For most EM sources the interference patterns are washed out due to temporal or spatial incoherence \cite{Oguri:2019fix,Liao:2022gde}. 
Observing these effects on gravitationally-lensed EM sources would require a very abundant population of light, compact lenses and can be limited by the source's physical size \cite{Katz:2018zrn,Katz:2019qug}.
Coherence makes GW detectors sensitive to the field's amplitude (rather than its intensity) which decays as $1/\mathcal{D}_S$, rather than $1/\mathcal{D}_S^2$ (here $\mathcal{D}_S$ is the luminosity distance from the source):
this scaling will allow the next generation of ground detectors to observe every stellar-mass binary black hole merger in the Universe \cite{Kalogera:2021bya}.
Detecting the field amplitude has the additional advantage of making fainter images comparably easier to detect, as they are dimmed by a factor $\sqrt{\mu_I}$ (instead of $\mu_I$, larger when $\mu_I<1$). Here $\mu_I$ is the lensing magnification factor.

The negligible absorption of GWs allows the observation of images through dense or opaque regions, which would either block or outshine EM signals. 
Strong lensing by smooth matter distributions predicts the formation of faint images near the center of the lens \cite{Schneider:1992}. Combined with the comparably smaller demagnification suffered by GWs, the observation of these images may allow GWs to probe the centers of galactic and dark matter halos.
This type of detection is challenging for EM sources: only two doubly-lensed quasars have been observed with central images where the lens is a single galaxy, with magnification ratios $\mu_H/\mu_{\rm brightest} = 0.004$, $0.007$ (see \cite{Shajib:2022con}, Sec.~4.4), where $\mu_H$ is the magnification of the central image. 
The central regions of lenses are where differences between dark matter models become more apparent and may even allow the observation of super-massive black holes \cite{Mao:2000mt, Rusin:2000mm,Keeton:2001kg,Evans:2002ut,Keeton:2002ed,Rusin:2004mz,Li:2011js,Hezaveh:2015oya,Quinn:2016jte}. A central image in a cluster-scale lens has been used to constrain the mass of a central black hole \cite{Inada:2008zb}.

GWs can also be modelled accurately. Analytic and numerical methods enable accurate waveform predictions in terms of the source parameters (e.g.~masses and spins) \cite{Blanchet:2013haa, Buonanno:2014aza, Schmidt:2020ekt}. In contrast, EM sources can rarely be described from first principles. Lensing of EM radiation thus requires observing the time variation of these sources \cite{Venumadhav:2017pps,Diego:2017drh}, or using objects that can be calibrated through empirical relations, such as Type Ia supernovae \cite{Zumalacarregui:2017qqd}. 
The existence of accurate, well-understood models provides an additional handle to test GW lensing effects in general systems, without additional assumptions. In addition, GWs allow an exquisite timing, making time-delay measurements more precise and robust than for EM systems, such as quasars \cite{Birrer:2022chj} and supernovae \cite{Oguri:2019fix,Suyu:2023jue}.

Thanks to these properties, GW lensing offers a synergy to probe the matter distribution in the Universe. Detection of lensed GWs by LISA could be used to accurately reconstruct the lens parameters \cite{Takahashi:2003ix}. While strongly-lensed LISA sources with WO effects are unlikely \cite{Sereno:2010dr}, WO effects can be detected for sizeable impact parameters, substantially boosting the chance of detection \cite{Takahashi:2003ix,Gao:2021sxw,Gais:2022xir,Savastano:2022jjv}. These determinations benefit from accurate waveforms, with additional information from higher harmonics improving the lens reconstruction \cite{Caliskan:2022hbu}.
These studies focused on symmetric and singular lenses: the point lens and/or singular isothermal sphere (SIS).\footnote{Non-symmetric, singular lenses have been treated in studies of microlensing for ground-based detectors \cite{Lai:2018rto,Christian:2018vsi, Diego:2019lcd, Diego:2019rzc, Meena:2019ate, Cheung:2020okf, Mishra:2021xzz, Basak:2021ten, Biesiada:2021pzo, Seo:2021ucd, Yeung:2021roe, Suvorov:2021uvd,Meena:2022unp}. WO imprints have also been studied in a lens-model agnostic way \cite{dai2018detecting}.}
Work on a simple extension of the SIS \cite{Choi:2021jqn} and symmetric but non-singular lenses (Navarro-Frenk-White, NFW) \cite{Cremonese:2021ahz,Guo:2022dre} suggests that the mismatch between waveforms may allow future observations to distinguish among different lens models. However, a more detailed analysis of the lens reconstruction, e.g.~including degeneracies between parameters, needs to be performed.

GW lensing may offer constraints on DM models complementary to other gravitational probes \cite{Buckley:2017ijx, Bertone:2018krk}. A prime target has been compact objects such as primordial black holes and compact and/or light DM halos \cite{Jung:2017flg,Diego:2019rzc, Oguri:2020ldf,Basak:2021ten,Urrutia:2021qak,Fairbairn:2022xln,Guo:2022dre,Oguri:2022zpn,Zhou:2022yeo}. 
GW may be able to probe the properties of DM further. Long-range interactions of DM particles may allow them to form compact substructures \cite{Buckley:2017ttd, Savastano:2019zpr,Arvanitaki:2019rax,Chakraborty:2022mwu}.
Other DM scenarios have distinct predictions on small scales, such as a lower abundance of sub-structure or central cores in DM halos. Such is the case of ultra-light boson fields (also known as ULDM) \cite{Hu:2000ke,Hui:2016ltb, Ferreira:2020fam,Hui:2021tkt} or self-interacting DM (SIDM) \cite{Spergel:1999mh,Tulin:2017ara}, which also address discrepancies between $\Lambda$CDM and observations on small scales \cite{Bullock:2017xww}. 
Probing these features with GWs is complementary to lensed EM sources.

Current searches for lensed signals in LIGO-Virgo-Kagra (LVK) focus on strong lensing, where multiple lensed signals with the same morphology are expected to be detected \cite{Haris:2018vmn,Li:2019osa,McIsaac:2019use, Liu:2020par,Lo:2021nae,Janquart:2022wxc}, using both general parametrizations \cite{Ali:2022guz} and lens models \cite{Janquart:2022zdd}. 
Detecting GW lensing with a single image requires either modelling of diffraction effects 
\cite{dai2018detecting,Lai:2018rto, Hannuksela:2019kle, Kim2022:2206.08234v1} or the identification of a Type II image through its Morse Phase \cite{Dai:2017huk, Ezquiaga:2020gdt,Wang:2021kzt,Vijaykumar:2022dlp}.

No clear detection of lensed GWs has been reported yet \cite{LIGOScientific:2021izm,Hannuksela:2019kle}. However, Ref.~\cite{Dai:2020tpj} found an intriguing strong-lensing candidate. Ultimately, ruling out the presence of lensed events requires assumptions about the high-redshift merger rate~\cite{Diego:2021fyd}. Regardless of its current status, lensing of GWs is bound to become a reality as the number of signals grows with cumulative observing time and detector upgrades
\cite{Lai:2018rto,Ng:2017yiu,Oguri:2018muv,Mukherjee:2021qam, Wierda:2021upe,Cusin:2021rjp,Xu:2021bfn}. Estimates indicate event rates $\sim 1$/yr for LIGO A+ and $\sim 50$/yr for 3G detectors \cite{Xu:2021bfn}, with a caveat of false alarm events \cite{Caliskan:2022wbh} and large uncertainties on the merger rates at high redshift.
In addition to its potential to probe cosmic structures, GW lensing may bias the inferred source population \cite{Dai:2016igl} as well as distance measurements \cite{Cusin:2020ezb,LISACosmologyWorkingGroup:2022jok}.

The purpose of this work is to address the capacity of strongly-lensed GWs to constrain the properties of galactic and DM halos by exploiting their properties complementary to EM observations. 
We review gravitational lensing in the WO regime in Sec.~\ref{sec:lensing}. In Sec.~\ref{sec:lens_models} we study the phenomenology of two symmetric lenses that generalise the SIS by varying its slope and introducing an inner core. Section \ref{sec:forecasts} presents a forecast on the detectability of lens features by LISA and LIGO using a Fisher-information matrix approach. As a potential application, Sec.~\ref{sec:dark_matter} explores the prospect of recovering the lens' redshift and its virial mass, as well as probing the parameter space of two dark matter scenarios (self-interacting and ultra-light).
We summarise and discuss our results in Sec.~\ref{sec:conclusions}. Appendices contain further details about lensing, the Fisher forecast calculation and its validation.

\subsection{Summary and guide for busy readers}

Our main results encompass the following applications of GW lensing:
\begin{itemize}
\item \textbf{Detecting central images}: GWs can probe faint images forming near the center of strong gravitational lenses, for which WO corrections are the largest. This is explored in detail in Sec.~\ref{sec:lens_models} for lenses with variable slope and an inner core.
\item \textbf{Probing lens features}: In Sec.~\ref{sec:forecasts} we present a forecast of the sensitivity of lensed GWs to the lens' parameters, including the core size and density slope. Precise measurements are possible thanks to multiple images and WO effects. 
\item \textbf{Testing large-scale structure}: In Sec.~\ref{sec:lens_mass} we develop a probabilistic framework to constrain the lens' redshift and virial mass, given observations of projected quantities, an expansion history and halo mass function.
\item \textbf{Constraining dark matter}: we show the capacity of lensed GWs to set stringent limits on dark matter theories that predict the formation of cores. Self-interacting dark matter and ultra-light dark matter are discussed in Secs.~\ref{sec:forecast_sidm} and \ref{sec:forecast_uldm}.
\end{itemize}
Our analysis highlights the differences and synergies between lensing of EM and GW sources.

Our conclusions (Sec.~\ref{sec:conclusions}) summarize our analyses, highlight our main results and discuss implications.
Readers are encouraged to skip content they are familiar with. We have provided abundant cross-references and figures describing our results and analysis. Readers interested in a specific lens model can read only the relevant subsections. \\

\section{Gravitational Wave Lensing} \label{sec:lensing}

In this Section, we will review the equations governing gravitational lensing in the wave-optics (WO) regime, Sec.~\ref{sec:lensing_wo}. We will recap analytic expansions valid in the high-frequency limit in Sec.~\ref{sec:lensing_go}. The low-frequency expansion and numerical methods are discussed in Appendices \ref{subsec:wo_low} and \ref{subsec:wo_numerical}, respectively.

\subsection{Equations \& definitions}
\label{sec:lensing_wo}

\begin{figure}
 \includegraphics[width=\columnwidth]{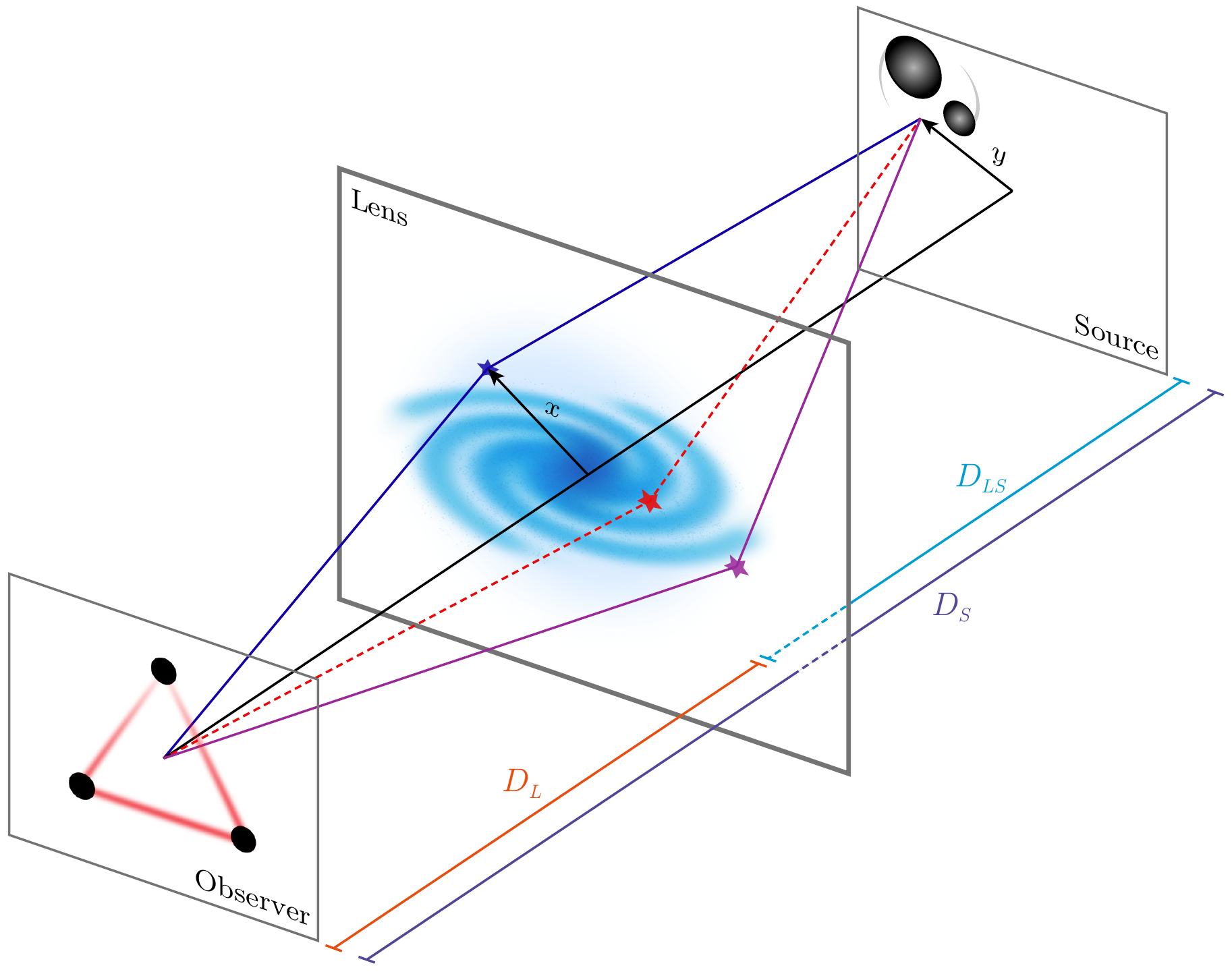}
 \caption{
 Setup for strong gravitational lensing. A lens located at (angular diameter) distance $D_L$ magnifies a source at $D_S$ with (dimensionless) impact parameter $\vect y$. In the geometric-optics limit, the lens produces multiple images (stars), whose positions are given by the dimensionless coordinates $\vect x_I$ in the lens plane.
 GWs allow us to observe all images, including the central image (red), whose EM counterpart is further suppressed ($1>\sqrt{|\mu|} > |\mu|$), and blocked or outshined by the lens matter.
}
 \label{fig:lensing_configuration}
\end{figure}

The WO regime of gravitational lensing is characterised by the \textit{amplification factor} $F(f)$, defined as the ratio between the lensed and unlensed waveforms in the frequency domain: 
\begin{equation}
    F(f) \equiv \frac{\tilde h(f)}{\tilde h_0(f)}\,,
\end{equation}
where $f$ is the frequency.\footnote{For EM signals one substitutes the waveform for the field strength in $F(f)$. In that case, the observable is not the field amplitude but its intensity, so the relevant quantity is $|F(f)|^2$.}
At the leading order in frequency ($f$ much higher than the typical background curvature), the polarization tensor of the wave is just parallelly transported along the null geodesics of the wave \cite{Harte:2018wni}. Therefore, the effect of lensing on the polarization is negligible, and the polarization tensor can be regarded as a constant, so it drops out in the definition of $F(f)$. Polarization-dependent corrections on the GW phase are suppressed by the curvature $\mathcal R$ and frequency by a factor $\sim \mathcal{R}/f^2$ \cite{Andersson:2020gsj,Oancea:2022szu}. 
Here and in the following we work in units where $c = 1$.

Let us consider a lens at redshift $z_L$ and define an effective distance
\begin{equation}\label{eq:deff_effective_distance}
    d_{\rm eff}\equiv \frac{D_L D_{LS}}{(1+z_L)D_S}\,.
\end{equation}
Here $D_L$, $D_S$ and $D_{LS}$ are the angular diameter distances to the lens, the source and between the lens and source. 
We will work within the thin-lens approximation, projecting the mass of the lens onto the lens plane \cite{Suyama:2005mx}. Moreover, we indicate positions on the lens plane with $\vect \xi$ and the impact parameter on the source plane with $\vect \eta$, where both are two-dimensional vectors.
The observer-lens-source configuration is summarised in Fig.~\ref{fig:lensing_configuration}.

The expression for $F(f)$ can be obtained from the simplified Fesnel-Kirchhoff integral \cite{Schneider:1992}
\begin{equation}\label{eq:F_f_def}
    F(f) 
    \equiv 
    \frac{-if}{d_{\rm eff}} 
    \int \de^2 \vect \xi 
    \exp\left[2\pi i f t_{d}(\vect \xi, \vect \eta)\right]\,,
\end{equation}
(a static configuration has been assumed in deriving this result). Here $t_d(\vect \xi, \vect \eta)$ is the \emph{time-delay function} of the lens, given by
\begin{equation}\label{eq:time_delay_def}
    t_d(\vect \xi, \vect \eta) 
    \equiv
    \frac{1}{2d_{\rm eff}}
    \left(\vect \xi - \frac{D_L}{D_S}\vect \eta\right)^2 
    - \hat{\psi}(\vect \xi) 
    -
    \hat{\phi}_m(\vect \eta)
    \,.
\end{equation}
The overall phase $\hat{\phi}_m(\vect \eta)$ is chosen such that the minimum time delay is zero, meaning that the first component of the lensed signal to be received (a type I image in the GO limit, cf.~Sec.~\ref{sec:lensing_go}) arrives at $t_d = 0$. The \emph{lensing potential} $\hat{\psi}(\vect \xi)$ is determined by the projected mass distribution of the lens. In particular, given a density $\rho(\vect r)$, the projected mass density $\Sigma(\vect \xi)$ is obtained by integrating along the direction $z$ perpendicular to the lens plane
\begin{equation}\label{eq:proj_mass_density_def}
    \Sigma(\vect \xi) 
    \equiv 
    \int_{-\infty}^{+\infty} \de z \, 
    \rho\left(\sqrt{z^2+\vect \xi^2}\right)
    \;,
\end{equation}
and $\hat{\psi}(\vect \xi)$ is the solution of the equation
\begin{equation}\label{eq:lens_pot_diff_eq}
    \nabla^2_{\vect \xi} 
    \hat{\psi}(\vect \xi) 
    =
    8 \pi G\,\Sigma(\vect \xi)
    \;,
\end{equation}
where $\nabla^2_{\vect \xi}$ is the 2-dimensional Laplacian and $G$ is Newton's constant.
It is convenient to recast the diffraction integral of Eq.~\eqref{eq:F_f_def} in terms of dimensionless quantities. To this end, we introduce two scales, $\xi_0$ and $\eta_0 \equiv D_S\xi_0/D_L$, that will be specified depending on the lens model. Then, we define the dimensionless quantities
\begin{equation}\label{eq:lens_plane_rescaling}
    \vect x 
    \equiv  
    \frac{\vect \xi}{\xi_0}
    \;,
    \quad 
    \vect y \equiv  \frac{\vect \eta}{\eta_0}
    \;.
\end{equation}
This allows one to construct dimensionless versions of the time delay \eqref{eq:time_delay_def} and of the lensing potential \eqref{eq:lens_pot_diff_eq} as follows
\begin{align}
    \phi(\vect x, \vect y)  &=\frac{d_{\rm eff}}{\xi_0^2} t_d(\vect x, \vect y)
    \;, \\
    \psi(\vect x, \vect y) &= \frac{(1+z_L)d_{\rm eff}}{\xi_0^2}\hat{\psi}(\vect x, \vect y)\;.
\end{align}

The rescaled lensing potential is given by the dimensionless version of Eq.~\eqref{eq:lens_pot_diff_eq}, $\nabla^2_{\vect x}\psi(\vect x)  = 2 \kappa(\vect x)$, in terms of the \emph{convergence}
\begin{equation}\label{eq:convergence}
 \kappa(\vect x) 
 \equiv
\frac{\Sigma(\xi_0\vect x)}{\Sigma_{\rm cr}}\,,
\end{equation}
where the \emph{critical density} is $\Sigma_{\rm cr} \equiv (4 \pi G (1+z_L)d_{\rm eff})^{-1}$.
The potential $\psi(\vect x)$ can be then obtained using the Green's function method
\begin{equation}\label{eq:dimensionless_lensing_pot}
    \psi(\vect x) 
    =
    \frac{1}{\pi}\int \de^2 \vect x' \kappa(\vect x') \log |\vect x- \vect x'| \;.
\end{equation}
Here and in the following, $\log z$ indicates the natural logarithm.
In Eq.~\eqref{eq:dimensionless_lensing_pot}, one needs to impose the proper boundary conditions by adding solutions of the homogeneous Laplace equation to the right-hand side.
In the context of the geometric-optics approximation discussed in the following subsection, it is also useful to define the reduced deflection angle 
\begin{equation}\label{eq:def_alpha}
    \vect \alpha 
    \equiv 
    \vect \nabla_{\vect x} \psi(\vect x)
    \,.    
\end{equation}

The dimensionless version of the time delay, also known as \emph{Fermat potential}, $\phi(\vect x, \vect y)$, takes the simple form
\begin{equation}\label{eq:fermat}
    \phi(\vect x, \vect y) 
    =
    \frac{1}{2}|\vect x - \vect y|^2
    - \psi(\vect x) 
    - \phi_m(\vect y)\;.
\end{equation}
From here on, we will suppress in our formulas the minimum of the Fermat potential $\phi_m(\vect y)$ and always assume that it is added to make the minimum arrival time equal to zero. When necessary, we will introduce it back.

All these definitions, when applied to the diffraction integral \eqref{eq:F_f_def}, lead to the following expression
\begin{equation}\label{eq:lensing_wave optics}
    F(w) 
    = 
    \frac{w}{2\pi i}\int \de^2 \vect x 
    \exp\left(i w \phi(\vect x, \vect y)\right)\,.
\end{equation}
Here we introduced the \emph{dimensionless frequency}
\begin{equation} \label{eq:lensing_freq_dimensionless}
 w \equiv 8\pi G M_{Lz} f
 \,, 
\end{equation}
where the redshifted \emph{effective lens mass} is given by 
\begin{equation}\label{eq:lens_mass_def}
    M_{Lz} 
    = 
    \frac{\xi_0^2 }{4 G d_{\rm eff}}
    \,.
\end{equation}
In the point-lens case, $M_{Lz}$ corresponds to the total (redshifted) mass $(1+z_L)M$, i.e.~setting the scale in Eq.~\eqref{eq:lens_plane_rescaling} to the Einstein radius
\begin{equation}\label{eq:einstein_radius_convention}
    \xi_0^2\to R_E^2 = 4G (1 + z_L) M d_{\rm eff}\,.
\end{equation}
However, for extended lenses, $M_{Lz}$ does not correspond to the total lens mass and may differ by several orders of magnitude from the virial mass, cf.~Eq.~\eqref{eq:virial_mass_sis}.\footnote{The value of $M_{Lz}$ depends on the choice of the scale $\xi_0$. However, all predictions are consistent once the value of $\xi_0$ is set, provided that the lens parameters are rescaled accordingly. One typically chooses either $\xi_0$ following Eq.~\eqref{eq:einstein_radius_convention} or to simplify the lensing potential.}
We will discuss the relationship between $M_{Lz}$ and the total halo mass for extended lenses in Sec.~\ref{sec:lens_models}. The role of the (unknown) lens redshift will be examined in Sec.~\ref{sec:dark_matter} (cf.~Fig.~\ref{fig:lens_projection}).

\begin{figure*}
    \centering
    \includegraphics[width=0.425\textwidth]{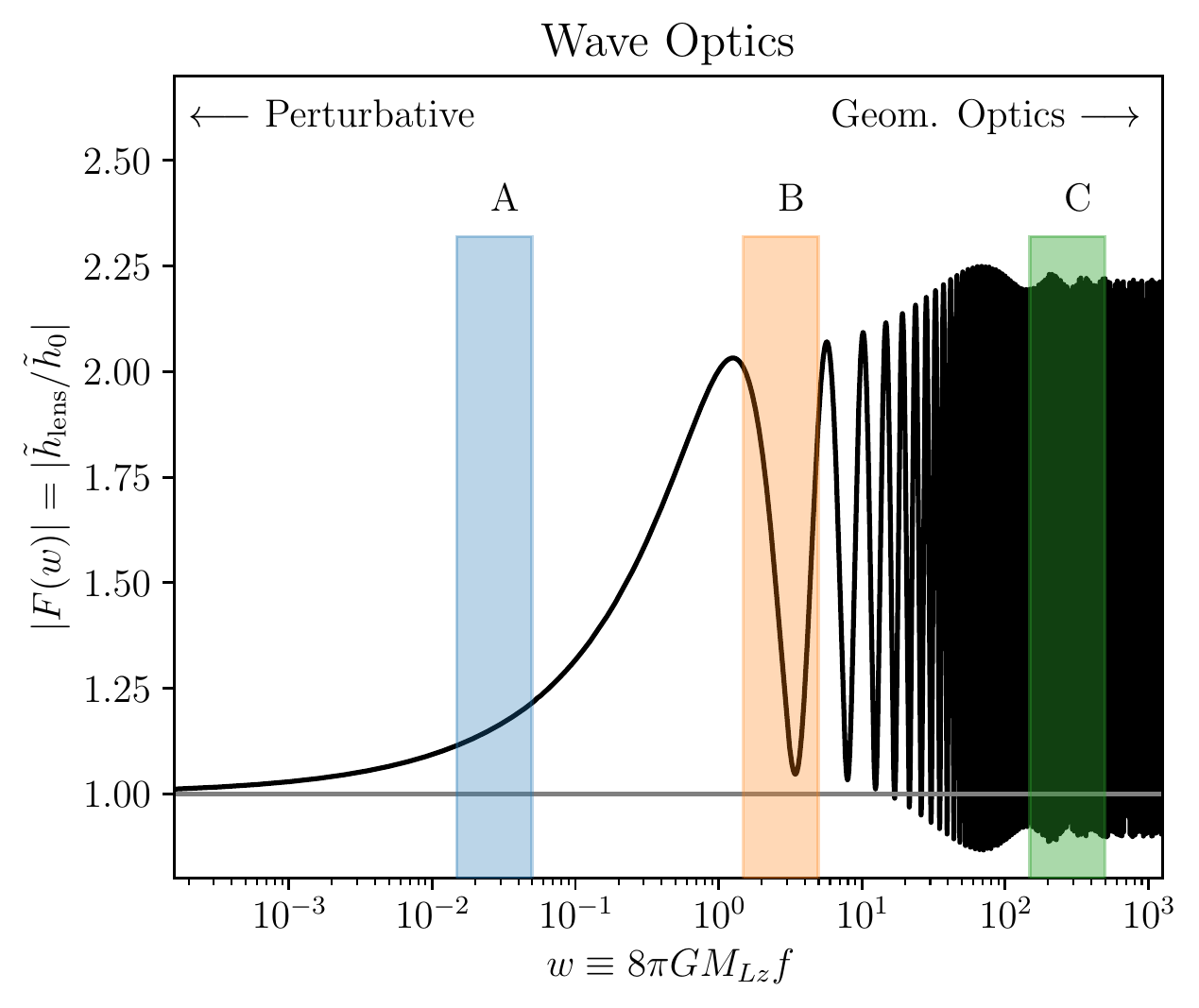} 
    \includegraphics[width=0.54\textwidth]{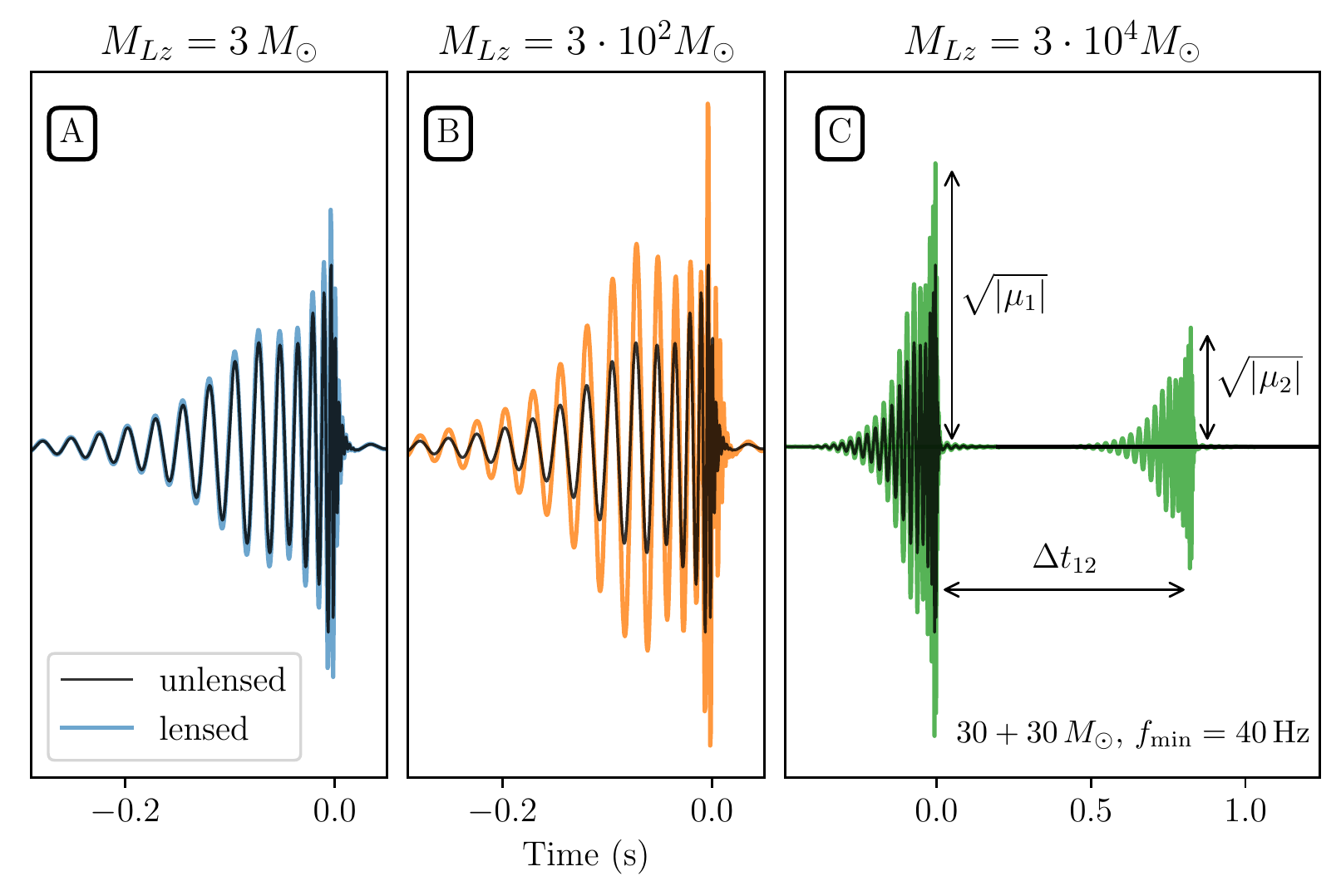}
    \caption{
    \textbf{Left}: dimensionless amplification factor for an SIS with impact parameter $y=0.7$. Different regions correspond to different lens masses for a given source, cf.~Eq.~(\ref{eq:lensing_freq_dimensionless}). 
    \textbf{Right}: time-domain waveforms for a typical LVK source ($30+30 \, M_\odot$) for three different effective lens masses representative of the perturbative (A), ``intermediate" (B) and geometric optics (C) regimes. Each panel corresponds to a shaded region in the $F(w)$ plot.}
    \label{fig:GW_lensing_pedagogical}
\end{figure*}

Once computed, $F(w)$ (for a given lens and impact parameter) can be compared at different source frequencies and lens masses. 
In typical situations, the WO lensing phenomena can be broadly characterised in three distinct regimes depending on the dimensionless frequency 
\begin{itemize}
    \item Perturbative: for $w \ll 1$ the signal undergoes a small amplification $F(w)\sim 1+w^{a}$, with $a$ determined by the asymptotic properties of $\psi(\vect x)$ (Sec.~\ref{subsec:wo_low}). This limit reflects how a wave is not affected by objects smaller than its wavelength.
    
    \item ``Intermediate" wave-optics: at finite $w$ the amplification receives contributions from large portions of the lens plane (Sec.~\ref{subsec:wo_numerical}). Around $w\sim 1$ we typically observe the \emph{onset of amplification}, i.e.~the first peak in $F(w)$.
    
    \item Geometric optics: for $w \gg 1$ the amplification factor is dominated by GO images (stationary points in the lens plane, where $\vect \nabla_{\vect x}\phi(\vect x, \vect y) = 0$). Each image gives a copy of the signal characterised by a magnification, time delay and phase factor (Sec.~\ref{sec:lensing_go}).
\end{itemize}
The three regimes are shown in Fig.~\ref{fig:GW_lensing_pedagogical} for a typical LVK source. Note that the accuracy of the perturbative and GO limits depends on the lens parameters and source's position, a dependence that will be discussed later. 
Specializing in GW detectors on the ground and in space, the typical order of magnitude for the dimensionless frequency is
\begin{align}
 w \sim &
 \left(\frac{M_{Lz}}{100\, M_\odot}\right)\left(\frac{f}{\rm{100 \,Hz}}\right) 
 \\ = & 
 \left(\frac{M_{Lz}}{10^6 \,M_\odot}\right)\left(\frac{f}{10\,\rm{mHz}}\right) \;.  \label{eq:w_typical_LISA}
\end{align}
The onset of magnification corresponds to lenses in the range of massive stellar objects and intermediate-mass black holes ($M_{Lz}\sim 10-1000 \,M_\odot$) for typical LVK sources, while for LISA it corresponds to sub-halos and massive black holes $M_{Lz}\sim 10^5 - 10^7 \,M_{\odot}$. Proposed lower-frequency detectors such as $\mu$ARES in the $\mu \rm{Hz}$ range would push the onset of magnification to even higher masses $M_{Lz}\sim 10^{10}\,M_\odot$ \cite{Sesana:2019vho}, while for pulsar timing arrays in the $\rm{nHz}$ band it would reach $\sim 10^{13}\,M_\odot$. 

\subsection{Geometric optics \& high-$w$ expansion}
\label{sec:lensing_go}
In the high-frequency limit, only the neighbourhoods of extrema of the Fermat potential \eqref{eq:fermat} contribute to the amplification factor \eqref{eq:lensing_wave optics}. 
Each extremum is associated with an image $J$, located at a position $\vect x_J$ in the image plane where the lens equation
\begin{equation}\label{eq:lens_eq}
 \vect \nabla_{\vect x} \phi(\vect x_J, \vect y) 
 = 
 \vect x_J - \vect y - \vect\alpha(\vect x_J) 
 = 0
\end{equation}
is satisfied.
The \textit{geometric-optics} regime emerges from a quadratic expansion of the Fermat potential around each image, so the diffraction integral can be performed analytically. 

The GO amplification factor \eqref{eq:lensing_wave optics} receives contributions from each image $J$
\begin{equation}\label{eq:lensing_geometric_optics}
    F(w) 
    = 
    \sum_J 
    |\mu_J|^{1/2}
    \exp(i w \phi_J - i\pi n_J)
    \,,
\end{equation}
where the \emph{magnification}
\begin{equation}\label{eq:go_magnification_def}
    \mu^{-1} 
    \equiv 
    \det\left(\phi_{,ij}\right)
    =
    \left(1-\frac{\alpha(x)}{x}\right)\left(1-\frac{\de \alpha(x)}{\de x}\right)\,,
\end{equation}
is evaluated on the image position $x_J$ (the second equality above applies to the specific case of axially-symmetric lenses).
In the above expressions, $\phi_J$ is the time delay (in units of $4GM_{Lz}$, Eq.~\eqref{eq:fermat}) of the $J$-th image and $\phi_{, ij} \equiv \partial_i\partial_j \phi$ is its Hessian matrix. The \textit{Morse Phase} \cite{Schneider:1992,Takahashi:2003ix} depends on the type of image as
\begin{equation}
 n_J = \left\{
 \begin{array}{lll}
  0 & \;\text{if} \det\left(\phi_{,ij}\right), \tr\left(\phi_{,ij}\right)>0 & \text{(minima)} \\[3pt]
  \frac{1}{2}& \;\text{if} \det\left(\phi_{,ij}\right) <0 & \text{(saddle)} \\[3pt]
  1 & \;\text{if} \det\left(\phi_{,ij}\right) >0\,,\, \tr\left(\phi_{,ij}\right)<0 & \text{(maxima)} 
 \end{array}
 \right.\,.
\end{equation}
Minima, saddle points and maxima of the time delay function are also known as type I, II and III images, respectively.

\emph{Beyond geometric optics} (bGO) corrections can be obtained as a series expansion in $1/w$.  We now present the expressions derived in Ref.~\cite{Tambalo:2022plm} (see also Ref.~\cite{Takahashi:2004mc}), focusing on axially-symmetric lenses.
The amplification at order $1 / w$ reads
\begin{equation}
    F(w) 
    = 
    \sum_{J} |\mu_J|^{1/2} 
    \left(1+ i\frac{\Delta_J}{w}\right)
    e^{i w \phi_J - i \pi n_J} 
    + \mathcal O(1/w^2)\;,
\label{eq:bGO}
\end{equation}
where the real number $\Delta_{J}$ characterizes the bGO correction for each image and is given by
\begin{equation}
    \Delta_{J}
    \equiv
    \frac{1}{16}\left[
    \frac{\psi_{J}^{(4)}}{2 a_{J}^{2}}
    +\frac{5}{12 a_{J}^{3}} (\psi_{J}^{(3)})^{2}
    +\frac{\psi_{J}^{(3)}}{a_{J}^{2} x_{J}}
    +\frac{a_{J}-b_{J}}{a_{J} b_{J} x_{J}^{2}}
    \right]\;.
\label{eq:Delta_bGO}
\end{equation}
Here $\psi^{(n)}\equiv \frac{\de^n}{\de x^n}\psi$, $a_J \equiv (1-\psi_J^{(2)})/2$ and $b_J \equiv ( 1-\psi_J^{(1)}/x_J)/2$. 
Equation \eqref{eq:bGO} shows that the leading order GO result is a good approximation provided that $\Delta_J/w\ll1$ for all images.

On top of the bGO corrections, originating from the locations of the images, subleading terms in $1/w$ also appear from \emph{non-analytic points} in the Fermat potential
(e.g.~cusps), without the presence of a corresponding image \cite{Takahashi:2004mc,Tambalo:2022plm}.\footnote{These features are typically associated with cusps in the Fermat potential or singular behaviours in the matter density profile. Notice they are not related to the cusp catastrophe.}
In general, these effects need to be accounted for at intermediate frequencies.

Let us summarise the results obtained in \cite{Tambalo:2022plm} regarding contributions from cusps at high $w$ (see also \cite{Takahashi:2004mc}). We can focus on a particular lens model, the gSIS, that we will discuss at length in this work (see Tab.~\ref{tab:lenses_summary}).
Following the same logic as in the stationary-phase approximation (that leads to the GO and bGO amplification factors), one can isolate the contribution of a cusp by expanding $\phi(\vect x, \vect y)$ around the location of the singular point and then integrating in a small neighbourhood around it (taking a small interval is justified when $w\gg1$).
In this case, the integral is not simply approximated by a Gaussian since the location of the cusp is not a stationary point of the Fermat potential. 
Let us focus on the gSIS lens for $k \geq 1$, which has a cusp at the origin in the lens plane. We indicate with $F_c(w)$ the contribution to the amplification factor from the cusp. For small impact parameter and high $w$, $F_c(w)$ is found to be
\begin{equation}\label{eq:F_cusp}
  F_c(w)
  \simeq
  - e^{i w \phi_c}
  \left(\frac{i w}{2 - k}\right)^{-\frac{k}{2-k}}
  \Gamma\left(\frac{2}{2-k}\right)
  \;,
\end{equation}
where $\phi_c \equiv y^2 / 2 - \phi_m$ is the time delay of the center of the lens
and $\Gamma(z)$ is the Gamma function. 
Notice that in the SIS case, $k = 1$, the cusp contribution becomes $F_c(w) \simeq i\, e^{i w \phi_c} / w$ while for $k > 1$ Eq.~\eqref{eq:F_cusp} decays faster than $1/w$.

We will indicate with rGO the approximation for $F(w)$ that includes both bGO terms and $F_c(w)$ at high $w$.

\section{Models of lens features}\label{sec:lens_models}

\begin{table*}[t]
    \centering
    \setlength{\tabcolsep}{10pt}
        \begin{tabular}{l c  c  c  c} \vspace*{0.1cm}
         Name                             & $\rho(r)$             & $\psi(x)$                 & Parameters & Section               \\ \hline \hline \vspace*{0.1cm}
         Point Lens                       & $\delta_D(r)$         & $\log(x)$                  & -          & -                     \\ \hline        \vspace*{0.1cm}
         Singular Isothermal Sphere (SIS) & $\frac{1}{r^{2}}$     & $x$                       & -          & \ref{sec:lens_models} \\               \vspace*{0.1cm}
         Generalised SIS (gSIS)           & $\frac{1}{r^{(k+1)}}$ & $\frac{x^{(2-k)}}{(2-k)}$ & Slope $k$  & \ref{sec:slope_lens}  \\               \vspace*{0.1cm}
         Cored Isothermal Sphere (CIS)    & $\frac{1}{r^2+r_c^2}$ & $\sqrt{x_c^2+x^2}+x_c \log \left(\frac{2 x_c}{\sqrt{x_c^2+x^2}+x_c}\right)$ 
         & Core size $x_c$ &  \ref{sec:cored_lens}
         \\ \hline \hline \vspace*{0.1cm}
    \end{tabular}
    \setlength{\tabcolsep}{6pt}
    \caption{Summary of lens models used in this work. Further details about these lenses (e.g.~normalisation, virial masses) and their phenomenology are discussed in the corresponding sections. The point lens is shown for comparison.}
    \label{tab:lenses_summary}
\end{table*}
\begin{figure*}
    \centering
    \includegraphics[width=\textwidth]{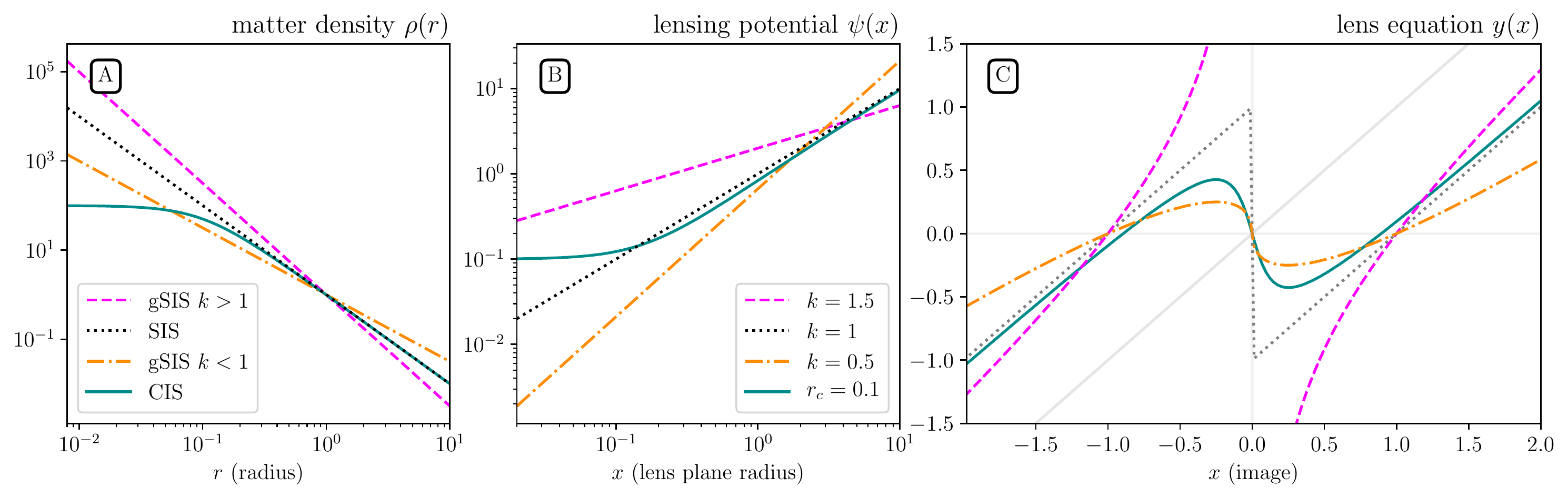}
    \caption{Overview of the lens models considered: SIS (dotted), gSIS with narrow (dashed) and broad (dash-dotted) profiles and CIS (solid). 
    \textbf{Panel A:} radial density profile (arbitrary units). All except the CIS have a divergent density at the center. 
    \textbf{Panel B:} lensing potential from the projected density. Narrow (broad) density profiles correspond to shallow (deep) lensing potential.
    \textbf{Panel C:} lens equation. Solutions are given by the set of points where the curves intersect $y=\text{const}$. Extrema of the lens equation separate regions with different numbers of solutions. Magnification is related to the slope of the curves. 
    }
    \label{fig:lens_models}
\end{figure*}

We now address the GW lensing phenomenology of the extended lens models. We focus on two 1-parameter extensions of the singular isothermal sphere (SIS, briefly reviewed in Appendix \ref{sec:lens_sis}). 
The generalised SIS (gSIS) allows a variable slope of the density distribution (Sec.~\ref{sec:slope_lens}). The cored isothermal sphere (CIS) introduces a central core with finite density (Sec.~\ref{sec:cored_lens}). 
In both cases, we will present the relation between the halo mass and $M_{Lz}$, the GO structure of the lens as well as bGO and WO features.
In Sec.~\ref{sec:lens_detecability} we address parameter reconstruction, detectability of central images and the mismatch with GWs lensed by an SIS.

The lens models are summarised in Table \ref{tab:lenses_summary}.
Figure \ref{fig:lens_models} shows the matter profile, lensing potential and lens equation for a characteristic example within each lens model, highlighting some of their differences. For the gSIS the narrow and wide cases are qualitatively different and are shown separately.

\subsection{Power law generalised SIS}\label{sec:slope_lens}

We now consider a generalization of the SIS lens with a generic power-law profile. This model, which we refer to as the $\rm gSIS$ model, is discussed in Refs.~\cite{Schneider:1992, Keeton:2001ss, Choi:2021jqn}. We will present the lens characteristics, its GO/bGO properties and WO signatures. For an illustrative comparison with the SIS and CIS lenses see Fig.~\ref{fig:lens_models}.

\subsubsection{Mass profile and scales}
The density profile of the gSIS model is given by
\begin{equation}\label{eq:gsis_rho}
    \rho(r) 
    = 
    \rho_0 \left(\frac{r_\star}{r}\right)^{k+1}
    \;,
\end{equation}
where $\rho_0$ is a typical value for the density, $r_\star$ is a radial scale and $k$ represents the slope of the halo: $k=1$ recovers the SIS, and larger/smaller values correspond to steeper/shallower profiles, respectively. Taking the range $0<k<2$ ensures both finite central densities and that the lensing potential grows more slowly than the quadratic part of the Fermat potential.
From Eq.~\eqref{eq:proj_mass_density_def}, the projected mass density of the lens is
\begin{equation}
    \Sigma(\xi)
    = 
    \beta_k
    \rho_0 r_\star 
    \left(\frac{r_{\star}}{\xi}\right)^{k}
    \;,
\end{equation}
where for convenience we defined the constant $\beta_k \equiv \sqrt{\pi}\Gamma(k/2)/\Gamma\left(\frac{k+1}{2}\right)$.
We make the following choice for the scale $\xi_0$:
\begin{equation}\label{eq:gsis_xi_0}
    \xi_0 
    = 
    \left(\frac{2\beta_k}{2-k}\frac{\rho_0 r_{\star}}{\Sigma_{\rm cr}}\right)^{1/k} r_{\star}
    \;.
\end{equation}
Then, using the relations of Sec.~\ref{sec:lensing_wo} we obtain that the lensing potential is
\begin{equation}\label{eq:slope_lens_normalized}
    \psi(x)
    =
    \frac{x^{2-k}}{2-k}\;.
\end{equation}
For all the allowed values of $k$, the total mass of the lens is divergent, as in the case of the SIS lens. Therefore, a cutoff in the radius is needed in order to define a virialized mass. In this case, $M_{\rm vir}$ is obtained as
\begin{equation}\label{eq:gsis_M_vir}
    M_{\rm vir} 
    = 
    \frac{4\pi}{2-k}
    \left(\frac{\rho_0}{\rho_{\rm vir}}\right)^{\frac{2-k}{1+k}}
    \rho_0 r_\star^3
    \;,
\end{equation}
where, as for the SIS, $\rho_{\rm vir} \equiv 200 \rho_c$.
This expression allows us to connect the effective lens mass $M_{Lz}$ with $M_{\rm vir}$. To do so, we can use Eq.~\eqref{eq:gsis_M_vir} to obtain $\rho_0$ in terms of $M_{\rm vir}$, and then replace this quantity in the expression for $M_{Lz}$. We can write this relation as follows
\begin{equation}
\label{eq:gsis_MLz}
    M_{Lz} 
    =
    \gamma_k
    \left(\frac{M_{\rm vir}}{M_0}\right)^{\frac{2-k}{3k}}
    M_{\rm vir}
    \;,
\end{equation}
where the dimensionless coefficient $\gamma_k$ and the mass scale $M_0$ are defined as
\begin{align}\label{eq:def_gsis_scales}
    \gamma_k 
    &\equiv
    \left[16\sqrt{2}(2-k)/\pi\right]^{2(2-k)/3k}
    \left[\beta_k (1+z_L)/ 2\right]^{k/2}
    \;,
    \nonumber \\
    M_0
    &\equiv
    \frac{1}{2 \pi^8 d_{\rm eff}^3 G^3 \rho_{\rm vir}^2} 
    \nonumber \\
    &= 
    6.5 \cdot 10^{17}\, M_\odot\left(\frac{1\text{Gpc}}{d_{\rm eff}}\right)^{3}\left(\frac{0.7}{h}\right)^{4}
    \;.
\end{align}
This generalises the SIS result, Eq.~\eqref{eq:MLz_sis}.
Notice that our choice for the normalisation scale $\xi_0$ and the effective lens mass $M_{Lz}$, Eqs.~\eqref{eq:gsis_xi_0} and \eqref{eq:gsis_MLz}, depend on the slope $k$. When comparing lensing results for different $k$s, one has to keep in mind that these scales are not kept fixed.

\subsubsection{GO structure \& bGO corrections}

\begin{figure}
\centering
 \includegraphics[width=0.45\textwidth]{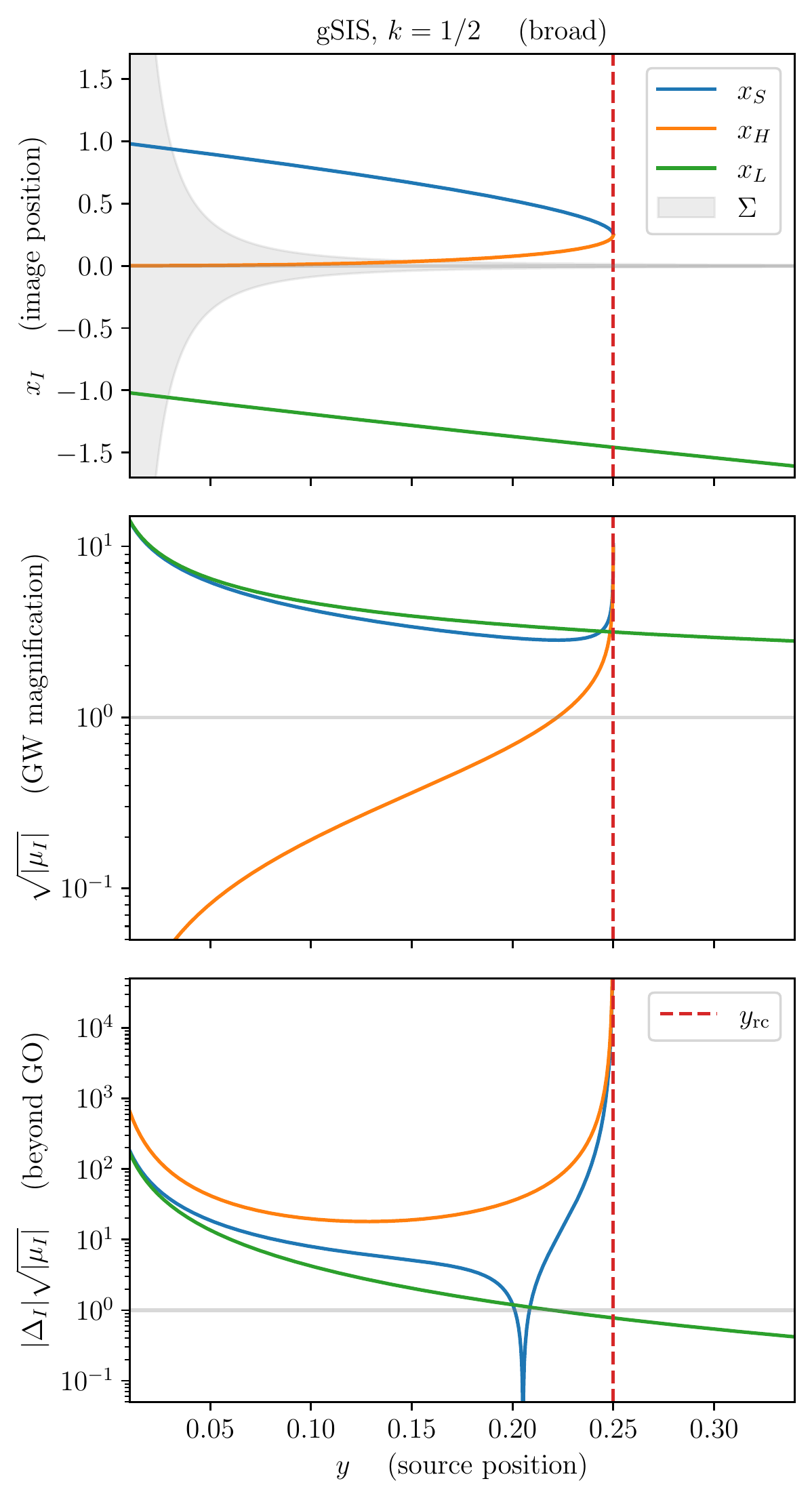}
 \caption{Geometric optics image positions and magnifications for broad gSIS lens (Eq.~\eqref{eq:slope_lens_normalized} with $k = 1/2$).
 \textbf{Top:} Image positions associated with saddle point (blue), maximum (orange) and minimum (green) of the Fermat potential, with the projected density (shaded region).
 The saddle and maximum images exist only within the caustic (red dashed vertical line, Eq.~\eqref{eq:lens_slope_critical_y}). 
 The shaded area represents $\Sigma$ as a function of $x$. Its height is rescaled to arbitrary units.
 \textbf{Middle:} Magnification for the GW amplitude for the different images from Eq.~\eqref{eq:general_k_magn}. The central image magnification goes to zero at low $y$ and diverges near the caustic.
 \textbf{Bottom:} Beyond geometric optic correction times the magnification for different images, obtained from Eq.~\eqref{eq:lens_slope_Delta} and Eq.~\eqref{eq:general_k_magn}. The central image has a large contribution that overcomes demagnification at low $y$. 
 }\label{fig:slope_IS_GO}
\end{figure}

\begin{figure}
\centering
 \includegraphics[width=0.45\textwidth]{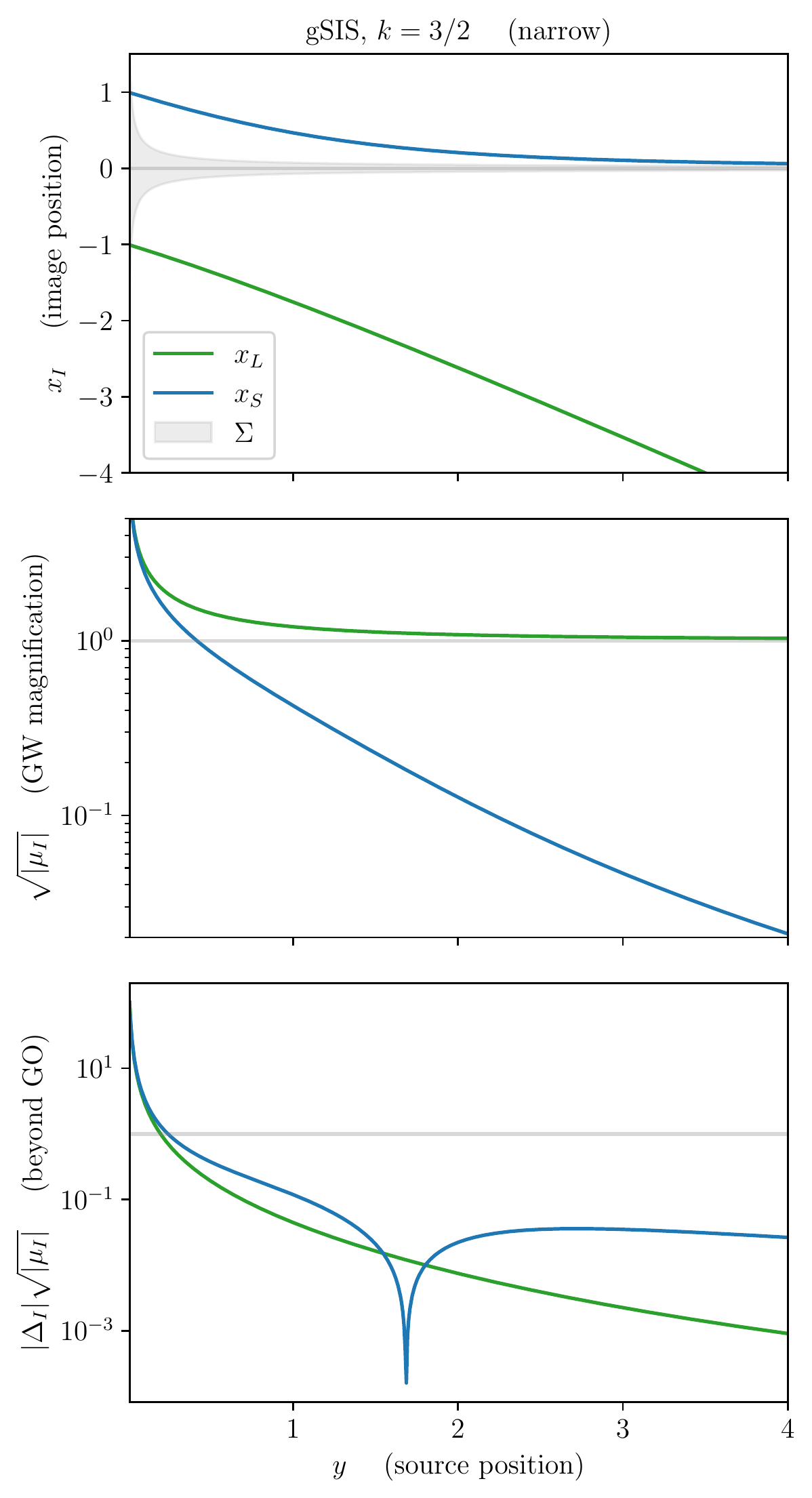}
 \caption{Geometric optics image positions and magnifications for the narrow gSIS lens Eq.~\eqref{eq:slope_lens_normalized} with $k = 3/2$. Here only the saddle point (blue) and minimum (green) exists, for any $y$. Same as Fig.~\ref{fig:slope_IS_GO} otherwise. 
 }\label{fig:slope_IS_GO_narrow}
\end{figure}

The steepness parameter $k$ leads to two distinct regimes:
\begin{enumerate}
    \item A broad matter profile ($0<k<1$)
    leads to the formation of a third GO image. This \textit{central image} is the closest to the lens center. It is associated with the maximum of the Fermat potential and has finite magnification. 
    \item A narrow lens profile ($1<k<2$) has only two GO images. However, both exist for arbitrarily large impact parameters, with the type II image becoming very faint for large $y$.
    This is similar to a point lens and is due to the compactness of the lens.
\end{enumerate}
Note that a broad matter profile is associated with a steep lens potential, while a narrow matter profile is associated with a shallower lens potential; compare Eqs.~\eqref{eq:gsis_rho} and \eqref{eq:slope_lens_normalized}. The SIS is the limiting case between the two.
The image positions, magnifications and bGO corrections for the broad/narrow lenses are shown in Figs.~\ref{fig:slope_IS_GO} and \ref{fig:slope_IS_GO_narrow}, respectively.


Let us see these differences in detail in the geometric-optics limit, first focusing on the critical curves (i.e.~points where the magnification Eq.~\eqref{eq:go_magnification_def} diverges, see also Ref.~\cite{Schneider:1992}).
Inspecting Eq.~\eqref{eq:go_magnification_def}, one sees that one family of solutions, the so-called tangential curves, are given by
the solutions to the equation $\alpha(x)/x = 1$. 
For the gSIS lens, using Eq.~\eqref{eq:def_alpha}, we have $\alpha(x) = x^{1-k}$.
Thus, for general $k$ the tangential curves are given by $x_{\rm tc} = 1$. This is also the case for the SIS lens, so we have no difference for what concerns the Einstein's ring.
On the other hand, the other family of solutions, the radial critical curves, are given by the equation ${\de}\alpha(x)/\de x = 1$. They have a more interesting dependence on $k$. The solution is indeed
\begin{equation}
    x_{\rm rc} = (1-k)^{1/k}\;,
\label{eq:lens_slope_radial_critical}
\end{equation}
and exists only for $k<1$ (otherwise Eq.~\eqref{eq:lens_slope_radial_critical} leads to a complex solution). 
The SIS corresponds to the limiting case $k=1$, where this curve approaches the origin and vanishes. 
In the case $k < 1$, the expression Eq.~\eqref{eq:lens_slope_radial_critical} together with the lens equation \eqref{eq:lens_eq} leads to a caustic in the image plane
\begin{equation}
    y_{\rm rc} = k (1-k)^{\frac{1-k}{k}}\;.
\label{eq:lens_slope_critical_y}
\end{equation}
Equation \eqref{eq:lens_slope_critical_y} gives us the region where multiple images form if $k<1$. 
In the SIS case, this reduces to $y_{\rm rc} = 1$ as expected. For $k<1$ instead, the value of $y_{\rm rc}$ is always less than $1$.

At this point, we can consider the magnification of the different images. The magnification for a given image $x_{I}$ is 
\begin{equation}
    \mu_{I} = \frac{x_I^{2 k}}{\left(x_I^k-1\right) \left(x_I^k+k-1\right)}\;.
\label{eq:general_k_magn}
\end{equation}
From this relation, we see that the critical curves are infinitely magnified as expected. Moreover, images close to the centre are highly demagnified since $\mu_I$ scales as $x_I^{2 k}$. 
Central images vanish when $x_I \to 0$. However, even in this limit, they can be observed via bGO corrections.

\begin{figure*}
\includegraphics[width=0.985\textwidth]{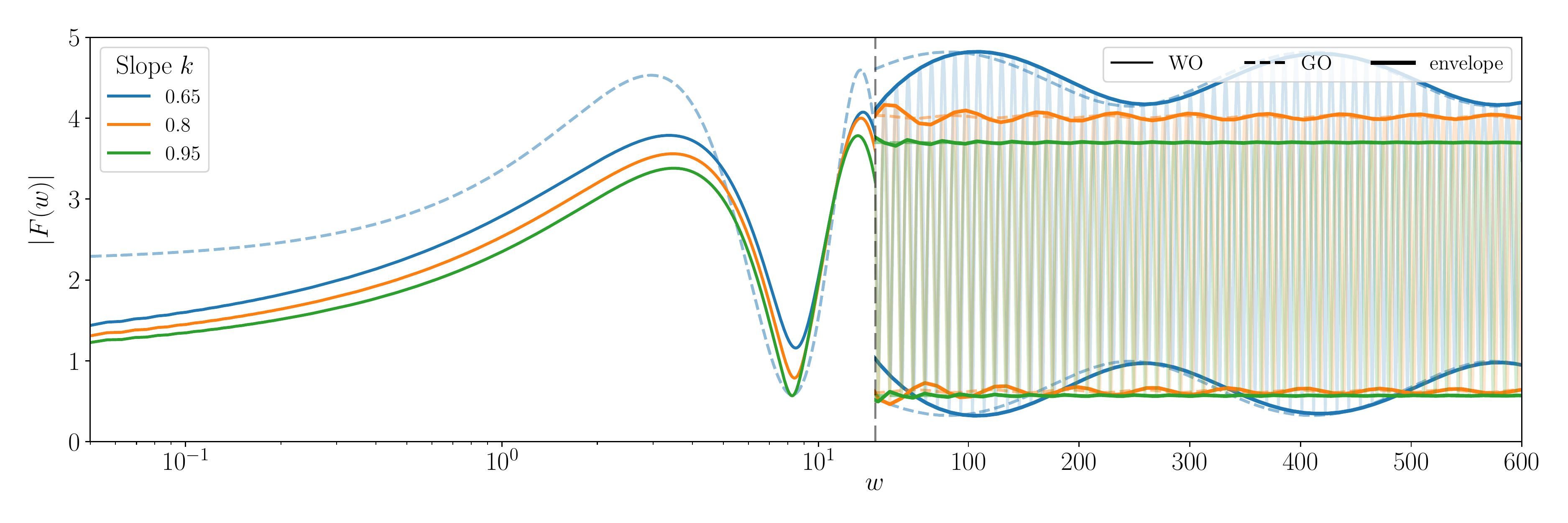}
\includegraphics[width=\textwidth]{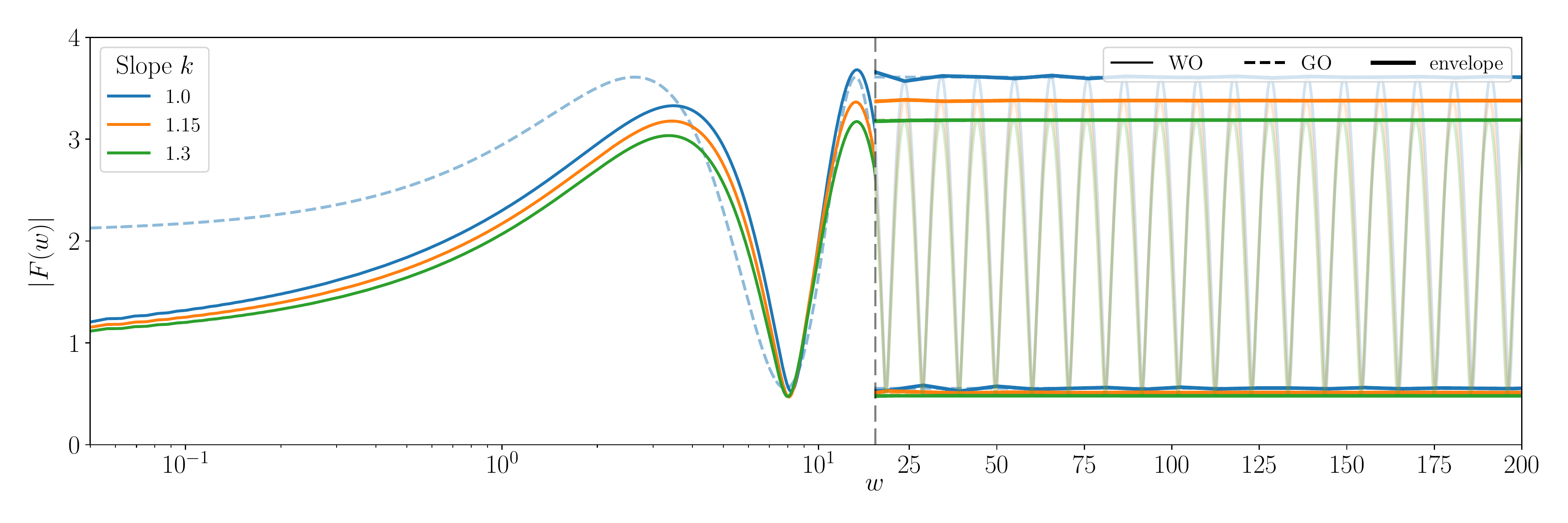}
\caption{Amplification factor of a gSIS for different values of the slope $k$, for a fixed value of the impact parameter $y = 0.3$ (in the strong-lensing regime). The WO curves are obtained with the numerical method of Sec.~\ref{subsec:wo_numerical} and are shown with solid lines. 
The GO result is shown in dashed lines (on the left-hand side GO is shown only for one curve).
On the right-hand side, the plots are in linear scale. 
Here, the envelope of the WO (GO) $|F(w)|$ is shown as thick solid (dashed) line.
\textbf{Top}: case $k<1$ where three images form within the caustic and give additional modulations in $w$ (see right panel). \textbf{Bottom}: case $k\geq1$ where only two images form. In this case, no large modulations appear, and the curves quickly approach the GO approximation.
\label{fig:gSIS_different_k}
}
\end{figure*}

The bGO corrections (parametrised by the coefficient $\Delta$ given in Eq.~\eqref{eq:Delta_bGO}) for a generic image reads
\begin{equation}
    \Delta_{I} = 
    \frac{
    k^2  
    \left[
    2 \left(k^2+k-2\right) x_I^k
    -3 (k-2) x_I^{2 k}
    -2 (k-1)^2
    \right]
    }
    {
    24 x_I^{2 - k}
    \left(x_I^k-1\right) \left(x_I^k+k-1\right)^3} \;.
\label{eq:lens_slope_Delta}
\end{equation}
Interestingly, for a central image this correction can become very large given that $\Delta_I \sim 1/x_{I}^{2-k}$. In order to assess whether bGO corrections from this image overcome the other two (brighter) terms one should look at the combination $\Delta_{I}|\mu_I|^{1/2}$ since, as we previously saw, the magnification tends to zero in the same limit. This combination scales as $\sim 1/x_{I}^{2(1-k)}$, hence is large at small enough $x_I$ when $k<1$ (which is the same situation where there is a central image in the first place). Moreover, the bGO correction is suppressed at high frequencies. Therefore, we expect this central image to be important in the wave-optics regime $w \sim 1$, at least in a certain range of values for the impact parameter.

We will now explore these results explicitly for some values of $k$ where the lens equation can be solved analytically. 

\paragraph{\textbf{Broad profiles} \texorpdfstring{$0<k<1$}{0<k<1}.}
We consider the specific case $k = 1/2$, where analytic solutions for the image positions exist. Although significantly broader than the SIS, this case has the same qualitative features found for all the models with $0<k<1$.
The geometric-optics results for $k = 1/2$ are plotted in Fig.~\ref{fig:slope_IS_GO}. Let us discuss this case in more detail.
The lens equation Eq.~\eqref{eq:lens_eq} is solved for\footnote{Here, with an abuse of notation, we consider $x_I$ as the image position along the direction of $\vect y$. Thus, it differs from the radial position $x$ as it can also take negative signs.}
\begin{align}
    x_{L} & =  \frac{1}{2}\left(1 + 2y + \sqrt{1+4y}\right) \;, \\
    x_{H} & = -\frac{1}{2}\left(1 - 2y - \sqrt{1-4y}\right) \;, \\
    x_{S} & = -\frac{1}{2}\left(1 - 2y + \sqrt{1-4y}\right) \;, 
\label{eq:k_1/2_images}
\end{align}
where, $x_L$, $x_H$ and $x_S$ are respectively the minimum, maximum and saddle points of the Fermat potential.
The maximum and the saddle merge at the caustic of Eq.~\eqref{eq:lens_slope_critical_y}, i.e.~at $y_{\rm rc} = 1/4$, or $x_{\rm rc} = 1/4$, and for larger impact parameters only the minimum image remains. 
The magnification, time delays and beyond geometric optics corrections can all be easily derived from these solutions in closed form. 
For instance, the magnification for the image $x_I$ is obtained by plugging the solutions Eq.~\eqref{eq:k_1/2_images} inside Eq.~\eqref{eq:general_k_magn}.

As already noted for general $k$, the central image $x_H$ has a vanishing magnification for $y = 0$. 
In the case at hand moreover, its magnification remains small relative to the others, unless we are very close to the caustic. However, its beyond geometric optic correction $\Delta$ is much larger than for the other two images. This can be seen by computing $\Delta$ from Eq.~\eqref{eq:lens_slope_Delta}.
Clearly, the central image has an enhanced value of $\Delta$ because of the term in the denominator $\sim x_I^{3/2}$. As in the case of $\mu$, we also have a divergence at the caustic.
We will come back to this issue after discussing the narrow lens.

\paragraph{\textbf{Narrow profiles} \texorpdfstring{$1<k<2$}{1<k<2}.}
Analytic solutions exist for a narrow lens with  $k = 3/2$. In this case, the lens equation reduces to a cubic equation in $\sqrt{x}$ with a simple closed-form solution:
\begin{equation}
    x_{L/S} 
    = 
    \frac{2}{3}
    \left(
    y \pm \frac{y^2}{2^{2/3}z_{\pm}^{1/3}} \pm \frac{z_{\pm}^{1/3}}{2^{4/3}} 
    \right)
    \;,
\label{eq:k_3/2_images}
\end{equation}
where $z_{\pm} \equiv \mp 2 y^{3} + 27 (1\pm \sqrt{1 \mp 4 y^3 /27})$. The minimum $x_L$ corresponds to $z_{+}$ while the saddle $x_{S}$ corresponds to $z_{-}$. The minimum image starts at $x_L = 1$ for $y = 0$ and moves to larger values of $x$ as $y$ increases. The saddle starts instead at $x_S = -1$ and moves towards zero.
Contrarily to the broad case, the two images exist for all values of $y$, and there is no caustic. 

We summarise the GO properties of this case in Fig.~\ref{fig:slope_IS_GO_narrow}.
The fainter image (saddle point) has a lower magnification than the minimum away from $y = 0$. This can be seen by inspecting the expression for the magnification \eqref{eq:general_k_magn}.
That expression indeed shows that the saddle gets very demagnified as we increase the impact parameter.

Even though the saddle-point image becomes undetectable at large $y$, bGO corrections may offer a window to probe compact lenses at high impact parameters. For narrow gSIS at $y\gtrsim 1$, the second image approaches the center of the lens and receives large bGO corrections due to the curvature 
of the lensing potential (its derivatives become large), and so $\Delta_{S} \gg \Delta_{L}$.
While for $y\gtrsim 1$ the second image is already very faint, its associated bGO correction remains sizeable. Equations \eqref{eq:general_k_magn} and \eqref{eq:lens_slope_Delta} evaluated on $x_S\to 0$ yield 
\begin{equation}\label{eq:slope_narrow_bgo_large_y}
    \sqrt{|\mu_S|}\Delta_S 
    \simeq 
    \frac{k^2(k-1)}{12}x^{2(k-1)}\propto y^{-2}\,,
\end{equation}
where the last relationship uses the position of the saddle point $|x_S|\sim y^{-1/(k-1)}$ for large $y$ (this holds only for $1<k<2$, as $x_S$ does not exist for $y>y_{\rm rc}$ otherwise). While Eq.~\eqref{eq:slope_narrow_bgo_large_y} tends to zero for large $y$, it remains sizeable for $y\gtrsim 1$. This is seen in the bottom panel of Fig.~\ref{fig:slope_IS_GO_narrow}, where $\sqrt{|\mu_S|}|\Delta_S|\sim \mathcal{O}(0.1)$ up to $y=4$, with the $1/y^2$ behaviour appearing at larger impact parameter. 
As the lensing probability is proportional to $y^2$, the bGO feature opens the possibility of detecting steep lenses in high SNR, low frequency (in terms of $w$) GW events via bGO corrections.

\subsubsection{Wave optics features}

Here we discuss the phenomenology of the gSIS lens and the effect of the slope $k$ on the full WO amplification.

\paragraph{\textbf{Imprints of the lens slope $k$:}}  
Let us now discuss the effect of the gSIS slope $k$ on the amplification factor. Figure \ref{fig:gSIS_different_k} shows the full WO predictions in the strong-lensing regime (fixed $y=0.3$) for both broad ($k<1$, top panel) and SIS/narrow ($k\geq 1$, bottom panel) lenses. For easier display, each plot is divided between small $w$ in log scale (left) and high $w$ in linear scale (right), where also the amplitude is shown.
Some of the features observed in the Figure depend on the relation between the impact parameter $y$ and the caustic $y_{\rm cr}$, which is a function of the slope via Eq.~\eqref{eq:lens_slope_critical_y}.
For $k=0.65$, $0.8$, $0.95$ and $1$ the caustic is located at $y_{\rm cr}= 0.369$, $0.535$, $0.811$ and $1$, respectively.

At low frequencies and fixed $y$, the effect of $k$ is to provide a different power-law behaviour in $w$, with $F(w)-1\propto w^{k / 2}$, Eq.~\eqref{eq:low_w_gsis}.
One can heuristically interpret these results as a wave not being disturbed by objects smaller than its wavelength: broader lenses (small $k$) converge more slowly to the free propagation case as $w \ll 1$, while the narrower lenses converge faster. In the narrowest possible gSIS ($k\to 2$), the convergence is as fast as a point lens $F-1\propto w$, cf.~Eq.~\eqref{eq:low_w_pt}.
The slope $k$ also affects the amplitude of the term $F(w)-1$, with smaller $k$ (broader lenses) producing larger $|F(w) - 1|$.
The larger amplitude is carried over also to the regime $w \gtrsim 1$. 

The high-frequency behaviour depends on the case under consideration. 
SIS and narrow lenses ($k>1$) display a beating pattern in $|F(w)|$, caused by the interference between the two GO images. Narrow lenses $k>1$ always form two images, and hence this pattern persists even for $y\gg1$ although with a small amplitude modulation. 
At intermediate frequencies there is a subtle amplitude modulation. 
This modulation is caused by the contribution from the center of the lens (a cusp in the Fermat potential), Eq.~\eqref{eq:F_cusp}, and decays roughly as $w^{\frac{-k}{2-k}}$. 
For the values $k \geq 1$, shown in Fig.~\ref{fig:gSIS_different_k}, the amplitude of the modulation is very small, and can only be appreciated for the SIS case
(notice the slight difference between the WO and GO curves for the SIS case in Fig.~\ref{fig:gSIS_different_k} at around $w\sim 20$).
These cases converge to GO at a relatively small $w$ (convergence to rGO is much faster). 

In broad lenses, the envelope of the beating pattern persists at arbitrarily high frequencies. This envelope modulation is caused by the central image, and its amplitude is determined by the magnification $\mu_{H}$. The associated sizeable bGO contribution, discussed after Eq.~\eqref{eq:lens_slope_Delta}, can be appreciated as a $w$-dependence of the modulation at intermediate frequencies (this is most appreciable in the $k=0.8$ case in Fig.~\ref{fig:gSIS_different_k}, where the amplitude of the modulation is largest at around $w\sim 20$ and then decreases, converging to a constant at larger frequencies).
The width of the modulation in frequency space is given by the difference in the time delay between the maximum and the saddle point, $\Delta w \sim 1/(\phi_H-\phi_S)$. As the impact parameter approaches the caustic, the amplitude and width of the modulation grow substantially (see the $k=0.65$ case in Fig.~\ref{fig:gSIS_different_k}). This regime is associated with a much slower convergence to GO.
Some of these features are qualitatively similar to the cored isothermal sphere, as we will see below.

\subsection{Cored isothermal sphere}\label{sec:cored_lens}

We will now generalise the SIS by introducing a core with finite density \cite{1987ApJ...320..468H,Flores:1995dc}, a feature shared by several DM scenarios that we will discuss in Sec.\ref{sec:dark_matter}.
We will present the cored isothermal sphere (CIS) lens characteristics, its GO/bGO properties and WO signatures.

\subsubsection{Mass profile and scales}

The CIS is characterised by a core size $r_c$ and finite central density $\rho_0$:
\begin{equation}\label{eq:rho_cored}
    \rho 
    = 
    \rho_0\frac{r_c^2}{r^2+r_c^2}
    \;.
\end{equation}
This axisymmetric lens model reduces to the SIS for large radii $r \gg r_c$ while providing a finite density at its centre. 

The projected mass density is obtained as
\begin{equation}
 \Sigma(\xi) 
 = 
 \frac{\pi \rho_0 r_c^2}{\sqrt{\xi^2+r_c^2}}
 \,.
\end{equation}
If we choose the normalisation scale $\xi_0$ to be
\begin{equation}\label{eq:cis_einstein_radius}
    \xi_0 
    = 
    \frac{2\pi \rho_0 r_c^2}{\Sigma_{\rm cr}}
    \;,
\end{equation}
then the lensing potential takes the form 
\begin{equation}
    \psi 
    =
    \sqrt{x_c^2+x^2}
    +x_c \log \left(\frac{2 x_c}{\sqrt{x_c^2+x^2}+x_c}\right)
    \;.
\end{equation} 
Here we have introduced a rescaled core radius $x_c \equiv r_c/\xi_0$ and added an unobservable constant factor $x_c$ to the lensing potential.
 
The virial radius for this lens is 
\begin{equation}\label{eq:r_vir_cis}
    r_{\rm vir}
    = 
    r_c \sqrt{\frac{\rho_0}{\rho_{\rm vir}}-1}
    \;.
\end{equation}
Contrary to the case of the gSIS lens, for the CIS the virial mass is just weakly affected by the new feature (the core radius in this case). Intuitively, for cores much smaller than the virial scale, $r_c \ll r_{\rm vir}$, the mass is dominated by the density away from the core. Hence, $M_{\rm vir}$ must have the approximate form of Eq.~\eqref{eq:virial_mass_sis}. To be more precise, the virial mass for the density profile \eqref{eq:rho_cored} is
\begin{equation}
    M_{\rm vir}
    = 
    4\pi r_c^2 \rho_0 \left[r_{\rm vir} - r_c \arctan (r_{\rm vir}/r_c)\right]
    \;.
 \end{equation} 
Then, in the limit $r_{\rm vir} \gg r_c$, the second term on the right-hand side gives a constant negative contribution to the mass, $\sim 2\pi^2 r_c^3 \rho_0$. From Eq.~\eqref{eq:r_vir_cis}, we can notice that this limit is equivalent to the limit $\rho_0 \gg \rho_{\rm vir}$.
At first order in such limit, we can obtain $\rho_0$ as a function of $M_{\rm vir}$. Following the procedure used for the other lenses, we can then relate the effective lens mass $M_{Lz}$ to the virial mass
\begin{equation}\label{eq:MLz_cis}
    M_{Lz}
    \simeq
    \gamma_c \left(\frac{M_{\rm vir}}{M_0}\right)^{\frac{1}{3}}\left[1+\frac{2}{3}\left(\frac{4 \pi^4 r_c^3 \rho_{\rm vir}}{M_{\rm vir}}\right)^{\frac{1}{3}}\right] M_{\rm vir}
    \;,
\end{equation}
where we defined $\gamma_c \equiv 2 (1+z_L)^2$ and the mass scale $M_0$ is given by Eq.~\eqref{eq:def_gsis_scales}.
The second term in the square bracket of Eq.~\eqref{eq:MLz_cis} represents a small correction from the result for the SIS \eqref{eq:MLz_sis} due to the mass removed in the core.

\subsubsection{GO structure \& bGO features}
\begin{figure}
\centering
 \includegraphics[width=0.45\textwidth]{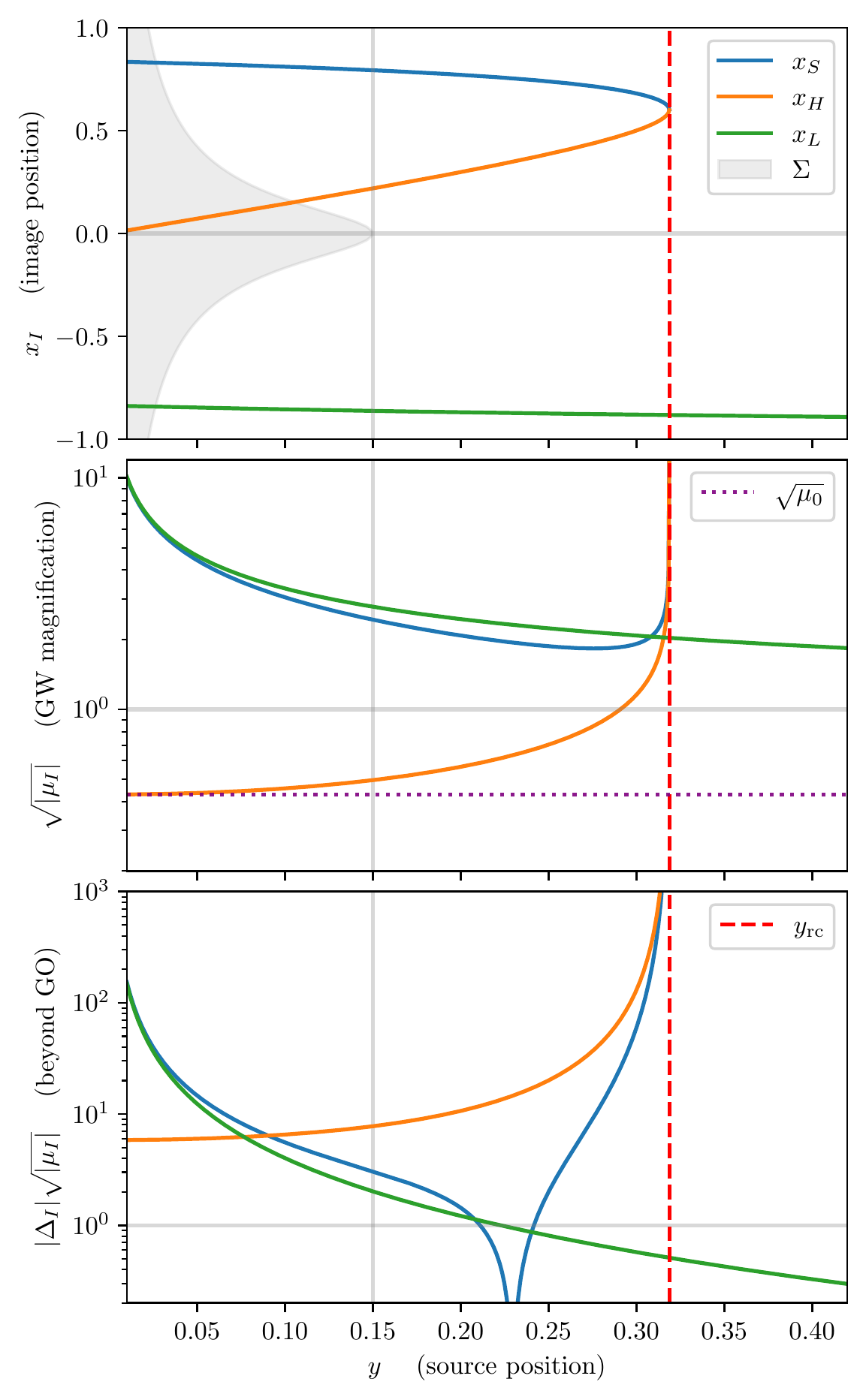}
 \caption{Geometric-optics quantities for a CIS with core size $x_c = 0.15$ as a function of the impact parameter. 
 \textbf{Top:} Image positions associated with saddle point (blue), maximum (orange) and minimum (green) of the Fermat potential, with the projected density (shaded region) and core size (vertical grey line). The saddle and maximum images exist only within the caustic (red dashed vertical line, Eq.~\eqref{eq:cis_yrc_caustic}). 
 \textbf{Middle:} Magnification for the GW amplitude for the different images. The central image magnification varies between a finite minimum ($\sim 0.429$ for $x_c=0.15$) and the divergent value near the caustic.
 \textbf{Bottom:} Beyond geometric optics correction times the magnification for different images, obtained from Eq.~\eqref{eq:lens_slope_Delta} and Eq.~\eqref{eq:general_k_magn}. The central image has a large contribution that can overcome the geometric-optics magnification of the other two images for $w \sim 1$. 
 }\label{fig:CIS_GO}
\end{figure}

Before discussing WO effects, let us examine the GO structure of the CIS. 
Depending on the core size and the impact parameter there can be one or three real solutions to the lens equation (\ref{eq:lens_eq}), similarly to the broad gSIS ($k<1$, Sec.~\ref{sec:slope_lens}).
As we will see below, multiple images can form only if $x_c < 1/2$, i.e.~if the density at the core is super-critical ($\Sigma > \Sigma_{\rm cr}$).
The GO images, their magnification and bGO corrections are shown in Fig.~\ref{fig:CIS_GO} as a function of the impact parameter for a CIS with $x_c=0.15$.

The critical curves are determined by the CIS deflection angle
\begin{equation}\label{eq:cis_deflection}
   \alpha(x) = \frac{x}{x_c+\sqrt{x_c^2+x^2}}\,.
\end{equation}
The tangential critical curve solves $\alpha(x)/x=1$ and is located at
\begin{equation}\label{eq:cis_crit_tangential}
x_{\rm tc}^2 = {1-2 x_c}\,,
\end{equation}
i.e.~the core reduces the enclosed mass, pushing the Einstein ring inward (if the standard normalisation is kept).
The tangential curve $x_{\rm tc}$ is associated with a caustic at $y_{\rm tc}=0$ when projected in the image plane via Eq.~\eqref{eq:cis_deflection}. The degeneracy of the caustic to a point is an artefact of the lens symmetry, any perturbation of the lens will spread the caustic \cite{Schneider:1992}.

Another radial critical curve exists under the condition $\de \alpha(x)/\de x=1$, or
\begin{equation}\label{eq:cis_crit_radial}
    x_{\rm rc}^2 = x_c -\frac{x_c^2}{2}-\frac{1}{2} x_c\sqrt{x_c (x_c+4)}\,,
\end{equation}
which is associated with a caustic in the image plane
\begin{equation}\label{eq:cis_yrc_caustic}
    y_{\rm rc}^2 = 1+5 x_c-\frac{1}{2}x_c^2-\frac{1}{2}\sqrt{x_c} (x_c+4)^{3/2}\,.
\end{equation}
Multiple images form only if
\begin{equation}\label{eq:cis_multiple_images}
    x_c<\frac{1}{2} \,\quad \text{and}\quad   y<y_{\rm rc}\, \,,
\end{equation}
with the two additional images corresponding to a saddle point and a maximum of the Fermat potential. The known SIS result $y_{\rm tc}\to \pm 1$ is recovered when $x_c\to 0$. In that case, the image associated with the maximum of the Fermat potential remains at the central cusp $x_H=0$ and is infinitely demagnified, $\mu_H\to 0$. 

The main GO effect of the core is to allow the central image to have a finite magnification $\mu_0$ above a certain threshold. 
In the limit $|x| \ll x_c$, the central image associated to the local maximum lies at
\begin{equation}
    x_H \simeq -\frac{2 x_c}{1-2 x_c}y\,.
\end{equation}
Note that this expression is valid for $|y|\ll 1/2-x_c$, in addition to the strong-lensing conditions $y<y_{\rm rc}$, $x_c<1/2$.
The magnification follows from Eq.~\eqref{eq:go_magnification_def}
\begin{equation}\label{eq:CIS_mag_central_img}
    \mu_H 
    \simeq 
    \frac{4 x_c^2}{(1-2 x_c)^2}+\frac{16 x_c^2 y^2}{(1-2 x_c)^5}+\mathcal{O}\left(y^4\right)
    \,.
\end{equation}
The first term on the right-hand side can be identified as a lower bound for the magnification
(note that the second term is positive if the lens is super-critical, Eq.~\eqref{eq:cis_multiple_images}). The magnification increases monotonically with $y$ until diverging at the caustic $y\to y_{\rm tc}$, see Fig.~\ref{fig:CIS_GO}.

The bGO factor $|\Delta_H|\sqrt{|\mu_H|}$ associated with the central image is similarly bounded from below and grows towards the caustic, where it diverges.
For low impact parameters, it is smaller than the bGO factors associated with the minimum and saddle point, both diverging for $y\to 0$. At intermediate $y$ the bGO correction from the central image dominates over the other images.

There are several differences between the CIS and the broad gSIS ($k<1$). The magnification of the central image approaches a constant $\mu_0$ for aligned lenses $y\to 0$, while it vanishes rapidly for the broad gSIS. In contrast, the bGO correction from the central image of the gSIS dominates over the entire strong-lensing regime, while for the CIS it only does at intermediate impact parameters. These features follow from the regularity of the CIS lens, compared to the cuspy gSIS.

\subsubsection{Wave optics features}

\begin{figure*}
 \includegraphics[width=0.99\textwidth]{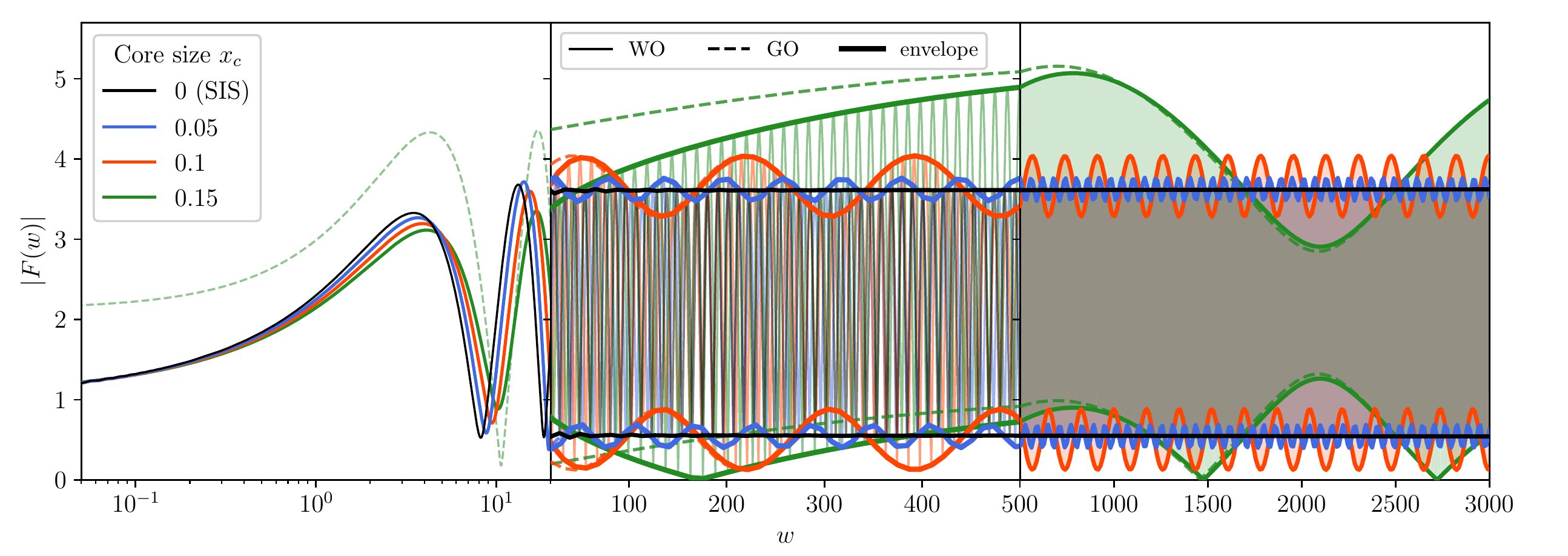}
 \caption{Amplification factor for a CIS with $y = 0.3$ and different values of the core size $x_c$. The panels correspond to dimensionless frequency $w\in (0.05,20)$ in logarithmic scale (left), $w\in (20,500)$ (center) and $w\in(500,3000)$ (right) in linear scale. The envelope of the WO (GO) $|F(w)|$ is shown as thick solid (dashed) lines (center and right). For $x_c = 0.15$ the impact parameter is close to the caustic $y = 0.3\simeq 0.3187 = y_{\rm rc}$. 
 \label{fig:cis_vary_xc}
 }
\end{figure*}

\begin{figure*}
 \includegraphics[width=0.99\textwidth]{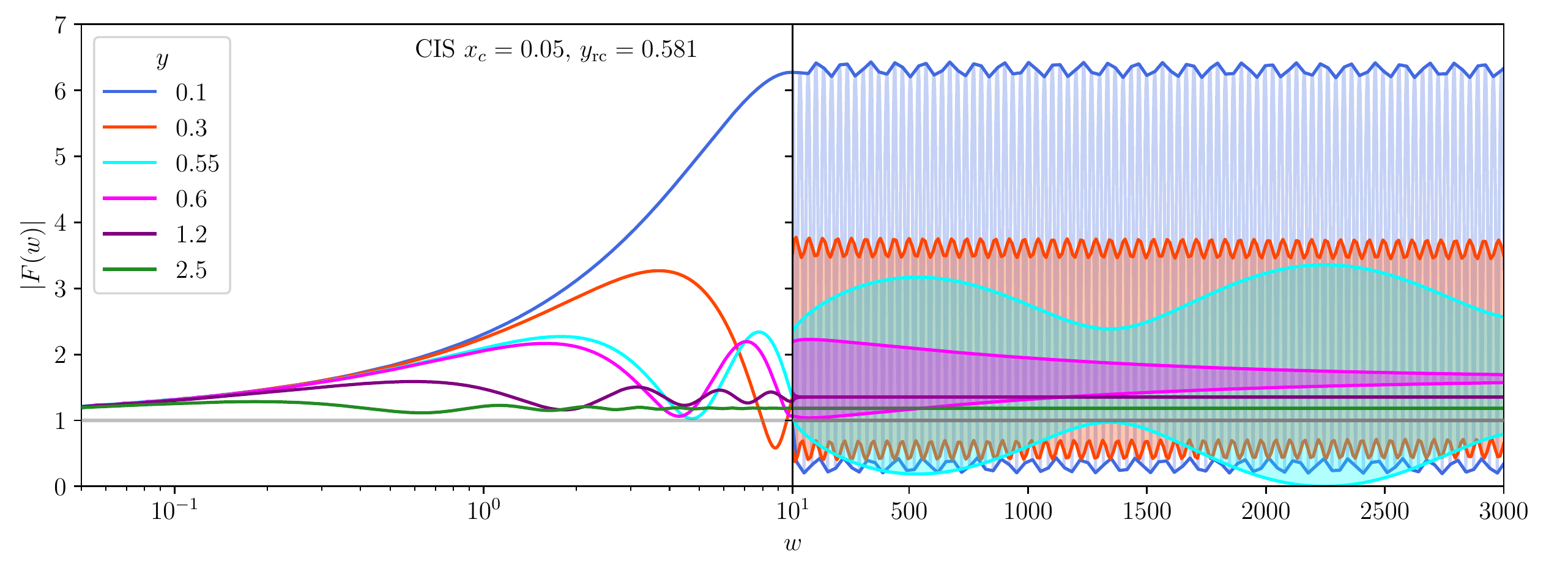}
 \caption{Amplification factor for a CIS with $x_c = 0.05$ and different values of the impact parameter $y$. The panels correspond to $w\in (0.05, 10)$ in logarithmic scale (left) and $w\in (10, 3000)$ (right) in linear scale. The envelope of the WO predictions is shown as thick solid lines (right panel). Note the different behaviour of the envelopes on both sides of the caustic ($y = 0.55, 0.6$, cyan and magenta respectively).
 \label{fig:cis_vary_y}
 }
\end{figure*}

Let us now discuss the effect of the CIS lens parameters in the WO regime. Figure \ref{fig:cis_vary_xc} shows the amplification factor at $y=0.3$ for different values of the core size $x_c$, between 0 (SIS) and 0.15, for which the source is right inside the caustic $y_{\rm rc}=0.3187$ (Eq.~\ref{eq:cis_yrc_caustic}). At low $w$, $x_c$ has a moderate effect on the onset of magnification, with smaller cores associated with higher first peaks at low frequency. This is expected, as a larger $x_c$ decreases the enclosed mass near the lens center.

The beating pattern sets in at intermediate frequencies $w\gtrsim 10$. Its width is given by the time delays between GO images and its amplitude by the relative magnification.
The primary, narrow oscillation is associated with interference between the minimum and the saddle point, $\Delta w_N \propto \phi_S^{-1}$ (recall $\phi_L=0$). This is the only oscillation seen for the SIS at high $w$, although a damped oscillation on the envelope due to the contribution from the cusp (similar to the gSIS, Sec.~\ref{sec:slope_lens} and Eq.~\ref{eq:F_cusp}) can be appreciated for $w<100$. In contrast, the CIS has a smooth profile.

A finite core size introduces a secondary modulation in $|F(w)|$. This broad envelope is due to the central image, with a characteristic period of oscillation $\Delta w_B \propto (\phi_H-\phi_L)^{-1}$. Lowering the value of $x_c$ increases the width and height of this effect because the location of the caustic approaches the source (recall $y$ is fixed). For the most extreme case shown ($x_c=0.15$), the magnification of the central image is sufficiently large to overtake the other two images. The central and right panels of Fig.~\ref{fig:cis_vary_xc} show the envelope of $|F(w)|$ to highlight the secondary modulation (this is similar to the broad gSIS, cf.~Fig.~\ref{fig:gSIS_different_k}, top panel).

The proximity to the caustic determines the convergence to GO. This can be seen by the difference between the WO (solid) and GO (dashed) envelopes in Fig.~\ref{fig:cis_vary_xc}. For WO and GO to agree well, a first requirement is to have negligible bGO corrections, $w\gg \sqrt{|\mu_I|}\Delta_I$ for all images $I$, Eq.~\eqref{eq:bGO}. In addition, it is necessary that $w \gg |\phi_I-\phi_J|^{-1}$ for all pairs of images with $I \neq J$. That is, the singular features in the time-domain integral are sufficiently far apart (relative to $1/w$) to be resolved by the Fourier transform at a given $w$, cf.~Sec.~IV and Fig.~2 of Ref.~\cite{Tambalo:2022plm}. The low/intermediate frequency behaviour of the $x_c=0.15$ curve in Fig.~\ref{fig:cis_vary_xc} can be heuristically understood as the two images (maximum and saddle point) acting as a single GO feature and gradually becoming distinguishable at higher frequencies. 

Let us now examine the effect of the impact parameter and the transition to the single-image regime. Figure \ref{fig:cis_vary_y} shows the amplification factor for $x_c=0.05$ and $y$ between 0.1 and 2.5, with a value within and closely outside the caustic $y_{\rm rc}=0.581$. The lowest impact parameters ($y=0.1$, $0.3$) show the same trends discussed above, with a narrow primary modulation and a broader secondary one due to the central image. The range of amplitudes $|F(w)|$ for the more aligned system, due to larger magnification.
The prediction for $y=0.55\lesssim y_{\rm rc}$ shows the same enhancement and broadening of the secondary modulation due to proximity to the caustic (but in the multiple-image regime) seen in Fig.~\ref{fig:cis_vary_xc}.

The predictions are qualitatively different in the single-image regime. At high frequencies there is only a single image, and therefore $|F(w)|\to \sqrt{|\mu_L|}$ when $w\to \infty$. For sizeable impact parameters, this limit is achieved at comparably low $w$. An additional feature is a weak pattern of the amplification at low frequencies $1\lesssim w\lesssim 10$ (e.g.~at $y=2.5$), which becomes more apparent for moderate impact parameters ($y=1.2$). As the source approaches the caustic, $y=0.6\gtrsim 0.581$, this oscillation not only becomes very pronounced at low $w$, but it also extends to very high frequencies $w\sim \mathcal{O}(10^3)$. In this regime, one observes a primary beating pattern, but with a damped envelope (rather than a modulated one, as in the multiple-image regime). This is a WO feature, absent in GO. 

\subsection{Qualitative comparison and detectability} \label{sec:lens_detecability}

After discussing the lens models in the previous sections, we would like to assess whether the CIS and gSIS lens profiles can be distinguished from SIS by future GW detectors. We first discuss the information contained in the GO/bGO limit (Sec.~\ref{sec:parameter_reconstruction_go}). Then, we address the central images and bGO signatures (Sec.~\ref{sec:lens_near_central_images}). Finally, we present the mismatch between the models and the SIS as a function of the lens parameters (Sec.~\ref{sec:mismatches}). A quantitative analysis of the lens parameter reconstruction will be presented separately, Sec.~\ref{sec:forecasts}.

\subsubsection{Parameter reconstruction for $w\gg1$}\label{sec:parameter_reconstruction_go}

Let us start our discussion by establishing how many lens parameters can be independently constrained for high-frequency sources.
The GO amplification factor \eqref{eq:lensing_geometric_optics} is fully characterised by two continuous parameters per image: the magnification $\mu_J$ and the time delay $\propto M_{Lz} \phi_J$.\footnote{The Morse phase can be determined in some situations \cite{Dai:2017huk,Ezquiaga:2020gdt}, but it takes only discrete values. Moreover, it is the same in all three lenses (SIS, CIS and gSIS) for the minimum and saddle images, and the maximum when present. Therefore, we will not consider it as a distinguishing feature in this subsection.}
However, one can only measure differences in time delays (the overall Fermat potential is degenerate with the coalescence time) and magnification ratios (the overall amplitude is degenerate with the source distance). 
Therefore, in GO one can determine at best
\begin{equation}\label{eq:parameter_reconstruction_GO}
    N_\text{lens par.} = 2N_{\rm images}-2 \quad \text{(GO)}
\end{equation}
lens parameters, even in the absence of measurement uncertainties. A known consequence of this relation is that weakly-lensed signals, $N_{\rm images} = 1$, do not allow any lens parameter to be inferred.
The situation is slightly better when including bGO corrections \eqref{eq:bGO}, as all additional $\Delta_I$ factors can in principle be measured. In that case,
\begin{equation}\label{eq:parameter_reconstruction_bGO}
    N_\text{lens par.} = 3N_{\rm images}-2 \quad \text{(bGO)}\,,
\end{equation}
and one can in principle reconstruct $N_{\rm images}$ additional lens parameters. Similar consideration can be applied to other $1/w$ corrections, such as the cusp contribution of the SIS and gSIS lenses, Eq.~\eqref{eq:F_cusp}.

An additional consideration pertains to the precision with which the GO parameters can be reconstructed. In general, the time delays can be determined with precision $1/M_{Lz}$, while the magnification ratios uncertainty scales as $1/{\rm SNR}$ \cite{Ali:2022guz}. Ultimately, the quality of the different parameters affects the GO limit of the reconstruction, as we will explore in detail in Sec.~\ref{sec:forcast_CIS}, \ref{sec:forcast_gSIS}.

Addressing lens-parameter reconstruction in the full WO regime is not as straightforward. Non-perturbative WO corrections are given by a general function of $w$ which cannot be captured in a finite number of parameters. This is in principle promising for reconstructing generic lensing potentials. However, important degeneracies still exist even for observations with ${\rm SNR} \gg 1$ (e.g.~the projected matter distribution is a two-dimensional function, while the WO amplification factor is one-dimensional). 
In practical situations, we are mainly limited by the precision of GW observations, both in the WO and perturbative regimes.

Let us now focus on the lens models discussed above. The distinction between GO and bGO reconstruction is starkest for the narrow gSIS. As these lenses form two images but have three parameters, a reconstruction is only possible for intermediate-mass lenses, when bGO effects are relevant. This will be shown explicitly in Sec.~\ref{sec:forcast_gSIS}. 
Both broad gSIS and CIS lenses predict the existence of a third image, which qualitatively changes the parameter estimation in the high-$w$ limit. Its detection allows us to reconstruct the three lens parameters even for very massive lenses, at least in principle, as we will see more in detail in Sec.~\ref{sec:forcast_CIS}.

\subsubsection{Properties of central images} \label{sec:lens_near_central_images}

\begin{figure*}
 \includegraphics[width=0.49\textwidth]{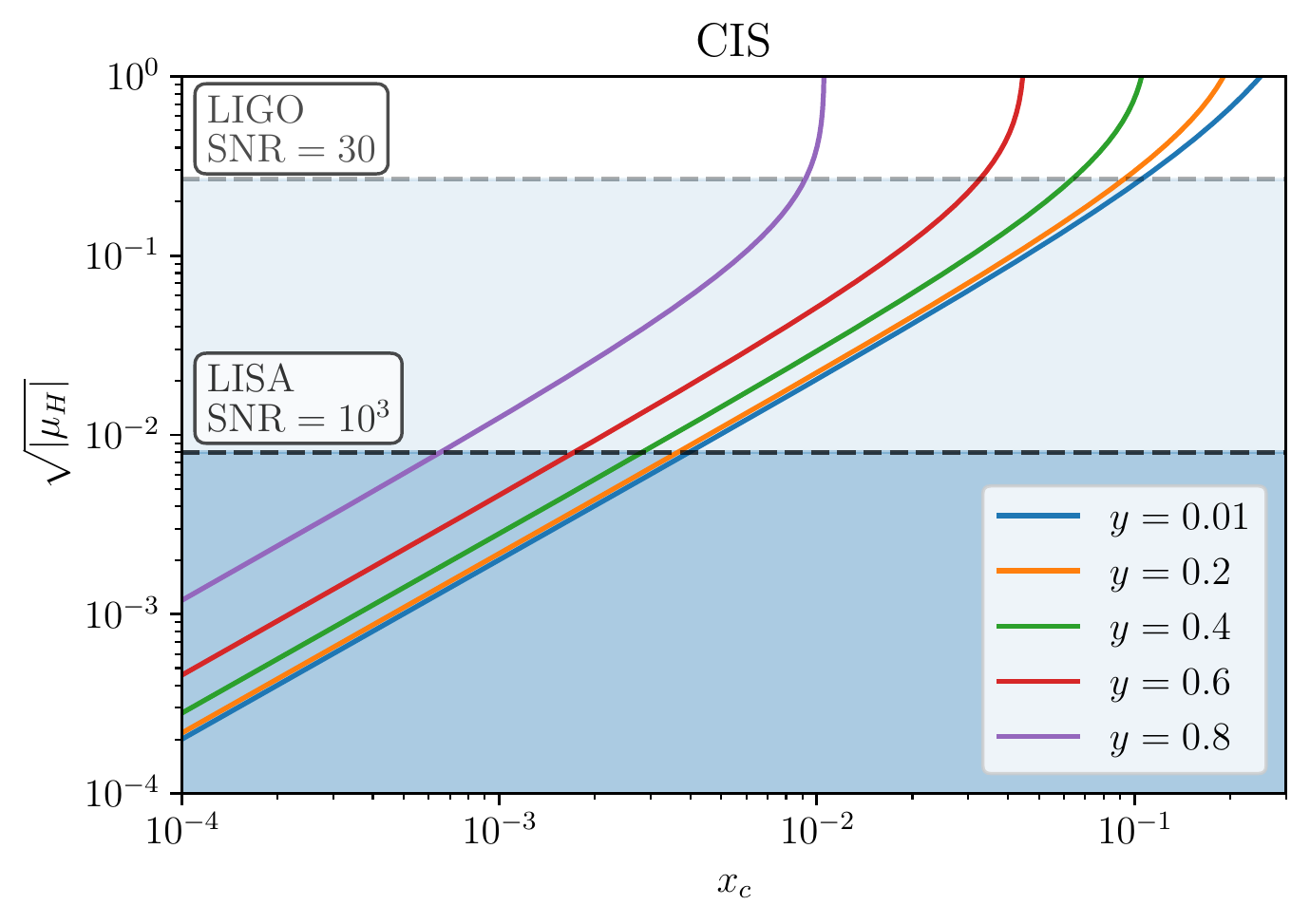}
 \includegraphics[width=0.49\textwidth]{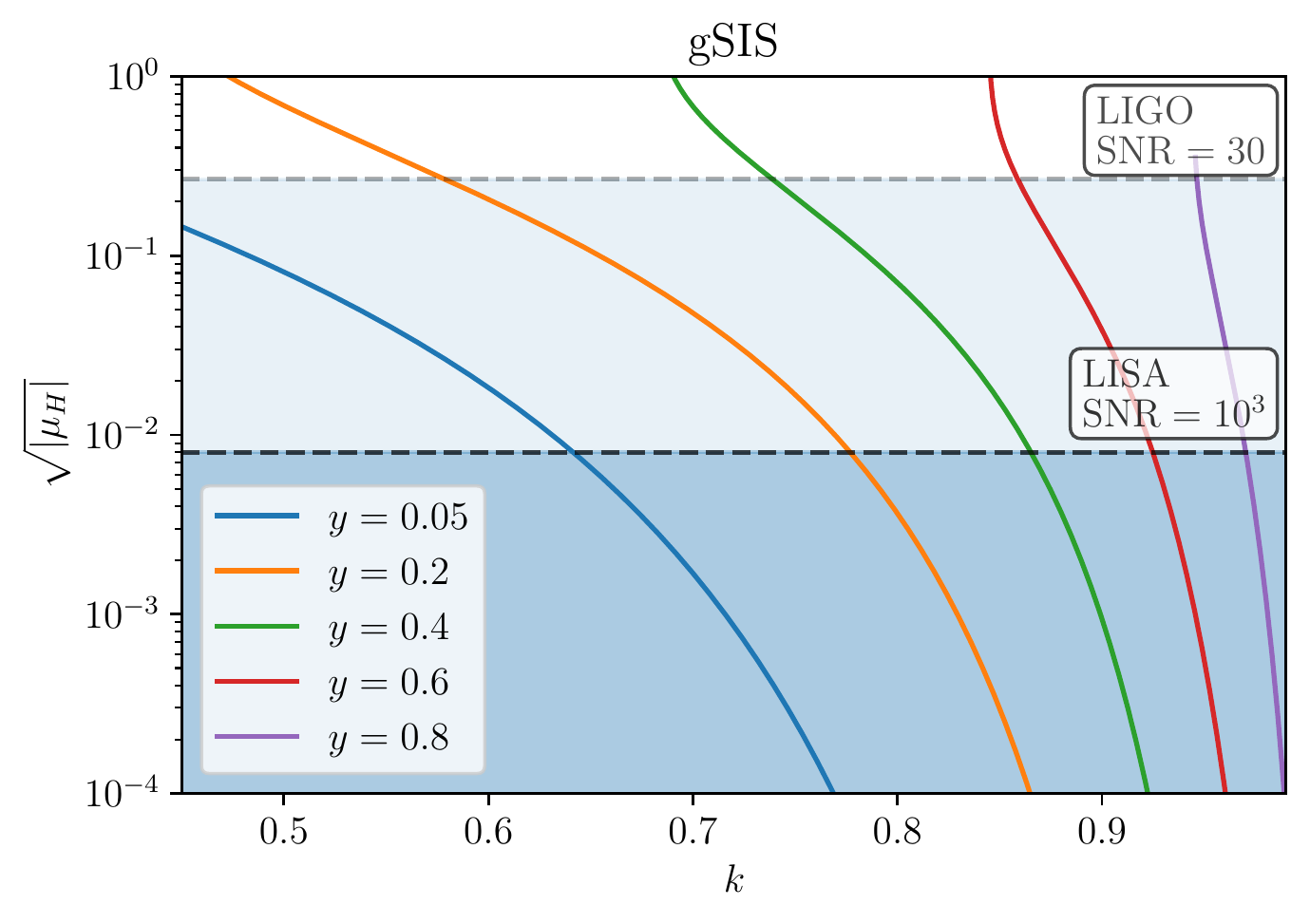}
 \caption{Detectability criterion in GO: magnification of the third image as a function of the lens parameter. Magnifications above the threshold values $\rm SNR_{th} = 8$ are detectable by GW experiments. Horizontal lines correspond to typical LIGO and LISA events, with SNR of $30$ and $10^3$ respectively.
\textbf{Left:} CIS case, as a function of the core size $x_c$.
\textbf{Right:} gSIS case, as a function of the slope parameter $k$.
\label{fig:detectability_third_img}
}
\end{figure*}

Both CIS and broad gSIS lenses form a central image $x_H$ in the strong-lensing regime. This image is associated with the maximum of the Fermat potential, with $|\mu_H|<1$. It is the closest to the lens' center. The central image has a large bGO correction (the largest if the lens is singular at the origin, as in the case of the gSIS, Figs.~\ref{fig:slope_IS_GO}, \ref{fig:slope_IS_GO_narrow}). 
Since it is absent for the SIS (or narrow gSIS), its detection provides important information about the lens.

For generic values of the impact parameter (not too close to the critical curve), the central image is fainter than the other two. Depending on the lens parameters and the SNR of the unlensed GW, it might not be possible to detect it.
For CIS, the magnification of the central image is given approximately by Eq.~\eqref{eq:CIS_mag_central_img} and is controlled by $x_c$. 
A similar reasoning applies to the broad gSIS lens with the parameter $k<1$ (here the 
relevant parameter controlling the magnification of the third image is $1-k$).
We first consider for what values of the lens parameters the central image has an SNR higher than a given detection threshold (which we call $\rm SNR_{th}$). For LIGO, the typical threshold value is $\rm SNR_{th} \gtrsim 8$ for an event to be considered detectable. The detectability of the central image can be increased by considering sub-threshold triggers $\rm SNR < \rm SNR_{th}$ \cite{Li:2019osa,McIsaac:2019use}, but this increases the chance of a noise fluctuation or other event mimicking the signature \cite{Caliskan:2022wbh}. 
We take ${\rm SNR} = 30$ for LIGO events (note that LVK events are volume-limited, hence typical SNR values for lensed events will be close to the detection threshold). 
Even in the case of LISA we consider the same threshold $\rm SNR_{th} \gtrsim 8$, although the typical $\rm SNR$ can be much higher than for LIGO. For LISA we take ${\rm SNR} = 10^3$.

Figure \ref{fig:detectability_third_img} shows the amplitude of the central image as a function of $x_c$ and $k$, for various values of the impact parameter. We can notice, in the CIS case, that $x_c \gtrsim 5\times10^{-3}$ is typically detectable with LISA signals and the dependence on $y$ is minimal (unless we are very close to the critical $y$). This results from having a finite minimum magnification for the central image, as obtained in Eq.~\eqref{eq:CIS_mag_central_img}. On the other hand, for gSIS detectability requires deviations $(1-k) \gtrsim 0.2$ and the dependence on $y$ is more pronounced than for CIS. 

As our analysis shows, central images are very sensitive to matter distribution near the center of the lens. In contrast, other images depend on the total projected mass, up to the radius where the image forms. This dependence is exact for axially-symmetric lenses in GO, as the reduced deflection angle $\alpha=M(x)/x$, where $M(x)$ is the total projected mass up to a radius $x$. Thus, finding central images of GWs might provide unique insights into the densest regions of galactic and DM halos \cite{Keeton:2001kg,Evans:2002ut, Li:2011js,Hezaveh:2015oya,Quinn:2016jte}. An application of this idea will be the prospects for DM tests, cf.~Sec.~\ref{sec:dark_matter}.

We note that observing a central image might be more difficult for lensed EM sources. First, the EM signal is demagnified by $|\mu_H| \ll 1$, which can be much smaller than the GW GO amplification $\sqrt{|\mu_H|}$. Second, because of its central location within the halo, the EM signal might be blocked by dust or gas, or outshined by other sources. Finally, because of the high frequency of EM signals, it is highly unlikely that additional information from bGO terms can be retrieved.
Thus, GWs provide a unique opportunity to probe central images and can be highly complementary to EM observations of lensed sources. See Sec.~4.4 of Ref.~\cite{Shajib:2022con} for a discussion of central-image searches with EM sources.

\begin{figure*}
 \includegraphics[width=0.45\textwidth]{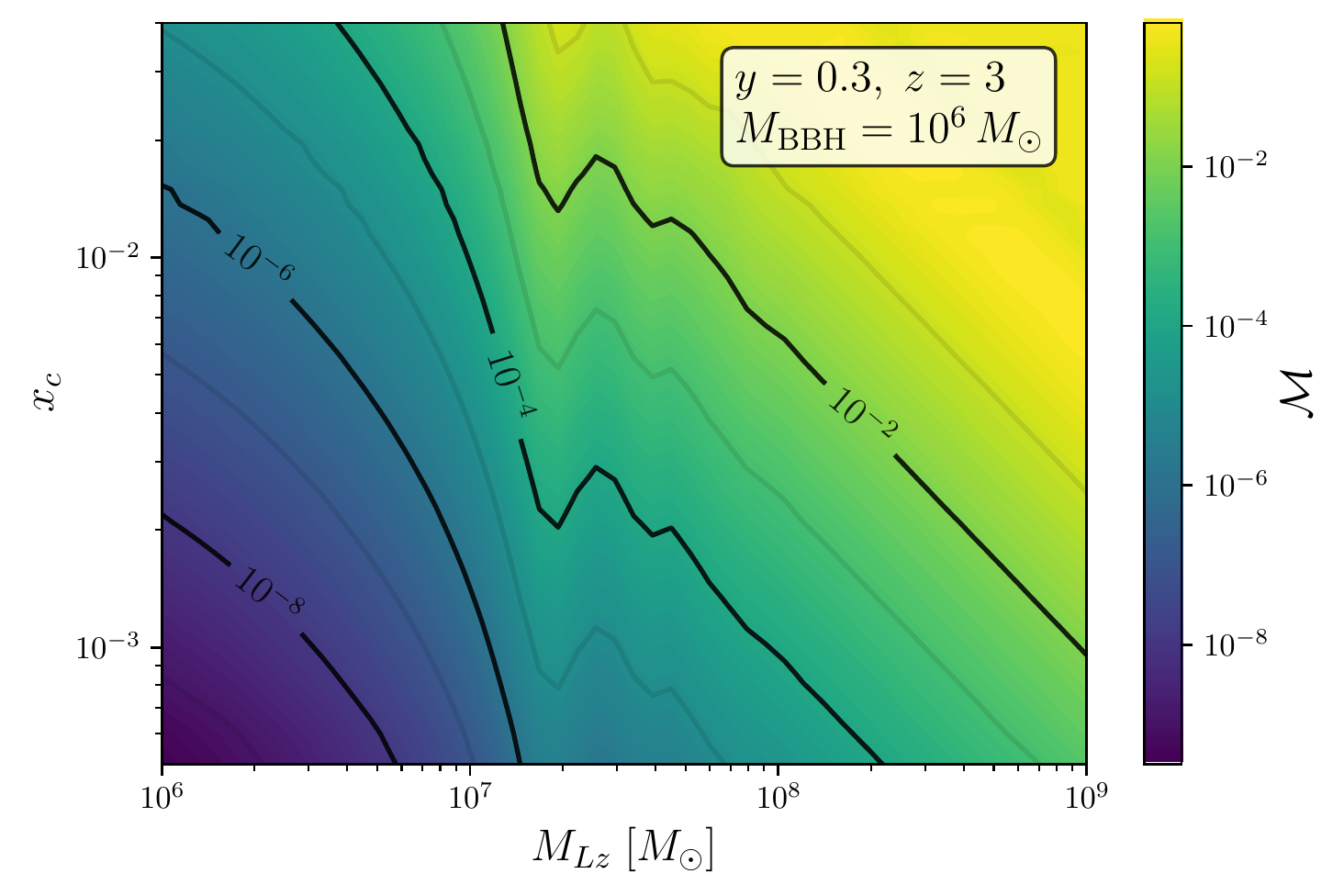}\quad
 \includegraphics[width=0.45\textwidth]{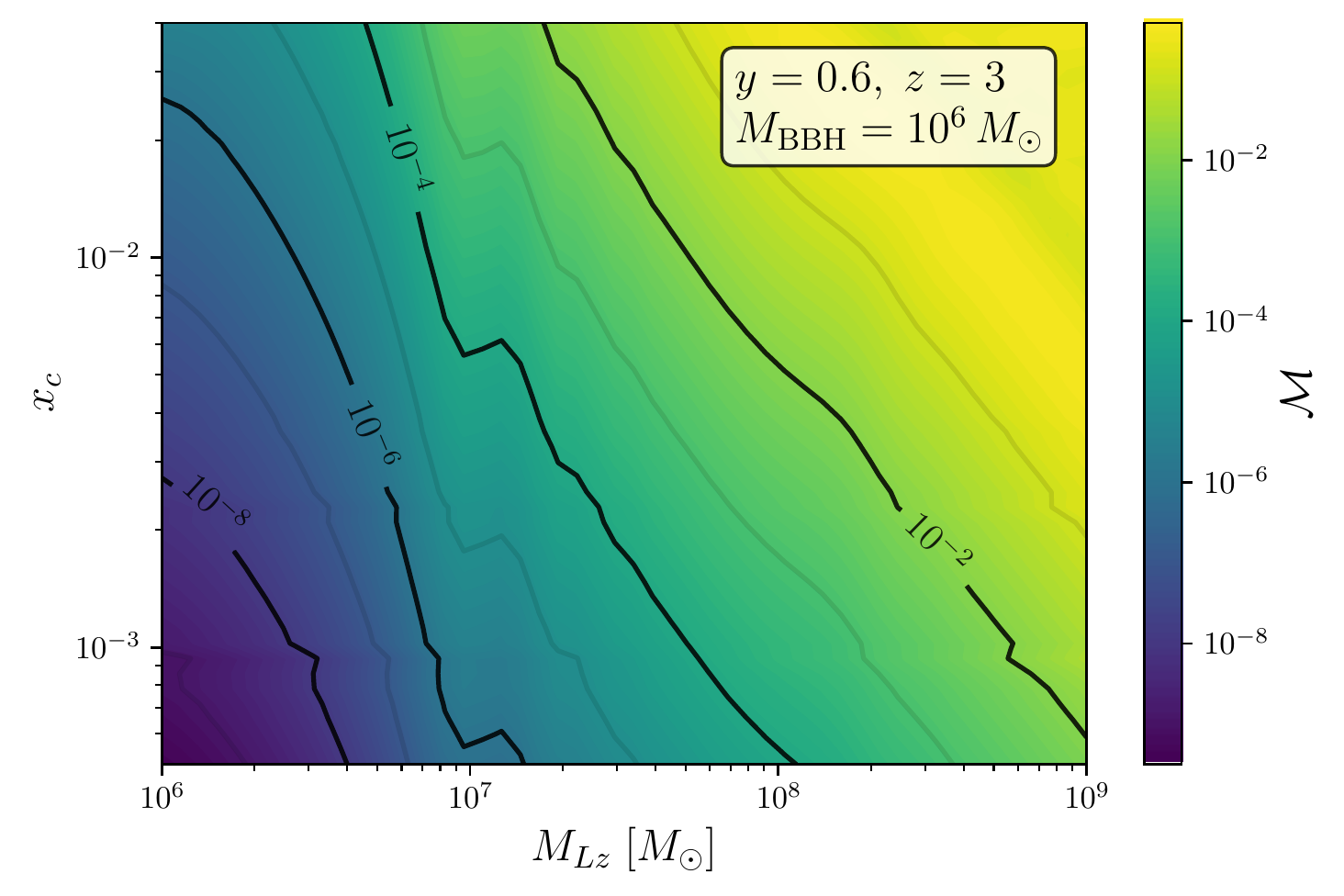}\\
 \includegraphics[width=0.45\textwidth]{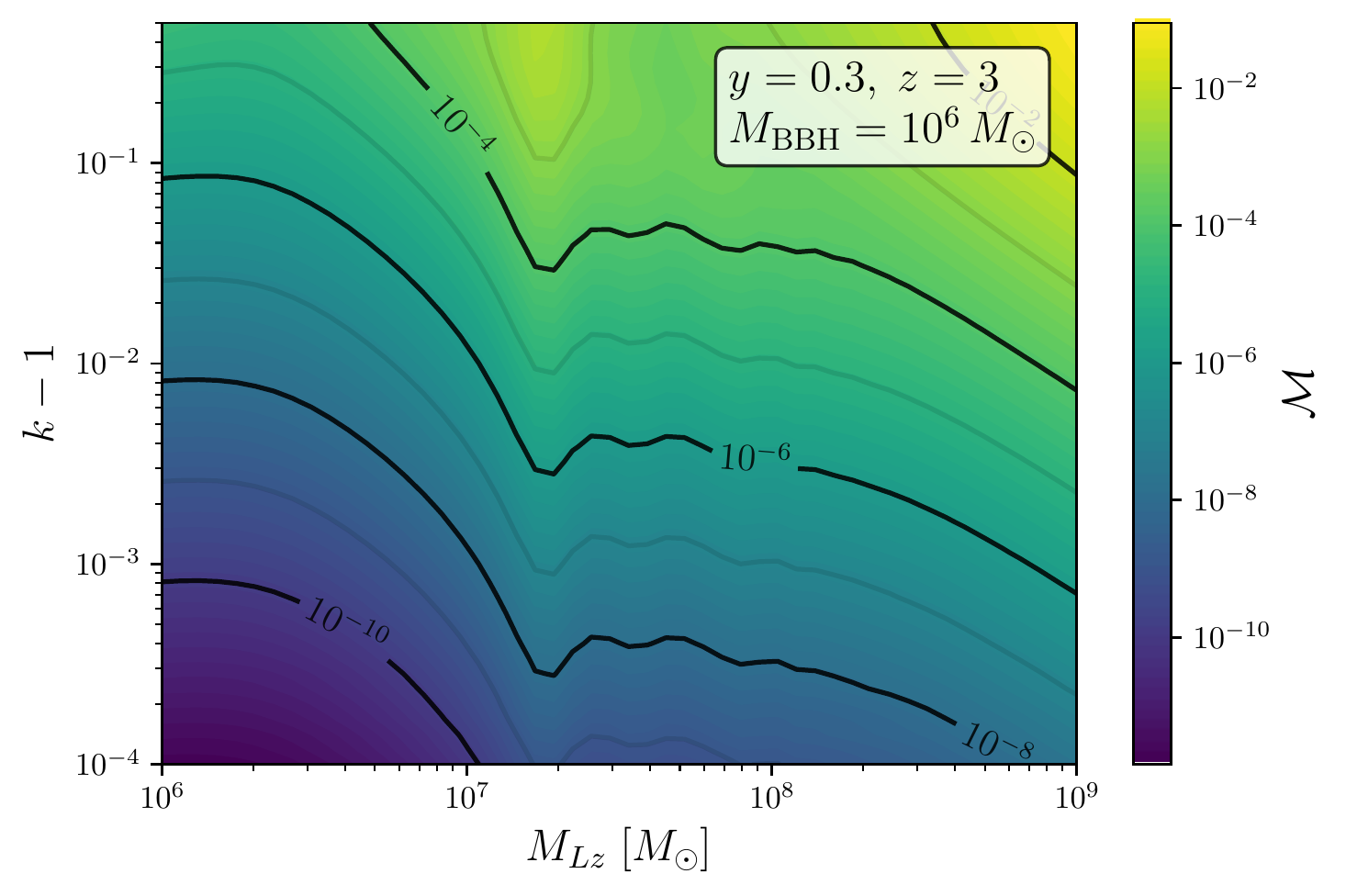} \quad
 \includegraphics[width=0.45\textwidth]{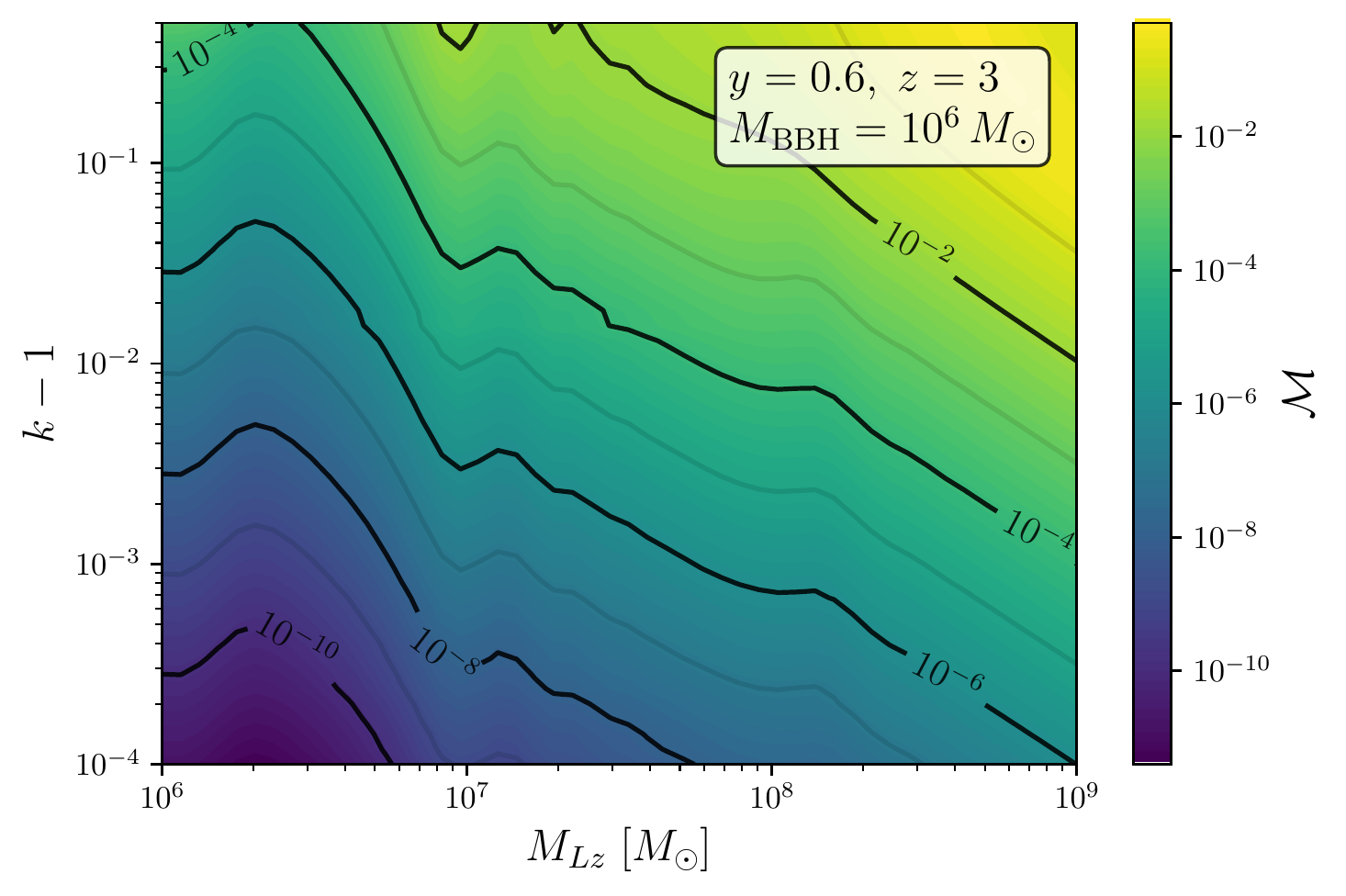}
 \caption{~Mismatch between CIS and SIS (top row), and gSIS and SIS (bottom row) for impact parameter $y=0.3$ (left column) and $y=0.6$ (right column) for a source mass $M_{\rm BBH} = 10^6\, M_{\odot}$. The mismatch is evaluated for different lens masses $M_{Lz}$ and by varying the lens parameter of the lens ($x_c$ for CIS and $k$ for gSIS).
\label{fig:mismatches}}
\end{figure*}

\subsubsection{Mismatch in the WO regime} \label{sec:mismatches}

In this Subsection, we discuss how well one can distinguish a GW waveform lensed by either CIS or narrow gSIS (with $k>1$) and a waveform lensed by an SIS. To assess the difference between the waveforms, we compute their mismatch, given a detector. For simplicity, we assume that both waveforms share the same parameters $M_{Lz}$ and $y$.
The mismatch between two waveforms $h_1$ and $h_2$ is defined as
\begin{equation}\label{eq:mismatch_def}
  \mathcal{M} \equiv 1 - \frac{(h_1|h_2)}{\sqrt{(h_1|h_1)(h_2|h_2)}} 
  \;.
\end{equation}
Here we introduced the noise-weighted inner product. For two signals $h_1(t)$, $h_2(t)$ with Fourier transforms $\tilde h_1(f)$, $\tilde h_2(f)$, it is defined as
\begin{equation}\label{eq:nwip}
 (h_1|h_2) 
 \equiv
 4~{\rm Re}
 \int_0^{+\infty}
 \frac{\tilde h_1(f)^* \tilde h_2(f)}{S_n(f)}
 \de f
 \,,
\end{equation}
where $S_n(f)$ is the sky-averaged detector strain sensitivity. 
The \emph{signal-to-noise ratio} (SNR) is defined in terms of this product as ${\rm SNR}^2 \equiv (h|h)$.

The right-hand side of Eq.~\eqref{eq:mismatch_def} is minimized over the phase and time difference between the two signals.
Notice also that the mismatch is invariant under rescalings of either waveform; therefore it is automatically minimized over the luminosity distance of, say, $h_1$ (which is indeed just a rescaling). 

A simple condition for detectability is that $\mathcal{M}\geq 1/{\rm SNR}^2$. This criterion is optimistic, as it neglects correlations between lens and source parameters that might make it more difficult to distinguish between lens models. Thus, the mismatch analysis is just a first step to addressing under which conditions two lenses can be distinguished.
We will address degeneracies when performing forecasts, in Sec.~\ref{sec:forecasts}.\footnote{Even though our mismatch criterion misses possible degeneracies, it does not make further assumptions about the difference between the two signals, $\delta h = h_1 - h_2$. As we will see instead, the Fisher-matrix approach assumes that this difference can be expanded in powers of the lensing parameters: this additional assumption is not obviously satisfied in our cases.}

In Fig.~\ref{fig:mismatches} we provide few examples of the mismatches between a CIS and SIS (top row) and gSIS and SIS (bottom row) for impact parameters of $y=0.3$ (left column) and $y=0.6$ (right column). The mismatch is given as a function of both the lens mass (horizontal axis) and the additional lens parameter ($x_c$ or $k-1$). We have assumed a LISA equal-mass binary with mass $M_{\rm BBH} = 10^{6}\, M_{\odot}$ at redshift $z=3$.  
This type of source has ${\rm SNR}\sim 10^{3}$, allowing the two waveforms to be distinguishable when the mismatch exceeds $\sim 10^{-6}$. 
See Sec.~\ref{sec:forcast_CIS} for more details on the waveforms we use.

The merger frequency of the signals enters the GO regime for lens masses larger than $M_{Lz}\sim 10^{7}\, M_{\odot}$ (that is, $w_{\rm ISCO}\gg 1$, where ISCO refers to the innermost stable circular orbit): one can notice a change of behaviour of the mismatch in all the plots around this value.
Increasing the impact parameter improves the mismatch slightly. This can be understood, at least in the GO regime, from the larger magnification of the third image for CIS. On the other hand, for gSIS the contribution of the cusp becomes larger.

In the case of the CIS, the mismatches are very sharp as a function of $M_{Lz}$, with a milder dependence on $x_c$. Thus, very small cores, even $x_c \sim 10^{-3}$, can be distinguished for sufficiently large lens masses. This result will be also confirmed by the Fisher analysis in the next section.
The case of the gSIS is less favourable for probing $k - 1$ close to zero, as the mismatch curves are flatter in $M_{Lz}$. Then, the difference in slope $k-1$ can be distinguished only up to $\sim 10^{-2}$ for $y=0.3$ ($10^{-3}$ for $y=0.6$).

\section{Probing lens features with LIGO and LISA}\label{sec:forecasts}

The existence of central images allows GW observations to probe the inner regions of matter halos. Together with the additional images and WO effects, they allow GWs to probe the matter distribution of gravitational lenses. In this Section, we perform a Fisher matrix forecast of the sensitivity expected from LISA and LIGO observations. Section \ref{sec:forecast_framework} presents the frameworks and assumptions. We present the results for CIS in Sec.~\ref{sec:forcast_CIS} and for gSIS in Sec.~\ref{sec:forcast_gSIS}. Appendices \ref{sec:forecast_implementation}, \ref{sec:forecast_GODA}, \ref{sec:fisher_validity} and \ref{sec:SPA} provide further technical details.

\subsection{Framework and assumptions}\label{sec:forecast_framework}

We will evaluate the sensitivity of different experiments to lens properties using a simplified Fisher matrix forecast \cite{Vallisneri:2007ev}, assuming Gaussian and stationary noise.
The probability distribution of the model parameters $\theta^i$ around their true values, $\theta^i = \bar \theta^i + \Delta \theta^i$, can be estimated with a linearised-signal approximation (LSA).
In the limit of high SNR, the leading contribution to the likelihood $\mathcal L$ only depends on the signal linearised around the fiducial values
\begin{align}
    - \log \mathcal L
    &\simeq 
    \mathcal F_{ij}\, \Delta \theta^i \Delta \theta^j
    \;,
\label{eq:likelihood_LSA}
\end{align}
where the \emph{Fisher matrix} $\mathcal F_{ij}$ is defined as
\begin{equation}
  \mathcal{F}_{ij} 
  \equiv 
  (\partial_i h|\partial_j h)
  \,.
\end{equation}
 
In the high-SNR limit, the standard deviations $\sigma^i$ for the parameters $\theta^i$ are obtained from the inverse Fisher matrix as 
$\sigma^i=(\mathcal{F}^{-1})_{ii}^{1/2}$ 
(where no summation is implied).
The Fisher matrix gives the unmarginalised error on a parameter, i.e.~keeping all parameters fixed $\sigma_i^{\rm fix}=(\mathcal{F}_{ii})^{-1/2}$.
In conclusion, for high SNR the leading contribution to the $\sigma^i$s is solely determined by the first derivatives of the signal. In this sense, the LSA is a valid approximation.

We consider the signal $h$ to be given by the lensed waveform $\tilde h_L(f) = F(w(f))\tilde h_0(f)$. This simplification assumes a single detector and neglects the time dependence of the antenna pattern. This is a good approximation if the signal's duration is much shorter than the timescale of the detector's motion 
\begin{equation}\label{eq:timescale_signal_requirement}
    t_{\rm signal}
    \ll
    t_{\rm motion}
    \sim 
    \left\{
    \begin{tabular}{l}
    1\text{ yr (LISA)} \\
    1\text{ day (LIGO)}
    \end{tabular}
    \right. 
\end{equation}
This criterion is always satisfied for LVK signals, for which $t_{\rm signal}$ can last as long as few minutes for neutron-star mergers, extending to tens of minutes for next-generation detectors.
For heavier sources, signals can be as short as a fraction of a second.
In the case of LISA, high-mass binaries can be observed months or years before coalescence. For this reason, we will therefore only consider the last month before the merger for LISA sources. The results change only minimally when considering the entire duration of the signal, as most of the SNR is located close to the merger \cite{Marsat:2020rtl}. 
For LIGO, the information about the source sky location (from multiple detectors) can be used to correct for the antenna pattern. Therefore, we do not expect it to be a limiting factor in our analysis.

Strictly speaking, the staticity requirement also sets a limit on the time delay between images, and thus on the effective lens mass
\begin{equation}\label{eq:timescale_lens_requirement}
    \Delta t_{IJ} 
    \sim 1 \, {\rm s}\, \frac{M_{Lz}}{10^5 M_\odot}
    \ll t_{\rm motion}
    \,. 
\end{equation}
Given the short-signal condition \eqref{eq:timescale_signal_requirement}, violations of the above relation imply observations in the GO regime, since $t_{IJ}\gtrsim t_{\rm motion}\gg t_{\rm signal}\gg 1/f$.
In practice, Eq.~\eqref{eq:timescale_lens_requirement} is necessary to ensure that the different antenna pattern is approximately constant for all GO images. This is a very stringent requirement. In practice we expect our results to be valid as long as all images can be detected and associated (this will be the case if the $\rm SNR$ of the different images is comparable). In this sense, gaps in observation, finite survey time \cite{Sereno:2010dr} and false-positive image association \cite{Caliskan:2022wbh} might be more stringent requirements.

We will further assume an edge-on source aligned with the detector, so $\tilde h_0 = \tilde h_+$. Because of this simplified setup, we will neglect the error on the source orientation and sky localization, whose determination requires including rotating antenna patterns or multiple detectors.  
Regarding the source parameters, we will marginalize over the coalescence phase $\phi_0$ and luminosity distance from the source $\mathcal{D}_{S} = (1+z_S)^2 D_S$.
We will assume equal-mass ratio, non-spinning sources, with all other parameters fixed. This follows the approach in Ref.~\cite{Takahashi:2003ix}, which focused on point-lens and SIS. A detailed analysis of the same lenses including detector motion, source properties and waveform models found that the results are robust against additional source parameters, and might even be more optimistic (e.g.~by the introduction of higher harmonics) \cite{Caliskan:2022hbu}.

We will normalize our results to a fiducial $\rm SNR$ of the lensed source ($10^2$ for LIGO, $10^3$ for LISA). The actual errors scale as $\sigma^i = \sigma^i_{\rm fid} {\rm SNR}_{\rm fid} / {\rm SNR}$. We expect some deviations from our assumptions to be approximately captured by a reduced $\rm SNR$: for instance, a sub-optimal sky localization or a different source orientation.
In our analysis, we will be primarily working with high-SNR signals. As we will see, however, the validity of the linearisation of Eq.~\eqref{eq:likelihood_LSA} is not always granted and needs to be explicitly checked. 
We will come back to this point below and when interpreting the results.

We compute the amplification factor and its derivatives using three methods, appropriate for different regimes. At high frequencies, $w > w_{\rm high} \gtrsim 100$, we use the bGO approximation summarised in Sec.~\ref{sec:lensing_go}. 
In the case of the gSIS lens, on top of the bGO contributions, we add the terms associated with the cusp, summarised in Eq.~\eqref{eq:F_cusp} (see also Sec.~III-A-2 of Ref.~\cite{Tambalo:2022plm}).
For intermediate frequencies $w_{\rm low}< w < w_{\rm high}$ we use instead the contour method summarised in Sec.~\ref{subsec:wo_numerical}. Derivatives of $F(w)$ with respect to the lens parameters are computed using the prescription defined in Appendix A of Ref.~\cite{Tambalo:2022plm}. 
For frequencies below $w_{\rm low} = 0.05$, we extrapolate using the analytic dependence described in Sec.~\ref{subsec:wo_low}.
The transition frequencies are defined so that the errors between methods are below 1\% for the amplification factor and 3\% for its derivatives with respect to the lens parameters, see Fig.~\ref{fig:forecast:high_freq_test}.
More details about the implementation of the Fisher matrix are given in App.~\ref{sec:forecast_implementation}.

The inversion of the Fisher matrix is very sensitive to numerical errors from the FFT truncation at high $w$. 
To test the high-$M_{Lz}$ limit calculation, we developed the \emph{Geometric Optics Diagonal Approximation} (GODA) of $\mathcal{F}_{ij}$, as well as its extension to bGO and rGO (the latter defined as including both bGO and cusp contributions). This approximation, described in App.~\ref{sec:forecast_GODA}, greatly speeds up the computation of the Fisher matrix by neglecting contributions that oscillate in frequency space which originate from products of amplification factors corresponding to different images $F_I^* F_J$. 
The result is a sum over images of terms in $|F_I|^2$ (hence diagonal in image space) times the unlensed SNR weighted by different powers of the frequency, such as $(h_0|h_0)$ and $(w h_0| w h_0)$.

At sufficiently high $M_{Lz}$ we expect the LSA approximation to break down. We diagnose this by evaluating the phase difference in the GO limit
\begin{equation}\label{eq:lsa_breakdown}
    \Delta_{\rm LSA}(w,\vect \theta)
    \equiv
    w(\vect \theta) \phi_J(\vect \theta)
    -w(\vect{\bar \theta}) \phi_J(\vect{\bar \theta})\,,
\end{equation}
Here $J$ labels the images while $\theta^i$ are the lens parameters.
The reason behind the LSA breakdown can be understood as follows. In the lensed waveform $h_{L}$, the lensing parameters appear in the magnification and time delays. The latter, in particular, are multiplied by $w$ in the phases of Eq.~\eqref{eq:bGO}. Linearisation is a good approximation if the changes in the phases are small. However, this is not necessarily the case for large $w$, or equivalently large $M_{Lz}$. 

We will indicate the values of the masses at which $\Delta_{\rm LSA} > 1$ in our plots (we check for this condition in the GO regime, $w > 100$).
We evaluate Eq.~\eqref{eq:lsa_breakdown} using Eq.~\eqref{eq:lensing_freq_dimensionless} to write $w$ as a function of $f$ and $M_{Lz}$, the latter displaced $1\sigma$ from the fiducial value (this is the parameter that gives the largest contribution to $\Delta_{\rm LSA}$ when varied from its fiducial value). 
The other lens parameters $\theta^i$ are taken as fiducial.
For the frequency we choose $f_*$, the value such that 90\% of the total SNR is given for $f\leq f_*$.
The image giving the largest contribution to $\Delta_{\rm LSA}$ is the central one (having the largest time delay).
Therefore, the expression we use for $\Delta_{\rm LSA}$ reduces to 
\begin{equation}\label{eq:LSA_cond2}
    \Delta_{\rm LSA}(w,\vect \theta)
    = 
    8 \pi G M_{Lz} f_{*} \, \phi_{H} \, \Delta \log (M_{Lz})
    \;,
\end{equation}
where $\Delta \log (M_{Lz})$ is the $1\sigma$ uncertainty on $\log (M_{Lz})$, as obtained from the Fisher analysis.
For the gSIS lens, with $k > 1$, $\phi_H$ is replaced by the time delay of the lens center (the location of the lens' cusp).

Above these lens masses, the breakdown of LSA indicates that the Gaussian approximation for likelihood does not hold, and the results need to be taken with caution. Nonetheless, whether the results are over or under-estimated will depend on the lens model: for the CIS (3 images) the Fisher approximation is correct, while for the narrow gSIS (2 images + 1 cusp) it leads to an over-estimation of the sensitivity at high $M_{Lz}$.
This result is consistent with the lack of available GO information (relative time delays, magnification ratios) to constrain all the lens parameters, as discussed in Sec.~\ref{sec:parameter_reconstruction_go}.
We will comment on the case of each lens in the corresponding Sections, with more details on Appendices \ref{sec:fisher_validity}, \ref{sec:SPA}.

Finally, we note that the LSA validity condition \eqref{eq:LSA_cond2} depends on the SNR of the signal (through the uncertainty $\Delta \log (M_{Lz})$, that scales as $\sim 1/ {\rm SNR}$). To evaluate this condition in our results, we use the fiducial $\rm SNR$ of the given experiment ($10^3$ for LISA and $10^2$ for LIGO). Events with lower $\rm SNR$ will violate the LSA at lower lens masses.

\subsection{Cored Isothermal Sphere}\label{sec:forcast_CIS}

\begin{figure*}
\centering
\hfill 
\includegraphics[width=1.0\columnwidth]{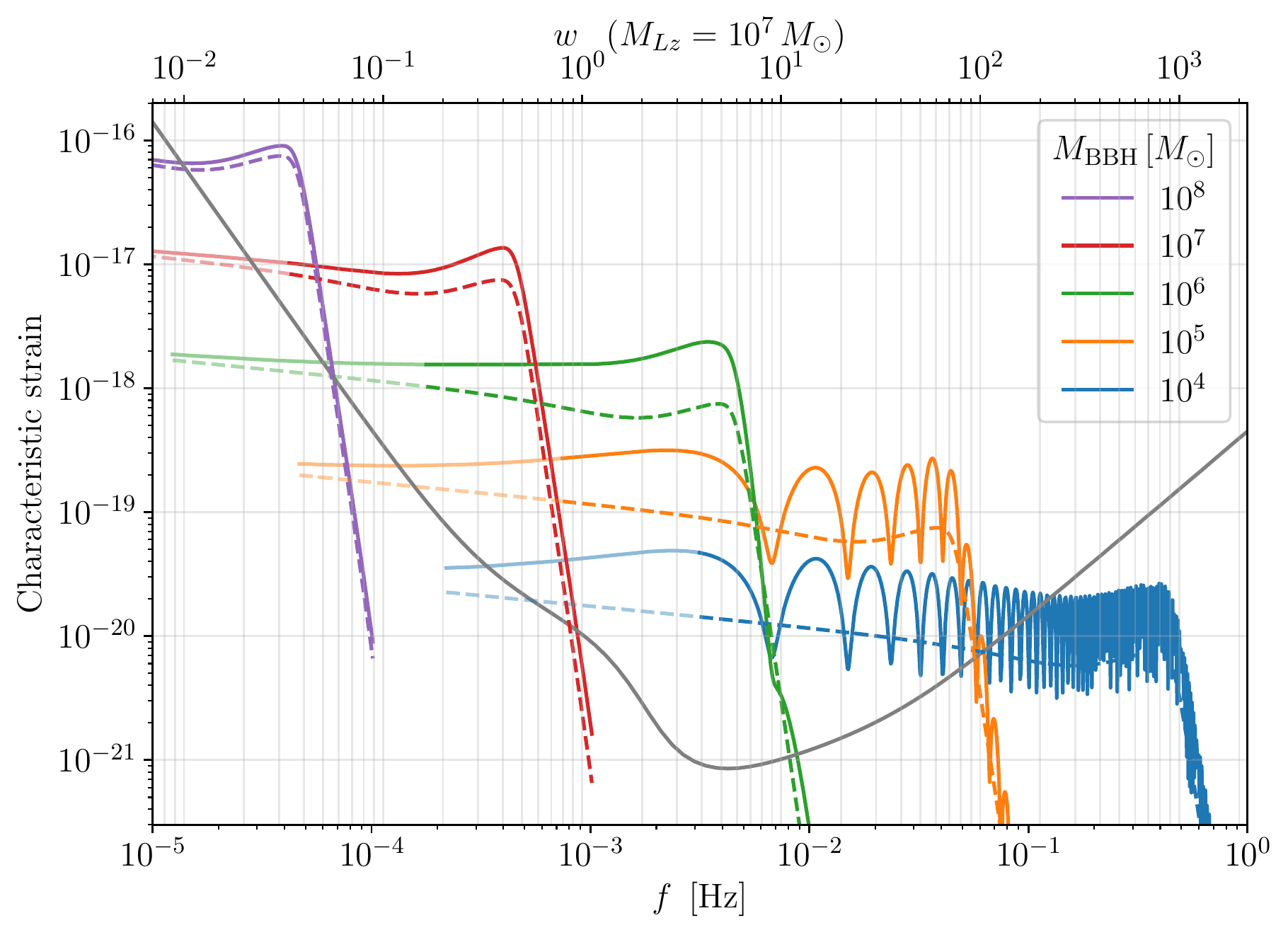}
\hfill
\includegraphics[width=0.91\columnwidth]{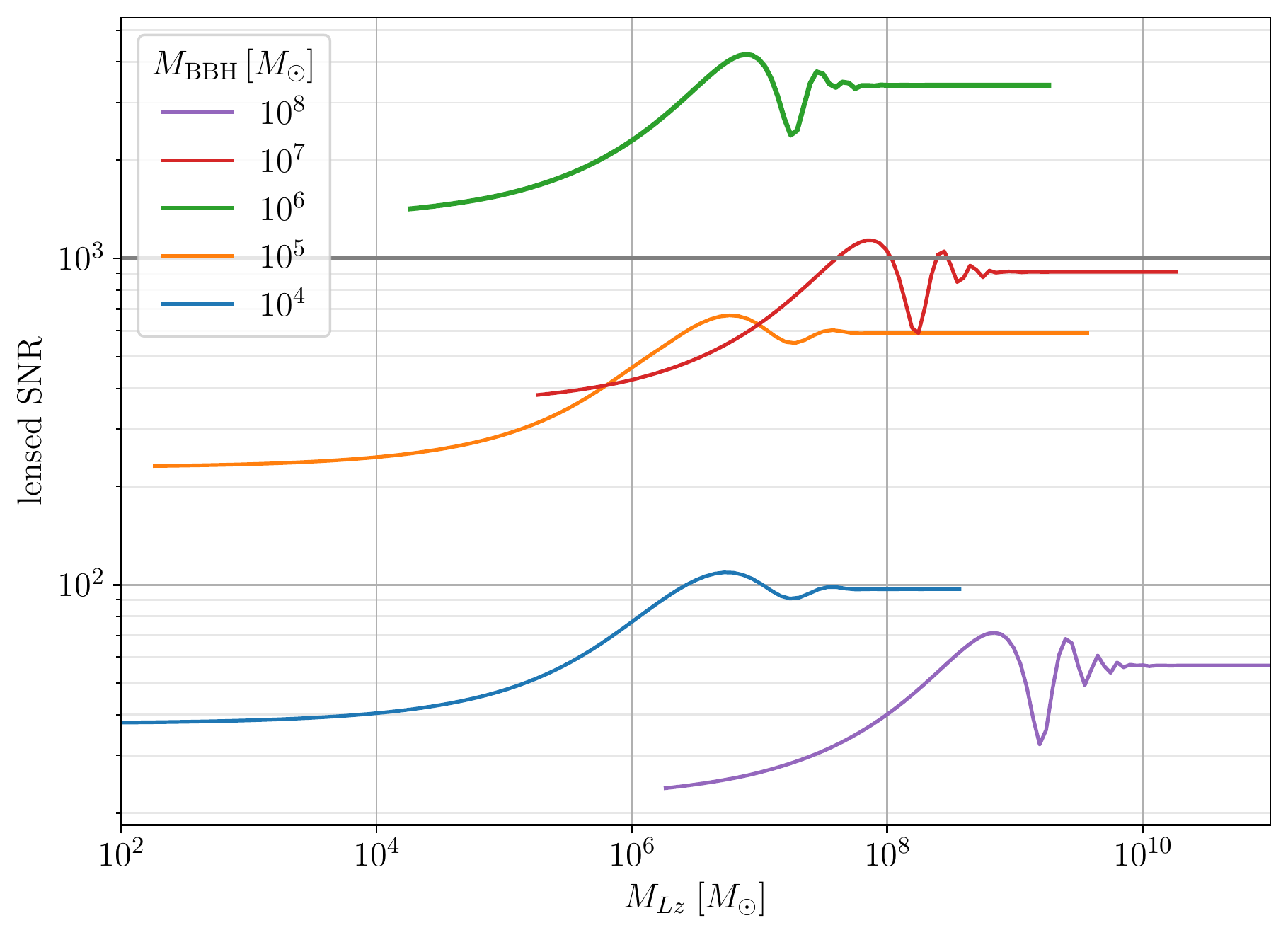}
\hfill 
\vspace{-10pt}
\caption{Summary of LISA sources. 
\textbf{Left:} Fourier-domain waveforms for different source masses. Dashed lines show the unlensed waveform, solid lines are lensed by a gSIS with $M_{Lz}=10^7\, M_\odot$, $y=0.3$, $x_c = 10^{-2}$.
Frequency components emitted before $t_{\rm merge}<1\, {\rm Mo}$, neglected in our analysis, are shown in lighter color.
\textbf{Right:} corresponding lensed SNR as a function of the effective lens mass, including only the last month of the signal before merger.
}
\label{fig:forecast_lisa_sources}
\vspace{10pt}

\centering
\includegraphics[width=0.95\textwidth]{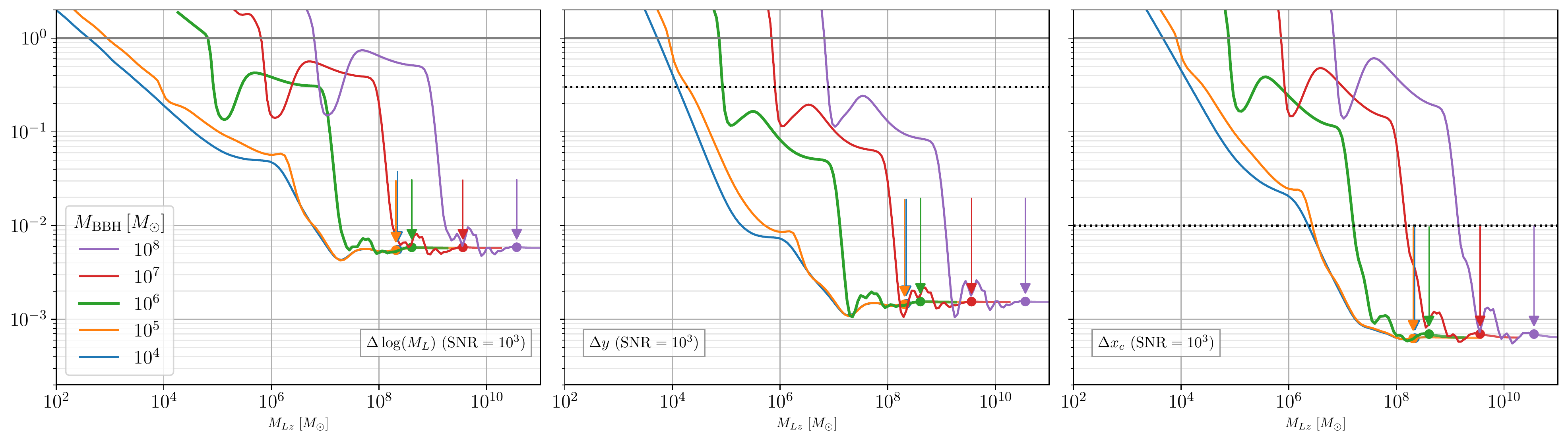}
\vspace{-10pt}
\caption{Constraints on the lens parameters ($\log(M_{Lz})$, $y$, $x_c$) 
as a function of the lens mass for LISA sources (cf.~Fig.~\ref{fig:forecast_lisa_sources}) lensed by a cored isothermal sphere with $y=0.3$, $x_c=10^{-2}$. 
Results are normalized to a lensed SNR of $10^3$. Fiducial values are shown as horizontal dashed lines. Vertical lines mark the breakdown of the LSA, condition \eqref{eq:LSA_cond2}.}\label{fig:forecast_LISA_source_mass_dependence}

\end{figure*}

\begin{figure*}

\centering
\includegraphics[width=0.95\textwidth]{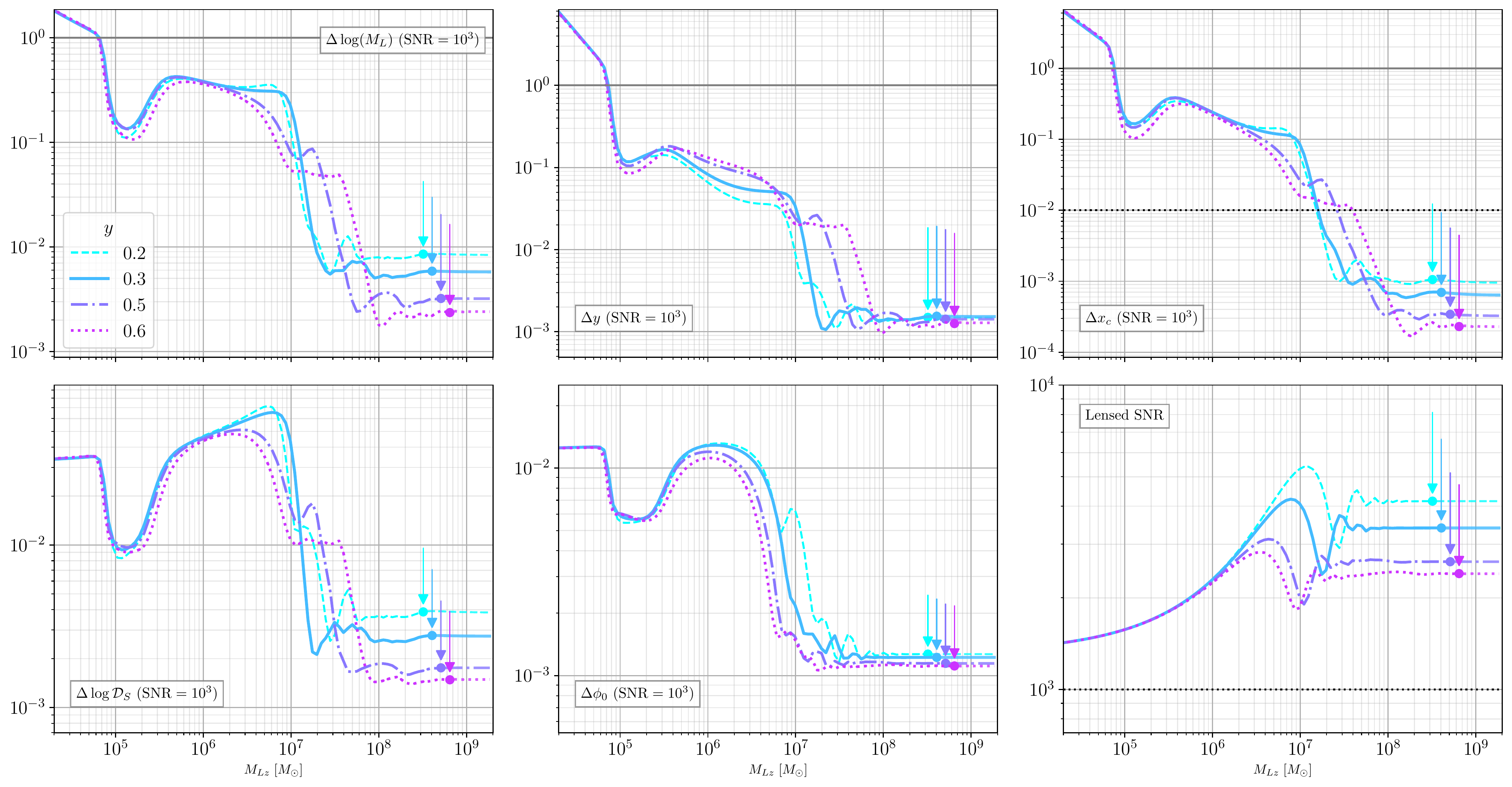}
\vspace{-10pt}
\caption{Constraints on parameters of the lens ($\log(M_{Lz})$, $y$, $x_c$) and source ($\log(\mathcal D_S)$, $\phi_0$) and lensed SNR
as a function of the lens mass and the impact parameter for a cored isothermal sphere with $x_c=10^{-2}$. The source is a $M_{\rm BBH}=10^6\, M_\odot$ equal-mass non-spinning binary observed by LISA. Results are normalized to a lensed SNR of $10^3$. Vertical lines mark the breakdown of the LSA condition \eqref{eq:LSA_cond2}.
}\label{fig:forecast_LISA_impact_parameter_dependence}

\vspace{10pt}

\centering
\includegraphics[width=0.95\textwidth]{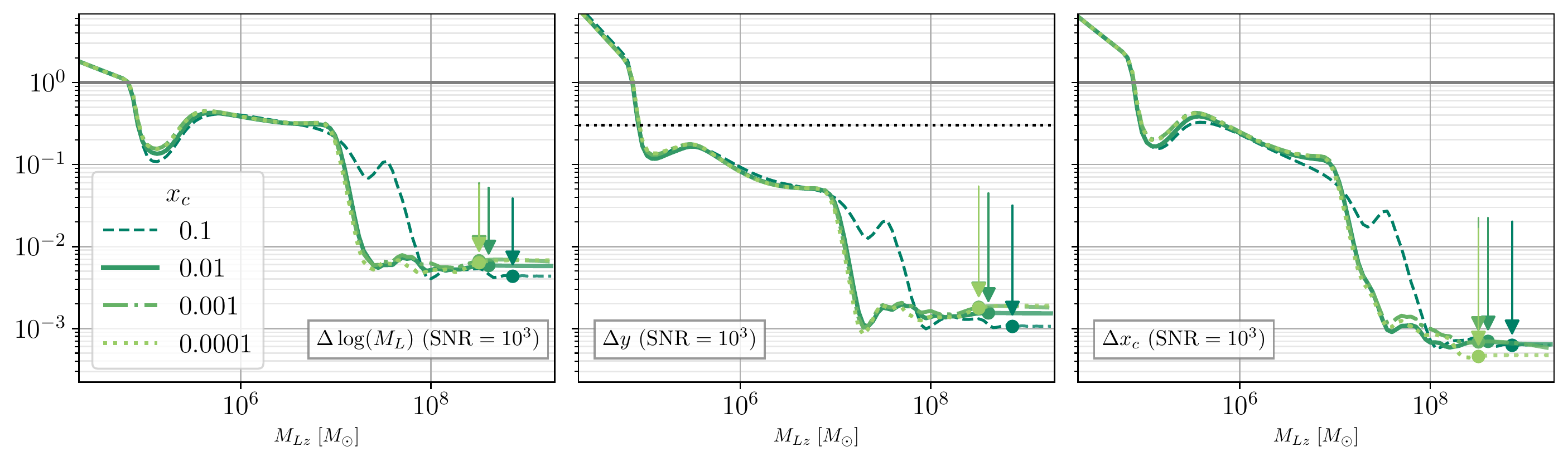}
\vspace{-10pt}
\caption{Constraints on the lens parameters ($\log(M_{Lz})$, $y$, $x_c$) as a function of the lens mass and the core size $x_c$ for a $M_{\rm BBH}=10^6\, M_\odot$ LISA source lensed by a CIS ($y=0.3$). Results are normalized to a lensed SNR of $10^3$. Vertical lines mark the breakdown of the LSA condition \eqref{eq:LSA_cond2}.
}\label{fig:forecast_LISA_core_size_dependence}
\end{figure*}

Let us now consider the capacity of different experiments to measure the lens parameters in the CIS lens. We will focus on the strong-lensing regime, with multiple GO images. We will present results for different source masses, impact parameters and fiducial slopes. We show the marginalised 68\% c.l.~posterior in the Fisher-matrix approximation as a function of the effective lens mass $M_{Lz}$, which can be obtained without recomputing $F(w)$. Our fiducial core size is $x_c=0.01$.

We start by considering LISA sources. We explore equal-mass, non-spinning binaries at $z=3$, with total mass $10^4 - 10^8 \, M_\odot$ (source frame), using \texttt{IMRPhenomD} waveform \cite{Husa:2015iqa, Khan:2015jqa}. Figure \ref{fig:forecast_lisa_sources} shows the lensed and unlensed strains along with the expected LISA sensitivity for $M_{Lz} = 10^7 \, M_\odot$. The figure highlights the last month of data before the merger, the only portion of the signal we included in the analysis, cf.~Eq.~\eqref{eq:timescale_signal_requirement}. We verified that the constraints vary negligibly when including longer data, except for the lightest sources $10^4 \, M_\odot$ for which the merger occurs off the LISA band. The second panel shows the lensed SNR for those sources as a function of $M_{Lz}$. The results of the marginalised errors will be rescaled to a fiducial SNR of $10^3$. 
This factors out the dependence on the distance exactly and sky localization and source inclination approximately.

\begin{figure*}
\includegraphics[width=0.99\textwidth]{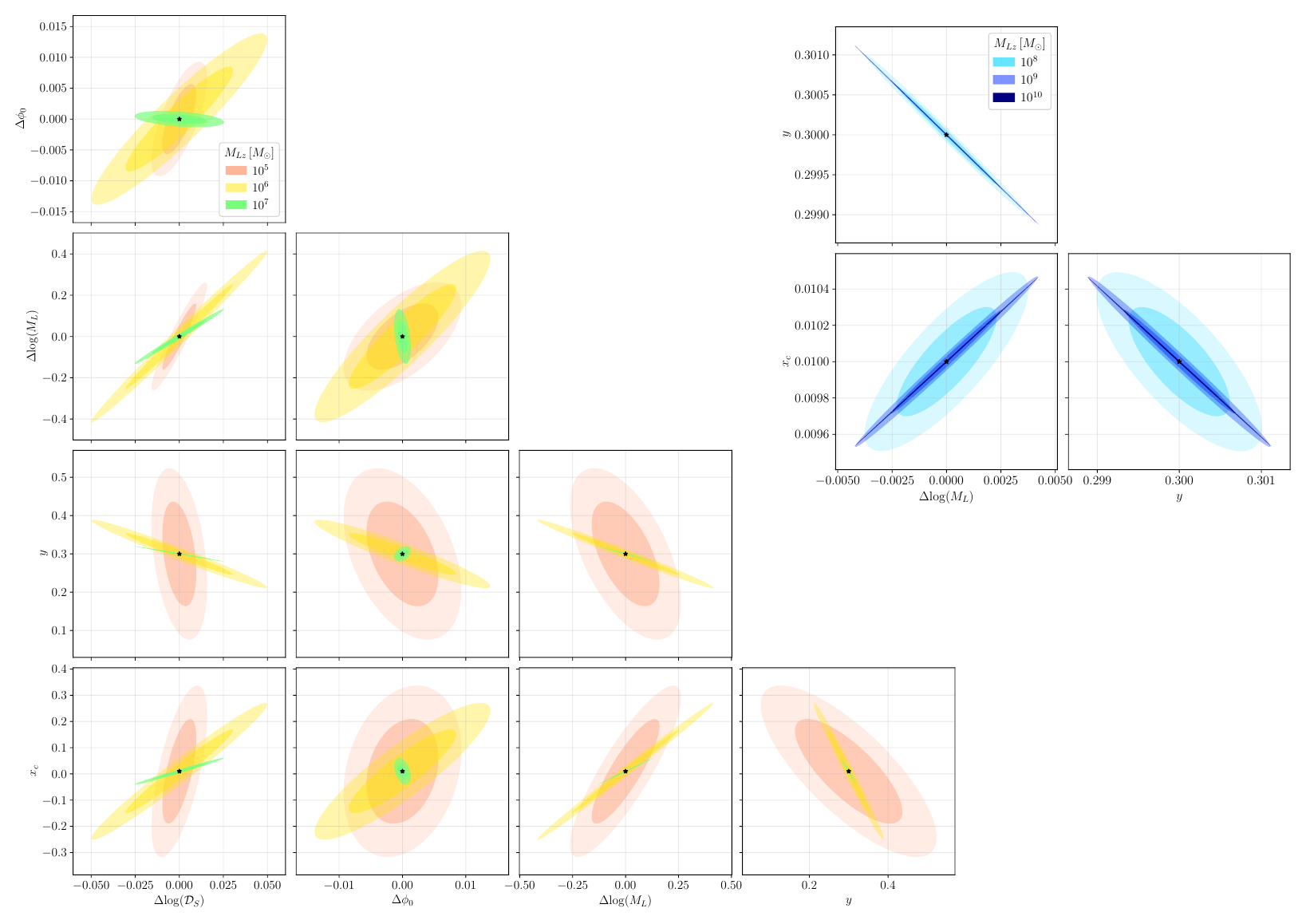}
\caption{68 and 95\% c.l.~marginalised posteriors for low (bottom left) and high (upper right) lens masses of a LISA source ($M_{\rm BBH}=10^6M_\odot$) lensed by a CIS with $y=0.3$, $x_c=10^{-2}$. All contours have been rescaled to a fiducial ${\rm SNR}=10^3$. In the high $M_{Lz}$ limit the lens parameter posteriors become highly degenerate.
} \label{fig:forecast_LISA_posteriors}
\vspace{10pt}
    \centering
    \includegraphics[width=0.4\textwidth]{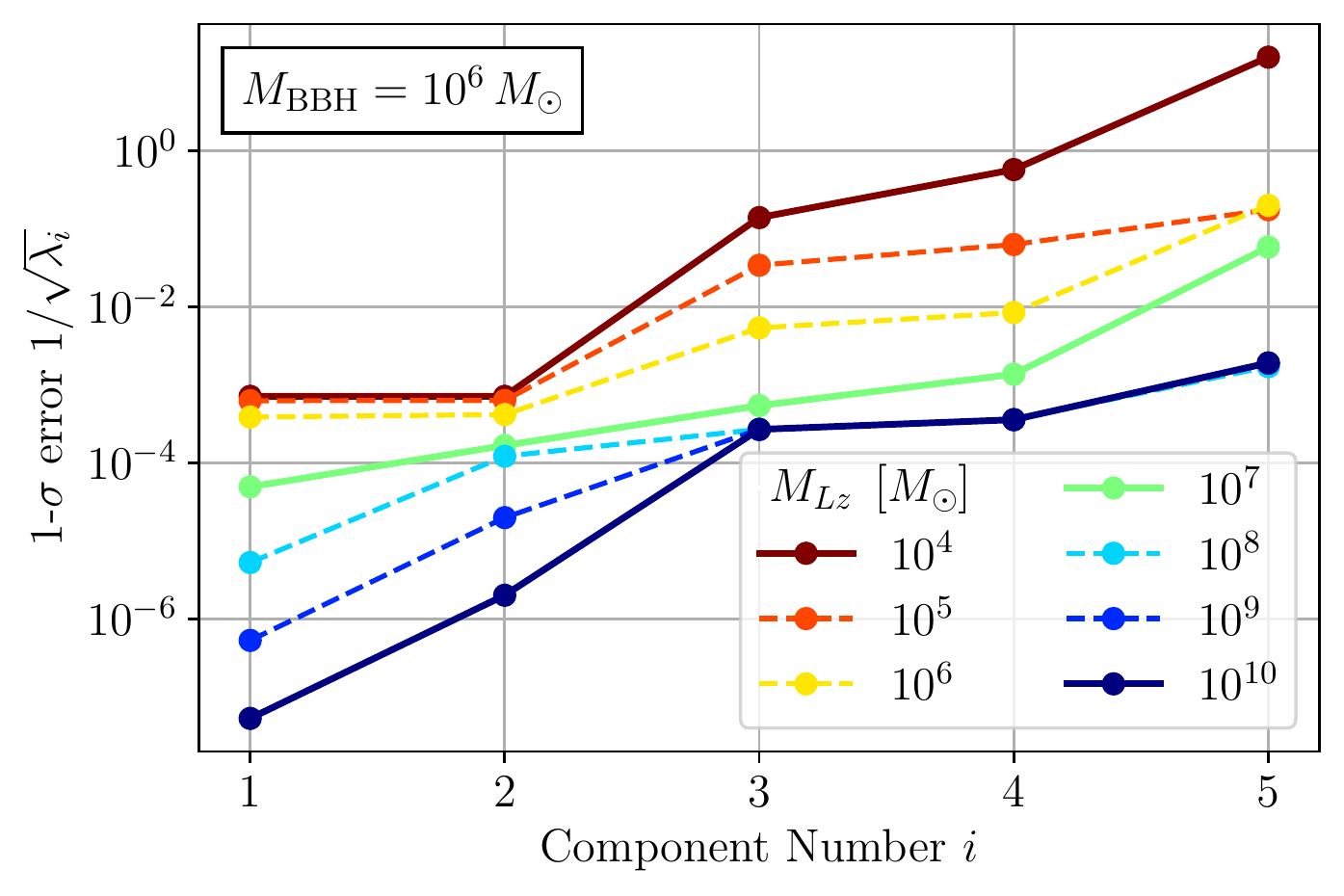}
    \hfill 
    \includegraphics[width=0.4\textwidth]{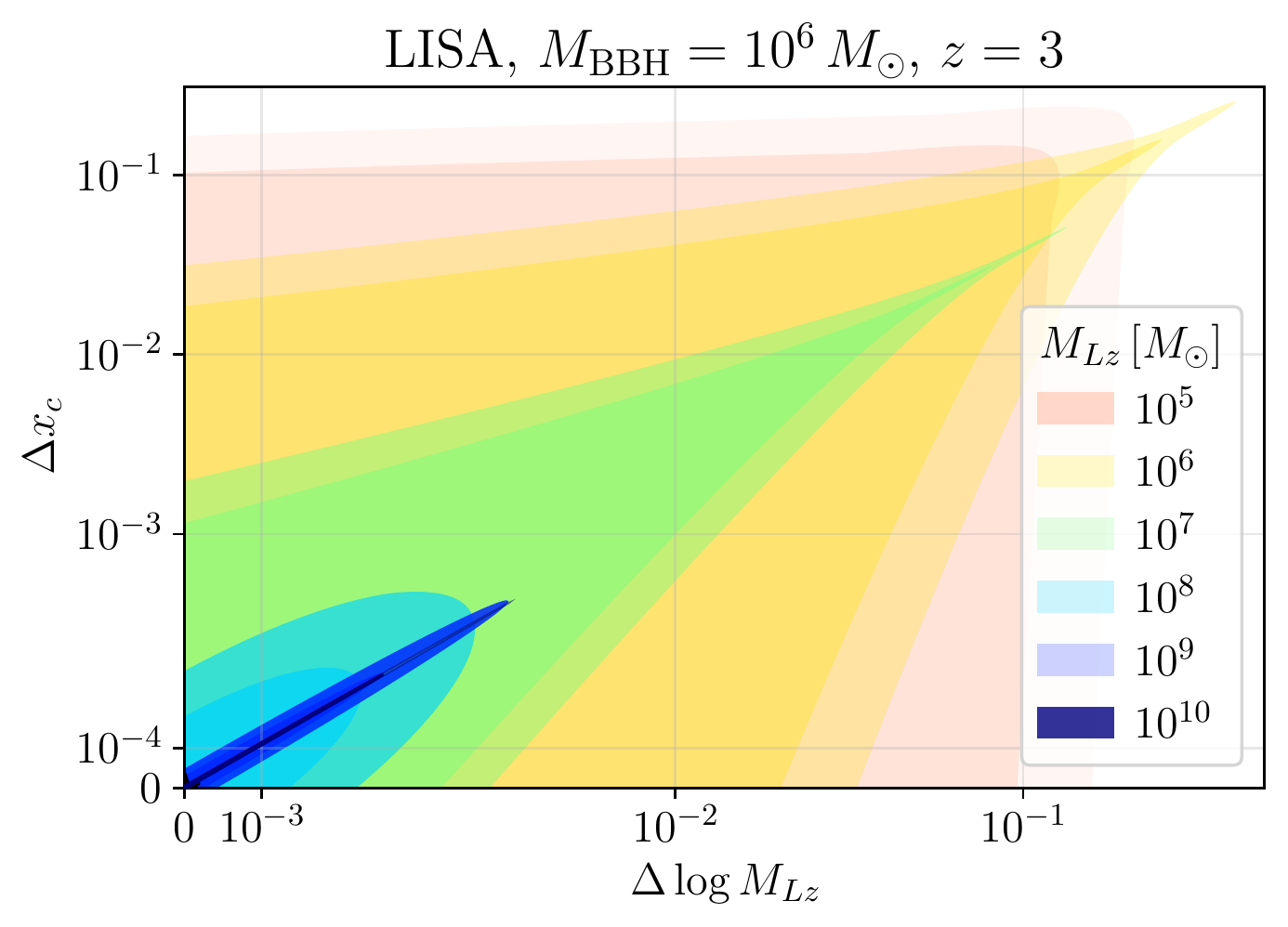}
    \hfill 
\vspace{-10pt}
\caption{
\textbf{Left:} $1\sigma$ error associated with the principal components of the Fisher matrix for the CIS lens, ordered from best to worst constrained (Fisher matrix, its inverse and eigenvectors for $M_{Lz}=10^4$, $10^7$, $10^{10}\, M_\odot$ are shown in Fig.~\ref{fig:forecast_LISA_fisher_matrix}).  
\textbf{Right:} 68\% and 95\% c.l.~$M_{Lz}-x_c$ marginalised posteriors for the largest masses. For clarity, we show one quadrant in symlog scale.
} \label{fig:forecast_LISA_fisher_degeneracies}
\end{figure*}

Figure \ref{fig:forecast_LISA_source_mass_dependence} shows the expected 68\% 1D marginalised posteriors for the lens parameters $\log(M_{Lz})$, $y$ and $x_c$. 
As just explained, the results are rescaled to a fiducial lensed SNR of $10^3$: the precision of a given source needs to be rescaled by $10^3/$SNR, e.g.~with the lensed SNR given in the right panel of Fig~\ref{fig:forecast_lisa_sources}. 
At fixed SNR and for high-mass lenses, the precision of all parameters converges to a constant value, of order $\sim 1/{\rm SNR}$ and independent of the source mass. The saturation of the sensitivity in the GO limit will be explained below in terms of parameter degeneracies (Figs.~\ref{fig:forecast_LISA_posteriors}, \ref{fig:forecast_LISA_fisher_degeneracies}).%
\footnote{The independence of the posteriors on the source mass can be understood from the convergence of the Fisher matrix to the Geometric Optics Diagonal Approximation, cf.~App.~\ref{sec:forecast_GODA}. In this limit, both the lensed SNR and each $\mathcal{F}_{ij}$ are a sum of terms that depend only on the lens parameters, all proportional to the unlensed SNR.}

Lighter sources are more effective at probing lower lens masses, at fixed SNR and when the signal is dominated by the merger. This is a consequence of the higher GW frequencies, emitted up to $\sim w_{\rm ISCO}\propto M_{Lz}/M_{\rm BBH}$ so the onset of magnification $w\gtrsim 1$ corresponds to lower $M_{Lz}$. 
In this case, the lens parameters can be independently determined for $M_{Lz} \gtrsim 0.1\, M_{\rm BBH} (10^3 / {\rm SNR})$.
For the lightest sources shown ($M_{\rm BBH} = 10^4$, $10^5\,M_\odot$) the SNR is not dominated by the merger, but by the waveform portion in the ``sweet spot'' of the LISA noise curve (cf.~Fig.~\ref{fig:forecast_lisa_sources}). In these cases, the higher frequencies of the source do not play a role, which explains their similar shape in Fig.~\ref{fig:forecast_LISA_source_mass_dependence}.
Note that the effect of different SNRs needs to be included when considering specific sources. For instance, a $10^6 \, M_\odot$ binary has a lensed SNR $\sim 40$ times larger than that of a $10^4\, M_\odot$ source.
In what follows we will show results for lensed $10^6 \, M_\odot$ LISA sources, as they are the most favourable case.

\begin{figure*}
\vspace{10pt}
\centering
\includegraphics[width=0.95\textwidth]{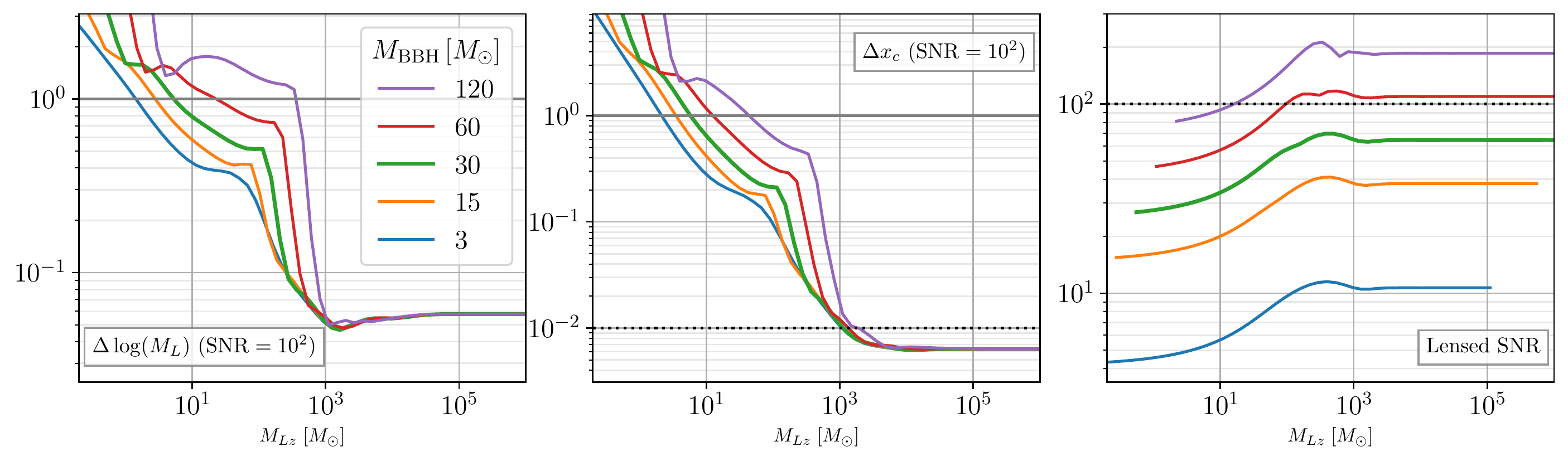}
\centering
\includegraphics[width=0.95\textwidth]{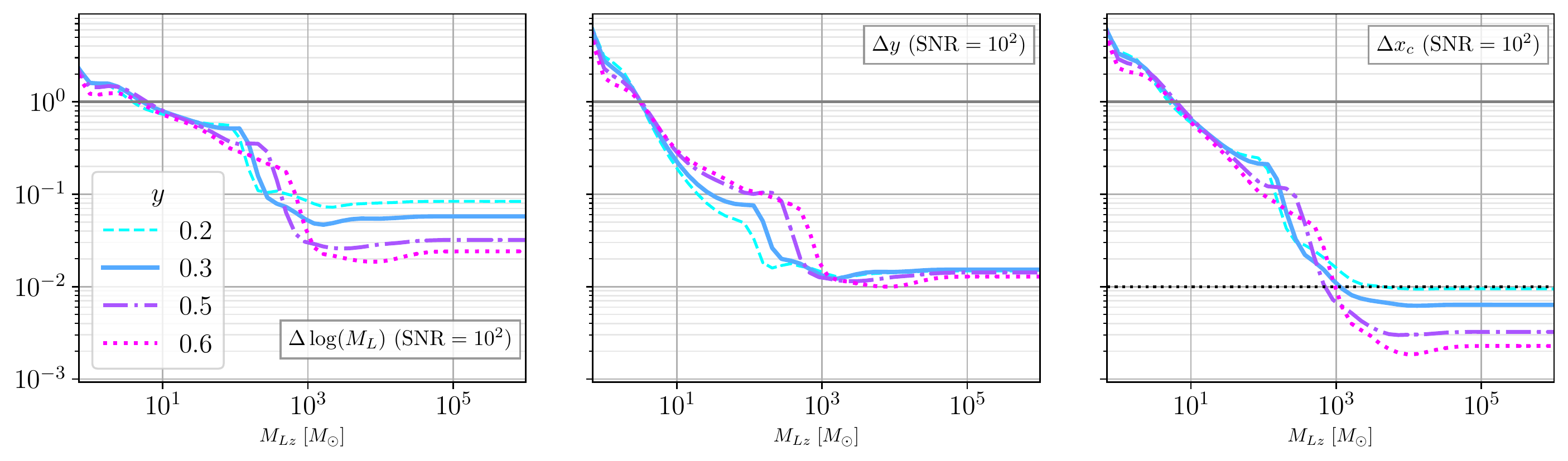}
\vspace{-10pt}
\caption{Same as Figs.~\ref{fig:forecast_LISA_source_mass_dependence} and \ref{fig:forecast_LISA_impact_parameter_dependence}, but for advanced LIGO Sources (see text).
In the case of LIGO, the LSA condition \eqref{eq:LSA_cond2} is typically violated as soon as we enter the GO regime given the larger uncertainties compared to the LISA case. Nonetheless, as we motivate in the main text, for the CIS lens we expect our results to be still reliable, as there are enough GO parameters to reconstruct all the lens parameters.
}\label{fig:forecast_LIGO_summary}
\end{figure*}

Varying the impact parameter changes the precision in the high $M_{Lz}$ limit, with higher precision for larger $y$, for fixed SNR. This is shown in Fig.~\ref{fig:forecast_LISA_impact_parameter_dependence} for $y=0.2$, $0.3$, $0.5$, $0.6$, sufficiently away from the caustic to avoid numerical problems when computing derivatives in the amplification factor, cf.~App.~\ref{sec:forecast_implementation}. 
The increase in precision when $y$ is increased is larger for $\log(M_{Lz})$ and $x_c$. This is explained by the lens mass being probed by time delays, which are larger for less aligned configurations. Similarly, the imprint of the core is more obvious for larger $y$, as both the amplitude of the central image and its bGO corrections increase as the source approaches the caustic, cf.~Fig.~\ref{fig:CIS_GO} (for our fiducial model $y_{\rm rc} \simeq 0.805$, Eq.~\eqref{eq:cis_yrc_caustic}).
This gain in sensitivity is partially offset when including the actual SNR, which is higher when the source and lens are aligned, at low $y$.

Varying the core size at fixed $y$ has a small impact on the constraints unless $x_c$ becomes large enough for $y$ to start approaching the caustic $y_{\rm rc}$. Figure \ref{fig:forecast_LISA_core_size_dependence} shows marginalised 1D posteriors for $x_c=10^{-4}$, $10^{-3}$, $0.01$ and $0.1$, for fixed SNR and $y = 0.3$. In these cases the caustic Eq.~\eqref{eq:cis_yrc_caustic} is located at $y_{\rm rc} = 0.98$, $0.937$, $0.805$, $0.427$, respectively, with a slow convergence to SIS ($x_c\to 0,\, y_{\rm rc}\to 1$) due to the $\sqrt{x_c}$ term in Eq.~\eqref{eq:cis_yrc_caustic}.
All forecast curves show qualitatively similar dependence on the lens mass, except for the larger cores $x_c=0.1$, for which the impact parameter approaches the caustic and the constraints are slightly enhanced.

The lens parameters $\log(M_{Lz})$, $y$, $x_c$ follow very similar trends regarding the dependence with source and lens mass (Fig.~\ref{fig:forecast_LISA_source_mass_dependence}), as well as with the impact parameter (Fig.~\ref{fig:forecast_LISA_impact_parameter_dependence}) and core size (Fig.~\ref{fig:forecast_LISA_core_size_dependence}). In contrast, the source parameters ($\log(\mathcal D_S)$, $\phi_0$) can always be determined, although their precision suffers at intermediate masses (cf.~lower panels in Fig.~\ref{fig:forecast_LISA_impact_parameter_dependence}) due to degeneracies with the lens parameters.

Let us now address the correlations between different parameters. Our previous discussion (Figs.~\ref{fig:forecast_LISA_source_mass_dependence}, \ref{fig:forecast_LISA_impact_parameter_dependence}, \ref{fig:forecast_LISA_core_size_dependence}) focused on the 1D marginalised posteriors, which quantify the ability to constrain the parameters separately. 
All these plots showed how the 1D posteriors saturate at high $M_{Lz}$ at fixed SNR. This can be explained by the amount of GO parameters available to reconstruct the lens (Sec.~\ref{sec:parameter_reconstruction_go}): two relative time delays, whose precision improves as $\sim 1/M_{Lz}$ and two magnification ratios, whose precision remains $\sim 1/{\rm SNR}$ \cite{Ali:2022guz}. Thus, one combination of $\log(M_{Lz})$, $y$ and $x_c$ remains poorly constrained (for fixed ${\rm SNR}$ but increasing $M_{Lz}$). 
This trend can be observed explicitly in Fig.~\ref{fig:forecast_LISA_posteriors}, representing the 2D marginalised posteriors in the Fisher-matrix approximation. The bottom-left panels show the posteriors shrinking as $M_{Lz}$ increases, for moderate lens masses. 
For high $M_{Lz}$ (top right panels) the ellipses become narrower but otherwise span the same range of lens parameters.

To further understand the degeneracies it is instructive to look at the principal components of the Fisher matrix, i.e.~its eigenvectors $\vec u_{(k)}$ and their associated eigenvalues $\lambda_{(k)}$
\begin{equation}
    \mathcal{F}_{ij}u^j_{(k)} = \lambda_{(k)}u^i_{(k)}\,.
    \label{eq:fisher_eigen}
\end{equation}
Each $\vec u_{(k)}$ represents a combination of parameters that can be constrained independently (i.e.~the eigenvectors are normal $\vec u_{(k)} \cdot \vec u_{(l)}\propto \delta_{kl}$) with precision $\lambda_{(k)}^{-1/2}$.
The left panel of Fig.~\ref{fig:forecast_LISA_fisher_degeneracies} shows the eigenvalues and eigenvectors of $\mathcal{F}_{ij}$, ranked by precision, as a function of the lens mass.
We note two distinctive regimes for light ($M_{Lz}\lesssim 10^6\, M_\odot$) and heavy ($M_{Lz}\gtrsim 10^8\, M_\odot$) lenses, with an intermediate trend for $M_{Lz}=10^7\, M_\odot$.
For lighter lenses ($10^4-10^6\, M_\odot$) the two most precise combinations are dominated by the source parameters $\log(\mathcal{D}_S)$, $\phi_0$, whose precision saturates for light lenses. The remaining three combinations are mainly combinations of lens parameters, whose precision decreases substantially as $M_{Lz}\to 0$, explaining the loss of sensitivity seen for light lenses in Fig.~\ref{fig:forecast_LISA_source_mass_dependence}, and expected from the convergence to $F(w) \to 1$ at low $w$. 

For heavy lenses, $M_{Lz}\gtrsim 10^8\, M_\odot$, Fig.~\ref{fig:forecast_LISA_fisher_degeneracies} shows how the precision of the two best-measured combinations increases with $M_{Lz}$, while the other three combinations remain independent of the lens mass. 1D marginalised posteriors are a combination of all eigenvalues and hence dominated by the least-precise ones, which saturate for $M_{Lz}/M_{\rm BBH} \gtrsim 10^2$. 
The higher precision in the first two eigenvalues for larger $M_{Lz}$ appears in parameter correlations as a thinning of the posteriors, seen in Fig.~\ref{fig:forecast_LISA_fisher_degeneracies}.
The large degeneracy would allow for substantial improvement in the constraints if one or more of the lensing parameters can be measured independently, for instance if the lens can be located via an EM counterpart or cross-correlating with optical catalogues \cite{Hannuksela:2020xor}.

Let us now consider the capacity of ground detectors to characterize CIS lenses, focusing on the advanced-LIGO sensitivity curve at design values \cite{ligo_noise_curve}. We will consider equal-mass non-spinning binaries with total mass $M_{\rm BBH}=3$, $15$, $30$, $60$ and $120$ solar masses observed respectively for $1\, {\rm h}$, $6\, {\rm min}$, $1\, {\rm min}$, $1\, {\rm min}$ and $10\, {\rm s}$ before merger. Reasonable variations of the observing time do not affect the final results.
We will show results up to $M_{Lz} < 10^6\, M_\odot$. For the lens parameters we consider (e.g.~away from caustics), this is sufficient to reach the GO saturation described above. 
This choice of masses also justifies considering a static detector, as typical time delays between GO images scale as $\sim 4GM_{Lz} = 19.4\, \text{s}\left(M_{Lz}/10^5\, M_\odot\right)$. As the time delays remain much smaller than changes in detector orientation, on the scale of $1\,{\rm day} = 8.6\cdot 10^4\, {\rm s}$, our results could be extended to slightly higher lens masses.

Figure \ref{fig:forecast_LIGO_summary} summarises the results for equal-mass non-spinning sources at $z=0.1$ as detected by one LIGO detector. In this case, the results are normalized to a fiducial lensed SNR of 100, which is plausible for lensed sources at low redshift (see the top-right panel in Fig.~\ref{fig:forecast_LIGO_summary}). 
The trends are qualitatively similar to LISA detections, shifted towards a lower range of $M_{Lz}$, as expected from the scaling of $w$, Eq.~\eqref{eq:lensing_freq_dimensionless}. 
Note that the curves are closer together because the hierarchy of $M_{\rm BBH}$ considered spans fewer orders of magnitude than in the LISA case.
The uncertainties on individual parameters are larger, but this is mostly due to the choice of fiducial lensed SNR.

To test the validity of the CIS Fisher matrix, we compared the full log-likelihood \eqref{eq:def_likelihood} to the quadratic expansion \eqref{eq:likelihood_LSA}, focusing on the GO limit for the LISA case for simplicity (see App.~\ref{sec:fisher_validity} for details on how these quantities are obtained). 

We sampled the full likelihood at $10^3$ random points located at $2\sigma$ from the fiducial, as defined by the Fisher-matrix eigenvectors.
We find that in $99.7\%$ of the cases, the full likelihood (evaluated at $2\sigma$) is smaller than the Fisher likelihood (evaluated at $1\sigma$). This shows that the true $1\sigma$ contours (as defined by the full likelihood) are bounded to be within the $2\sigma$ contours of the Fisher, indicating that the Fisher approximation is an adequate description, i.e. it captures the degeneracy directions and magnitude correctly.

\begin{figure*}
\centering
\includegraphics[width=0.95\textwidth]{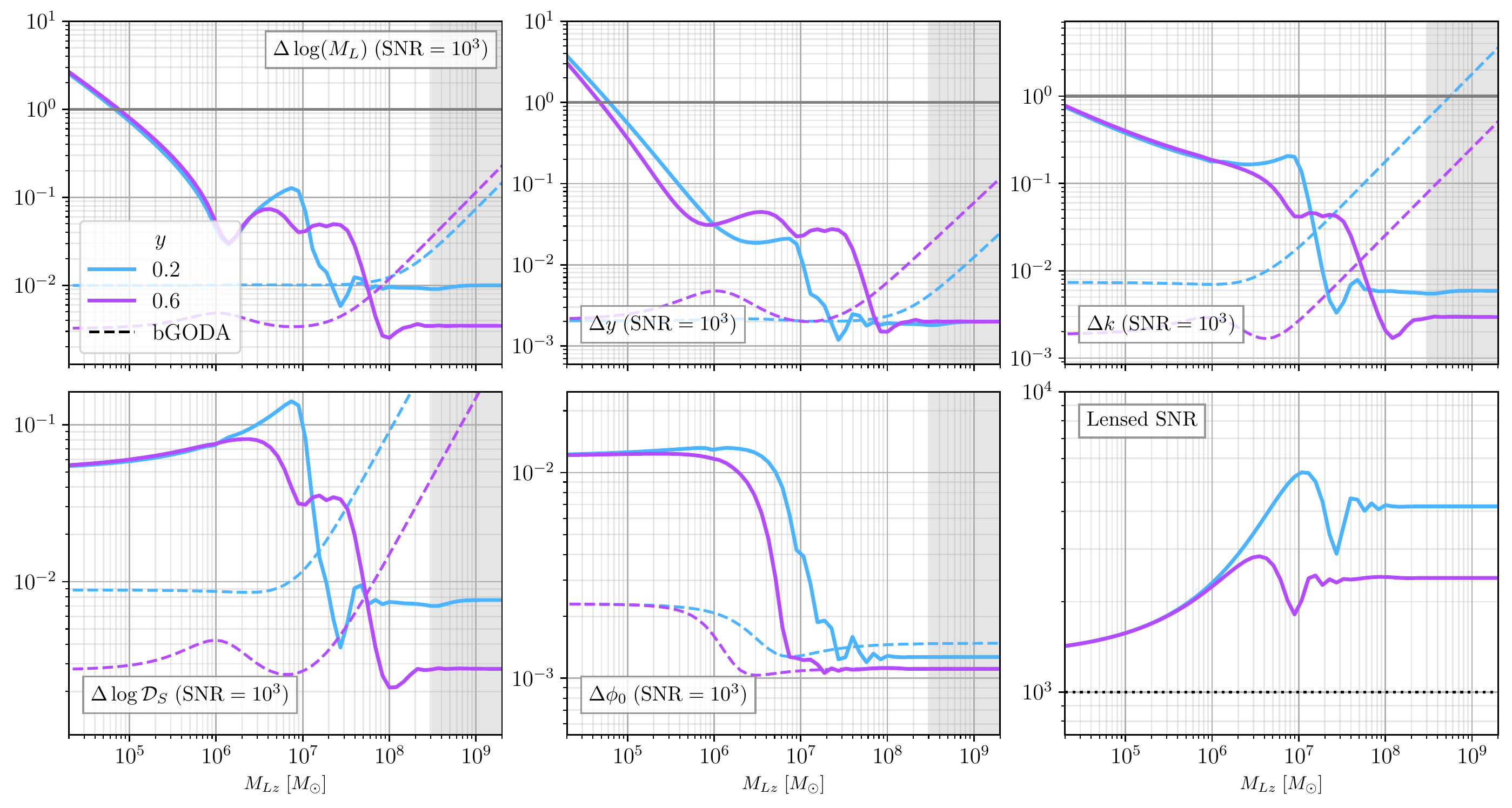}
\vspace{-10pt}
\caption{Constraints on the gSIS lens ($\log(M_{Lz})$, $y$, $k$) and source ($\log(\mathcal D_S)$, $\phi_0$) parameters, and lensed SNR for different impact parameter (fiducial slope $k = 1+10^{-4}$). The source is a $M_{\rm BBH}=10^6\, M_\odot$ equal-mass non-spinning LISA binary. Results are normalized to a lensed SNR of $10^3$. For values of $M_{Lz}$ in the grey-shaded area, the LSA criterion Eq.~\eqref{eq:lsa_breakdown} is violated. This criterion is obtained as $\Delta_{\rm LSA} < 1$ and also by requiring we are in the GO limit ($w_{\rm max} > 100$). Dashed lines are the results of the Fisher forecast using the bGODA approximation which neglects the cusp contribution. The correct constraint should lie between WO (which overestimates the contribution from the cusp) and bGODA (which neglects it completely). 
}
\label{fig:gsis_forecast_LISA_impact_parameter_dependence}
\end{figure*}

\begin{figure*}

\includegraphics[width=0.95\textwidth]{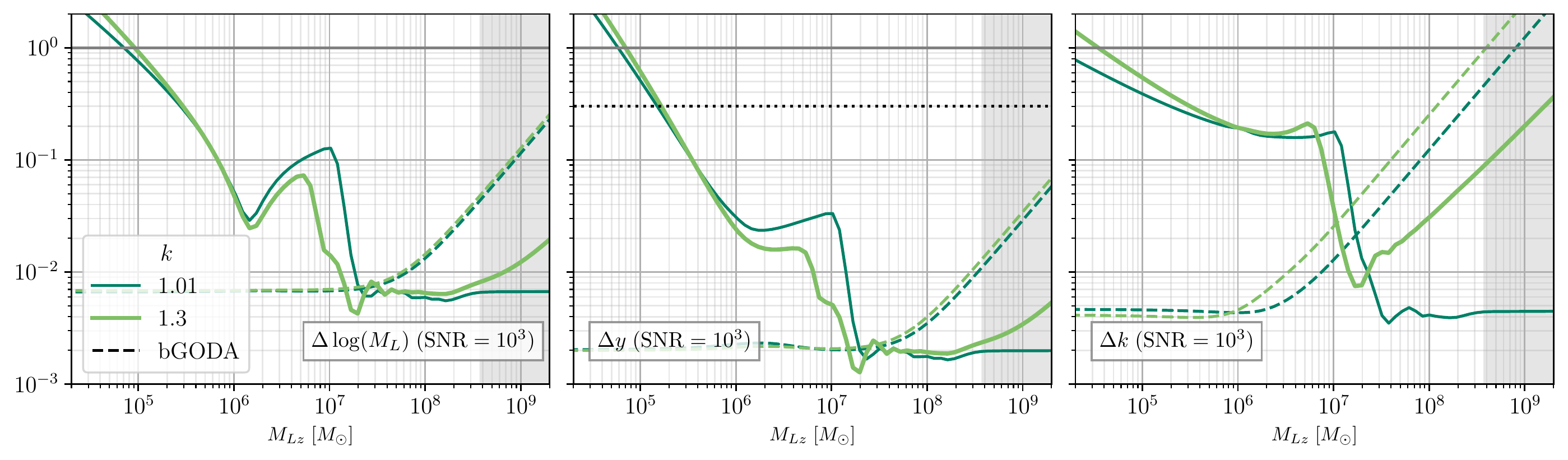}
\vspace{-10pt}
\caption{Constraints on the gSIS lens parameters ($\log(M_{Lz})$, $y$, $k$) for different fiducial lens slope $k$ values ($y=0.3$). The source is a $M_{\rm BBH}=10^6\, M_\odot$ LISA binary. Results are normalized to a lensed SNR of $10^3$. The grey area is the region where the LSA condition is violated and the dashed lines are the Fisher forecast results in the bGODA approximation (see caption of Fig.~\ref{fig:gsis_forecast_LISA_impact_parameter_dependence}). 
The solid dark-green lines ($k = 1.01$) are expected to start growing at large lens masses, following the qualitative trend of the light-green curves ($k = 1.3$). This ultimately follows from the scaling of the cusp contribution $F_c(w)$ as a function of $w$ (see Eq.~\eqref{eq:F_cusp}).
}
\label{fig:gsis_forecast_LISA_slope_dependence}
\end{figure*}

\subsection{Narrow generalised-SIS}\label{sec:forcast_gSIS}

Let us now consider the capacity of different detectors to measure the slope in the matter distribution using the gSIS lens.
We will focus on narrow gSIS ($k > 1$) with two GO images, as the broad gSIS is qualitatively similar to the CIS lens (Sec.~\ref{sec:forcast_CIS}). Having fewer images, its analysis is qualitatively different. Our fiducial value of $k$ is very close to the SIS, $k = 1 + 10^{-4}$, indistinguishable from the SIS for reasonable values of the lensed SNR ($\lesssim 10^4$).\footnote{This choice is made for simplicity: our numerical implementation treats the cases $k<1$, $k=1$, $k>1$ separately, but the forecast results need to be continuous in the limit of $k$ going to one. Indeed, notice that the magnification of the third image for $k<1$ goes to zero in this limit, while the cusp contribution $F_c(w)$ is the same as for $k \geq 1$.}

Before discussing our results, we note that the LSA breakdown leads to artificially optimistic results in the full WO calculation at high $M_{Lz}$. For the narrow gSIS, only two GO images exist, not providing enough information to recover $M_{Lz}$, $y$ and $k$, cf.~Sec.~\ref{sec:parameter_reconstruction_go} (unlike for CIS, where the existence of three images allows a full reconstruction in GO). Thus, information from bGO corrections $\propto \Delta_I/w$ and the cusp $F_c\propto 1/w$ plays a critical role in the narrow gSIS. While these terms become negligible at high $M_{Lz}$, the WO Fisher-matrix results predict a constant sensitivity. 

The breakdown of the LSA is due to the cusp contribution. It can be traced back to the linearisation of the phase $e^{i w \phi_c(k)}$ in $F_c(w)$: if linearly expanded around $k = \bar k + \Delta k$, it gives $\sim  i w \partial_k \phi_c \Delta k e^{iw \phi_c(\bar k)}$. Clearly, the linearisation is invalid if $M_{Lz}$ becomes too large and moreover, if trusted beyond this limit, this factor becomes much larger than the phase $e^{i w \phi_c(k)}$. Therefore, the linearisation seems to overestimate the sensitivity on $k$. 
In order to confirm our intuition, we have checked that indeed the contribution of the cusp to the Fisher-matrix log-likelihood (given by Eq.~\eqref{eq:likelihood_LSA}) 
becomes much larger than the corresponding contribution in the full likelihood (i.e.~not expanding in powers of $\Delta \theta^i$), for some sample points at high lens masses. 
We expand on the issue in App.~\ref{sec:SPA} with an analytical toy example and show different predictions (Fig.~\ref{fig:GODA_tests_gSIS}) that confirm the disproportionate effect of the cusp at high $M_{Lz}$.

To circumvent the cusp problem without abandoning the Fisher-matrix analysis, we will show two results for each case: the full WO Fisher forecast and the bGODA approximation. Since the former overestimates the contribution from the cusp and the latter neglects it, the truth should lie between both. 
In addition, we highlight the values of $M_{Lz}$ where the LSA approximation, Eq.~\eqref{eq:lsa_breakdown}, breaks down. In each plot, we show this for the model where the breakdown happens at a lower lens mass.

Given the premise above, we can now briefly discuss the results of our forecast, obtained using a combination of WO and bGODA, for LISA.
As in the CIS case, we conventionally give results normalized to a fiducial ${\rm SNR} = 10^3$. See the discussion in Sec.~\ref{sec:forecast_framework} on how to rescale the results for a given source.

Figure \ref{fig:gsis_forecast_LISA_impact_parameter_dependence} shows the dependence on $y$, at fixed $k$ for two fiducial values of $y = 0.2$ and $y = 0.6$. Larger values of $y$ give worse $\Delta y$ and better $\Delta k$, $\Delta \log \mathcal D_S$ instead.
From this Figure, one also notices how $\phi_0$ can always be recovered at high $M_{Lz}$, contrary to the recovery of lens parameters. These trends follow the CIS case already described.
Interestingly, the inclusion of WO effects breaks the parameter degeneracy, and the slope $k$ can therefore be measured, both in WO and bGO. The best accuracies for the lens parameters are expected when the onset of WO, $w \sim 10$, happens around the peak of the signal SNR. Typically, this corresponds with the ISCO frequency, or $w \sim 8 \pi G M_{Lz}f_{\rm ISCO}\sim M_{Lz}/M_{\rm BBH}\sim 10$. On the other hand, the results degrade when $M_{Lz}$ is increased or decreased (if we take the worst between the WO and bGODA results, which is conservative).

Fig.~\ref{fig:gsis_forecast_LISA_slope_dependence} shows the dependence on the fiducial value of $k$, at fixed $y$. We can notice slight differences at the onset of WO, around $M_{Lz}\sim 10^{7}\, M_{\odot}$: larger $k$ gives better results. For higher lens masses instead, the noticeable difference is in $\Delta k$, whose 1D posterior worsens for larger $k$s
(and for the other lens parameters but only in the region where the LSA fails). This trend is explained by the different scaling of the cusp corrections $\propto w^{-k/(2-k)}$. Due to this scaling, the cusp contribution decays faster for larger $k$s and does not lead to a saturation of the Fisher posteriors.

\section{Towards tests of large-scale structure and dark matter}\label{sec:dark_matter}

We will consider the potential of GW lensing to constrain dark matter (DM) properties. First, we will discuss the prospects of inferring the virial mass of the lens, given that its redshift is unknown, Sec.~\ref{sec:lens_mass}. Then we consider two scenarios that lead to the formation of cores: modelling DM halos as a CIS allows us to translate our forecasts into a range of DM parameters accessible to LIGO and LISA.
We discuss two theories proposed in the context of small-scale challenges to standard cosmology \cite{Bullock:2017xww}: self-interacting DM (SIDM, Sec.~\ref{sec:forecast_sidm}) and ultra-light DM (ULDM, Sec.~\ref{sec:forecast_uldm}). We conclude by discussing the assumption of axially-symmetric lenses, Sec.~\ref{sec:non_symmetric_lenses}.

\begin{figure*}
    \centering
    \includegraphics[width=.81\columnwidth]{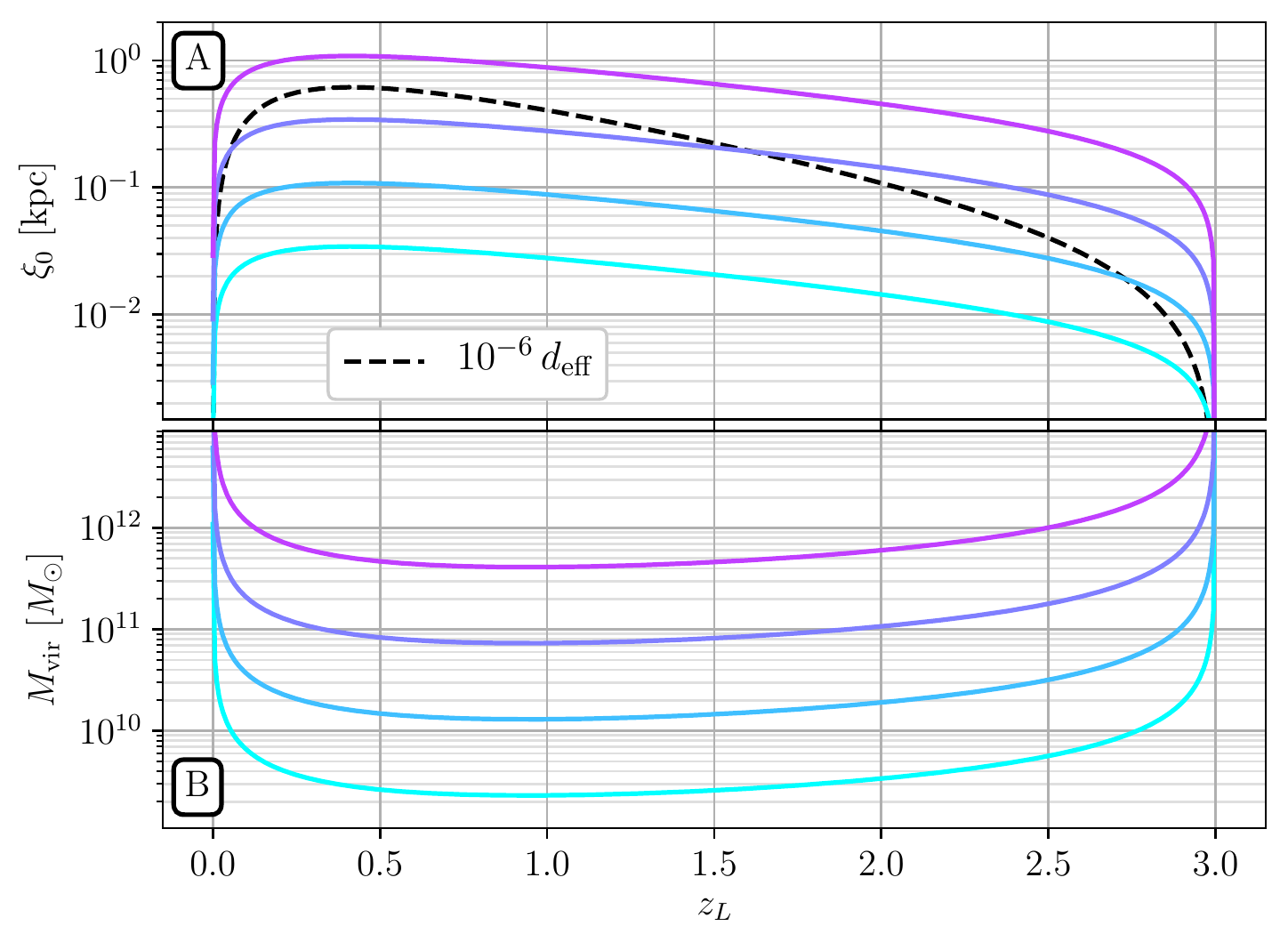}
    \includegraphics[width=.595\columnwidth]{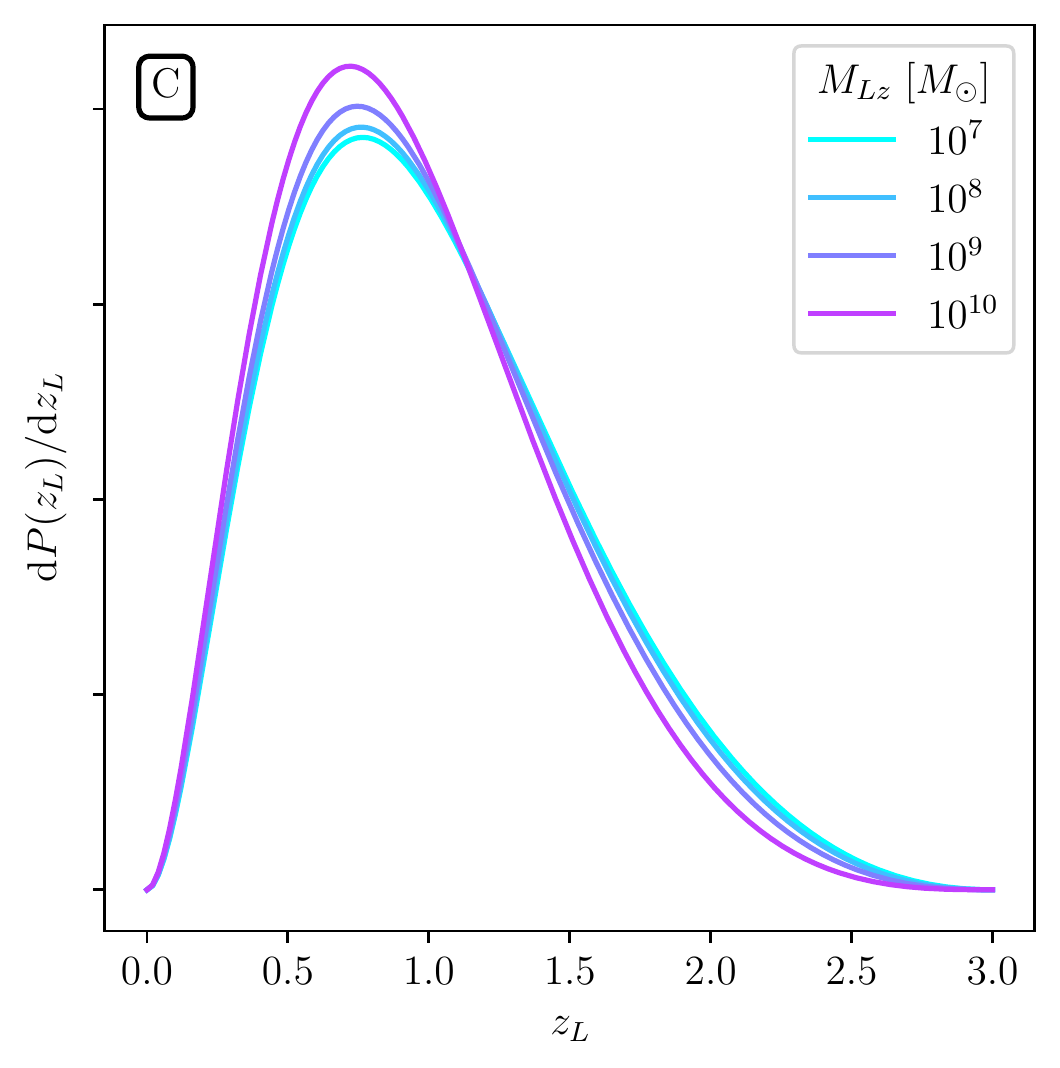}
    \includegraphics[width=.61\columnwidth]{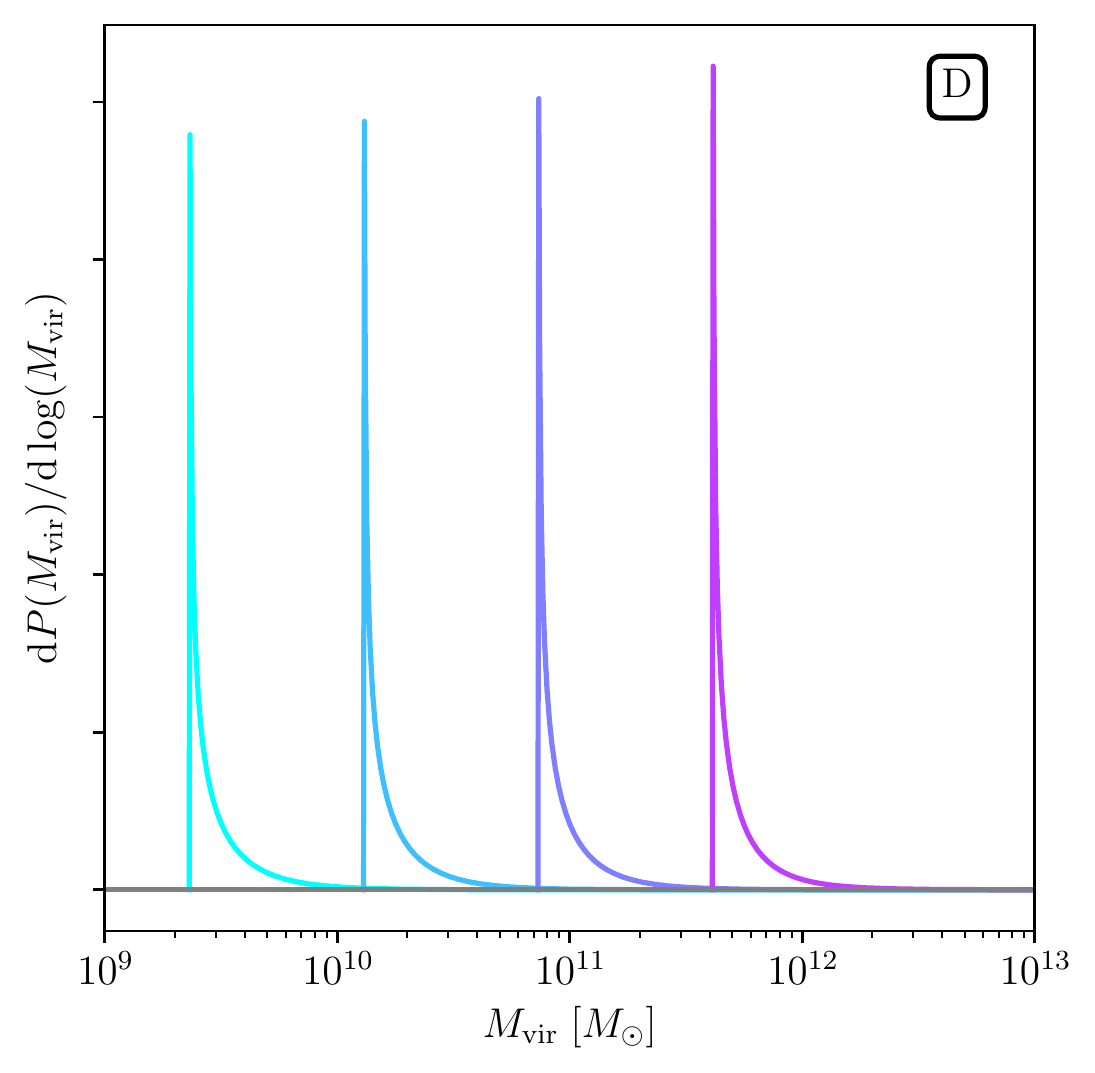}
    \caption{Conversion between the observed parameters ($M_{Lz}$, $y$, $x_c$) and the physical properties of the lens, $M_{\rm vir}$ and $r_c = x_c\xi_0$. 
    Here $z_S = 3$.
    \textbf{Panel A:} dependence on the projection scale $\xi_0$ on the unknown lens redshfit $z_L$, Eq.~\eqref{eq:lens_mass_def} ($d_{\rm eff}$, Eq.~\eqref{eq:deff_effective_distance} is shown for reference). 
    \textbf{Panel B:} Virial mass $M_{\rm vir}\propto M_{Lz}^{3/4}$ for an SIS lens, Eq.~\eqref{eq:MLz_sis}. 
    \textbf{Panel C:} Probability distribution of $z_L$ for fixed $M_{Lz}$ and $z_S$, including a halo mass function.
    \textbf{Panel D:} corresponding probability distribution for the virial mass.  
    }
    \label{fig:lens_projection}
\end{figure*}

\subsection{Lens redshift and halo mass}\label{sec:lens_mass}

Let us now address the prospect of relating the measured variables ($M_{Lz}$, $y$, $x_c$, $\mathcal{D}_S$) to the physical properties of the lens. 
The main obstruction is the lack of knowledge of the lens' redshift $z_L$, which is needed to determine the virial mass of the halo (Eqs.~\eqref{eq:MLz_sis}, \eqref{eq:gsis_MLz} and \eqref{eq:MLz_cis}) and the physical scales, i.e.~$r_c = x_c\xi_0(M_{Lz},z_L,z_S)$ (Eqs.~\eqref{eq:deff_effective_distance} and \eqref{eq:lens_mass_def}).
The dependence of physical scales (virial mass) on the lens redshift is shown on panel A (B) of Fig.~\ref{fig:lens_projection}: the maximum (minimum) value is reached at an intermediate redshift, while for $z_L\to 0$ and $z_L\to z_S$ the curves approach zero (infinity). Values several orders of magnitude away from the maximum (minimum) require lenses that are very close to either the source or the observer. While not impossible, this situation is unlikely.

We can now compute the probability distribution for $z_L$ and $M_{\rm vir}$, given $M_{Lz}$, $z_S$ and some reasonable assumptions about the cosmology and the halo mass function. 
The probability distribution for $M_{\rm vir}$ and $z_L$ describing the lens is $\frac{\de P}{\de z_L\de  M_{\rm vir}} \propto \sigma \frac{\de V}{\de z_L}\frac{\de n}{\de M_{\rm vir}}$.
The angular cross-section is $\sigma = 2\pi y\Delta y \left(\xi_0 / D_L\right)^2$, i.e.~the projected area of the annulus compatible with the measured impact parameter. The volume element $\frac{\de V}{\de z_L}=\frac{(1+z_{L})^2 D_L^2}{H(z_L)}$ and halo mass function are both in comoving coordinates. We will use the Tinker \textit{et al.}~form of $\de n/\de M_{\rm vir}$ \cite{Tinker:2008ff}, as implemented in the Colossus package \cite{Diemer:2017bwl}. We will assume fixed cosmological parameters, given by Planck $\Lambda$CDM best fit \cite{Planck:2018vyg} and focus on an SIS when concrete results are needed. 
Changes due to the cosmology or halo mass function do not qualitatively affect our analysis.\footnote{A non-standard halo mass function may have a significant impact on the recovered $z_L,M_{\rm vir}$, e.g.~if light halos are suppressed. We will not consider this possibility further.}

We will focus on $z_L$ and $M_{\rm vir}$, assuming that $M_{Lz},y$ and $z_S$ are known. The former can be inferred precisely, while the latter can be constrained from the source's luminosity distance $\mathcal{D}_s$ assuming an expansion history (including uncertainties on these and other parameters,e.g.~from parameter posteriors is straightforward). 
Then, the probability of a certain lens redshift is 
\begin{align}\label{eq:zL_probability_base}
    \frac{\de P}{\de z_L} 
    &= 
    \int \de M_{\rm vir} 
    \frac{\de P}{\de z_L \de M_{\rm vir}} 
    \delta\left(\bar M_{Lz}-M_{Lz}(z_L,M_{\rm vir})\right)
    \\ 
    \label{eq:zL_probability}
    &\propto 
    \bar{\xi_0}^2 
    \frac{(1+z_L)^2}{H(z_L)}
    \left|\frac{\partial M_{Lz}}{\bar{\partial M_{\rm vir}}}\right|^{-1}
    \frac{\de n(\bar M_{\rm vir},z_L)}{\de M_{\rm vir}}
    \,, 
\end{align}
where in the second line we omit all the constant terms.
Here $\bar M_{Lz}$ is the measured value, and a bar means that a quantity is evaluated on it, e.g.~$\bar \xi_0=\xi_0(\bar M_{Lz},z_L,z_S)$ in Eq.~\eqref{eq:lens_mass_def}. 
The jacobian stems from integrating the delta function over $M_{\rm vir}$ and is $\propto M_{\rm vir}^{-1/3}$ for the SIS, cf.~Eq.~\eqref{eq:MLz_sis}. 
The resulting probability distribution is shown on panel C of Fig.~\ref{fig:lens_projection}. Higher values of $M_{Lz}$ are more skewed towards lower $z_L$ because heavy halos are rare at high redshift.

A similar calculation allows us to find a probability distribution for the lens' virial mass. 
Starting from the r.h.s.~of Eq.~\eqref{eq:zL_probability_base} but integrating over $z_L$ gives
\begin{equation}\label{eq:Mvir_probability}
    \frac{\de P}{\de M_{\rm vir}}
    =
    \sum_k \xi_{0,k}^2\frac{(1+z_{k})^2}{H(z_k)}
    \left|\frac{\partial M_{Lz}}{\partial z_L}\right|^{-1}_k
    \frac{\de n(M_{\rm vir}, z_{k})}{\de M_{\rm vir}}
    \;.
\end{equation}
Here $k = 1$, $2$ labels the two solutions of $M_{\rm vir}(z_L,\bar M_{Lz})$, see panel A in Fig.~\ref{fig:lens_projection}. 
The probability distribution for the virial mass, Eq.~\eqref{eq:Mvir_probability}, is shown in panel D of Fig.~\ref{fig:lens_projection}. 
The probability is rather peaked at the lowest value due to the divergence of the jacobian at the minimum of $M_{\rm vir}$. The probability of $M_{\rm vir}$ being much larger than the minimum is very suppressed: it requires $z_L\to 0, z_S$, a limit in which the cross section ($\propto \xi_0^2$) vanishes very rapidly (see Fig.~\ref{fig:lens_projection}, A, B).

This framework allows one to constrain the virial mass to within a factor of a few. For $M_{Lz} = 10^7 \, M_\odot$, the 95\% c.l.~limits are $M_{\rm vir} \in [3.8 ,~11)\, 10^{9} M_\odot$ for $z_S = 1$ and $[2.3,~7.1)\,  10^{9} M_\odot$ at $z_S = 3$. The distribution becomes wider for higher masses and redshifts: for $z_S = 10$ the 95\% c.l.~ranges are $[1.7,~17)\, 10^{9} M_\odot$ for $M_{Lz} = 10^7 \, M_\odot$
and $[9.7 \cdot 10^{12},~5.3 \cdot 10^{14}) \, M_\odot$ for $M_{Lz} = 10^{12} \,M_\odot$. 
The virial mass cannot take values below the lower limits above, as $M_{\rm vir}$ has an absolute minimum as a function of $z_L$.
Note that in general, the posterior distribution of $M_{Lz}$, $z_S$ and other parameters will lead to an uncertainty on this lower bound, as well as a broadening of the probability density shown in panel D of Fig.~\ref{fig:lens_projection}. Nonetheless, given the precision in $M_{Lz}$, $\mathcal{D}_S$ (e.g.~Fig.~\ref{fig:forecast_LISA_impact_parameter_dependence}), $M_{\rm vir}$'s upper limit uncertainty will be dominated by lack of knowledge of $z_L$.

We note that the limitations discussed above may be lifted if the source/lens system can be accurately identified, e.g.~via EM follow-ups \cite{Hannuksela:2020xor,2022arXiv220408732W}. 

\subsection{Self-interacting dark matter} \label{sec:forecast_sidm}

SIDM has been proposed to solve the core-cusp and missing satellites problems \cite{Spergel:1999mh}. In this scenario, DM particles scatter elastically with each other, leading to deviations from CDM predictions regarding the inner halo structure \cite{Rocha:2012jg,Vogelsberger:2012ku,Elbert:2014bma,Tulin:2017ara,Brinckmann:2017uve,Nadler:2020ulu}. The self-interaction is described by a cross-section, which is in general velocity-dependent. For a recent summary of constraints and measurement claims see Table I of Ref.~\cite{Tulin:2017ara}.

We will estimate the size of the core through the condition \cite{Kaplinghat:2015aga}:
\begin{equation}\label{eq:sidm_core_condition}
    \frac{\langle\sigma v\rangle}{m}\rho(r_c)t_{\rm age}
    \simeq 
    1
    \;.
\end{equation}
Here $\rho(r_c)$ is the density evaluated at the core radius $r_c$, $\langle\sigma v\rangle$ is the velocity-averaged self-interaction cross-section, $m$ is the mass of the DM particle and $t_{\rm age}$ is the time since the formation of a given structure. 
Equation \eqref{eq:sidm_core_condition} establishes that the core forms where the DM density $\rho$ is high enough for DM particles to interact on average at least once since the formation of the halo. We will take $t_{\rm age} = 5 \, {\rm Gyr}$ as our fiducial choice. This corresponds to the age of clusters of galaxies. It is a conservative choice, as lighter structures will have higher $t_{\rm age}$, leading to larger cores and more stringent constraints.

Let us first give an order-of-magnitude estimate for the constraints on the DM cross-section $\langle \sigma v \rangle / m$ expected from a detection of a lensed GW event.
At large-enough radii, the density of DM halos is typically described by an NFW profile, $\rho_{\rm NFW}(r) \equiv 4 \rho_s (r / r_s )^{-1} (1 + r / r_s )^{-2}$ for $r > r_c$.
Self-interactions can instead lead to a constant density inside the core, $\rho(r) \simeq \rho_{\rm NFW}(r_c) \simeq 4 \rho_s / (r_c / r_s)$ for $r < r_c$ (with $r_c \ll r_s$). In this case, the core size can be characterised by the dimensionless variable $\tilde x_c \equiv r_c / r_s$ (analogous to $x_c$ for the CIS lens discussed previously).
Using the relation \eqref{eq:sidm_core_condition}, $\tilde x_c$ is approximately equal to
\begin{align}
\label{eq:nfw_xc_sidm}
    \tilde x_c 
    &=
    6.4 \cdot 10^2 
    \left( \frac{10^{12}\, M_\odot}{M_{\rm vir}} \right)^{0.23}
    \times 
    \nonumber \\
    &\hspace{2.5cm}
    \left( \frac{\langle\sigma v\rangle / m}{1\, {\rm cm^2 / g}}\right)
    \left( \frac{t_{\rm age}}{5\, {\rm Gyr}}\right)
    \;.
\end{align}
Here we replaced $\rho_s$ using its phenomenological relation with the virial mass (see e.g.~Eqs.~(42) and (43) of Ref.~\cite{Choi:2021jqn}).
This relation can be turned into a bound for $\langle \sigma v\rangle / m$. Indeed, using Eq.~\eqref{eq:nfw_xc_sidm}, a bound from GW lensing on the dimensionless core size $\tilde x_c \lesssim \mathcal O(1)$ for virial masses $M_{\rm vir} \simeq 10^{10}\, M_\odot$ would constrain the averaged cross section to be $\langle \sigma v\rangle / m \lesssim 5 \cdot 10^{-4} \, {\rm cm^2 / g}$.
In this estimate we made the assumption of having $\tilde x_c \lesssim \mathcal O(1)$ at $M_{\rm vir} \simeq 10^{10}\, M_\odot$. This is expected to hold, as virial masses in this range typically produce lensing signals in the GO limit, where lensing parameters are obtained accurately. 
More sophisticated lensing forecasts, for a cored NFW profile, are needed in order to make this estimate more precise.

Assuming instead that the matter density is described by a CIS profile, we can outline how to obtain concrete bounds using our forecast results. In this case, the projected core size $x_c\equiv r_c/\xi_0$ is given by 
\begin{align}\label{eq:sidm_cis_xc}
    x_c 
    &= 
    \frac{100.3}{(1+z_L)}
    \left(
    \frac{1 \, {\rm Gpc}}{d_{\rm eff}}
    \right)
    \left(
    \frac{10^{12} \, M_\odot}{M_{\rm vir}}
    \right)^{1/3} 
    \times
    \nonumber \\
    & \hspace{2.5cm}
    \left(
    \frac{t_{\rm age}}{5 \,{\rm Gyr}}
    \right)^{1/2}
    \left(
    \frac{\langle\sigma v\rangle/m}{1\, {\rm cm^2/g}}
    \right)^{1/2}
    \;.
\end{align}
This expression has been obtained by evaluating Eq.~\eqref{eq:sidm_core_condition} on the CIS density profile, substituting $\xi_0$ from Eq.~\eqref{eq:lensing_freq_dimensionless} in terms of the virial mass \eqref{eq:MLz_cis}, where we neglect terms $\sim r_c/r_{\rm vir} \ll 1$. Note that the dependence between the projected core size and the cross-section is quadratic, as given by the CIS dependence. Therefore, the capacity to probe very small cores grants access to very low SIDM cross-sections. This is in contrast to the NFW case, where this relation is only linear, Eq.~\eqref{eq:nfw_xc_sidm}. The different behaviour originates from the different slopes of the two profiles in the region $r_c \lesssim r < r_s$.

We will assume the lens' redshift that maximizes $(1 + z_L) d_{\rm eff}$, for fixed $z_S$. This choice turns the equality in Eq.~\eqref{eq:sidm_cis_xc} into a lower bound on the projected core size and is a pessimistic assumption in terms of using $x_c$ to constrain SIDM.
We will show our results as a function of $M_{\rm vir}$, obtained from $M_{Lz}$ with this choice of $d_{\rm eff}$. The choice of $z_L$ results in a plausible value for the virial mass given the uncertainties, see Fig.~\ref{fig:lens_projection}.

\begin{figure}
    \centering
    \includegraphics[width=\columnwidth]{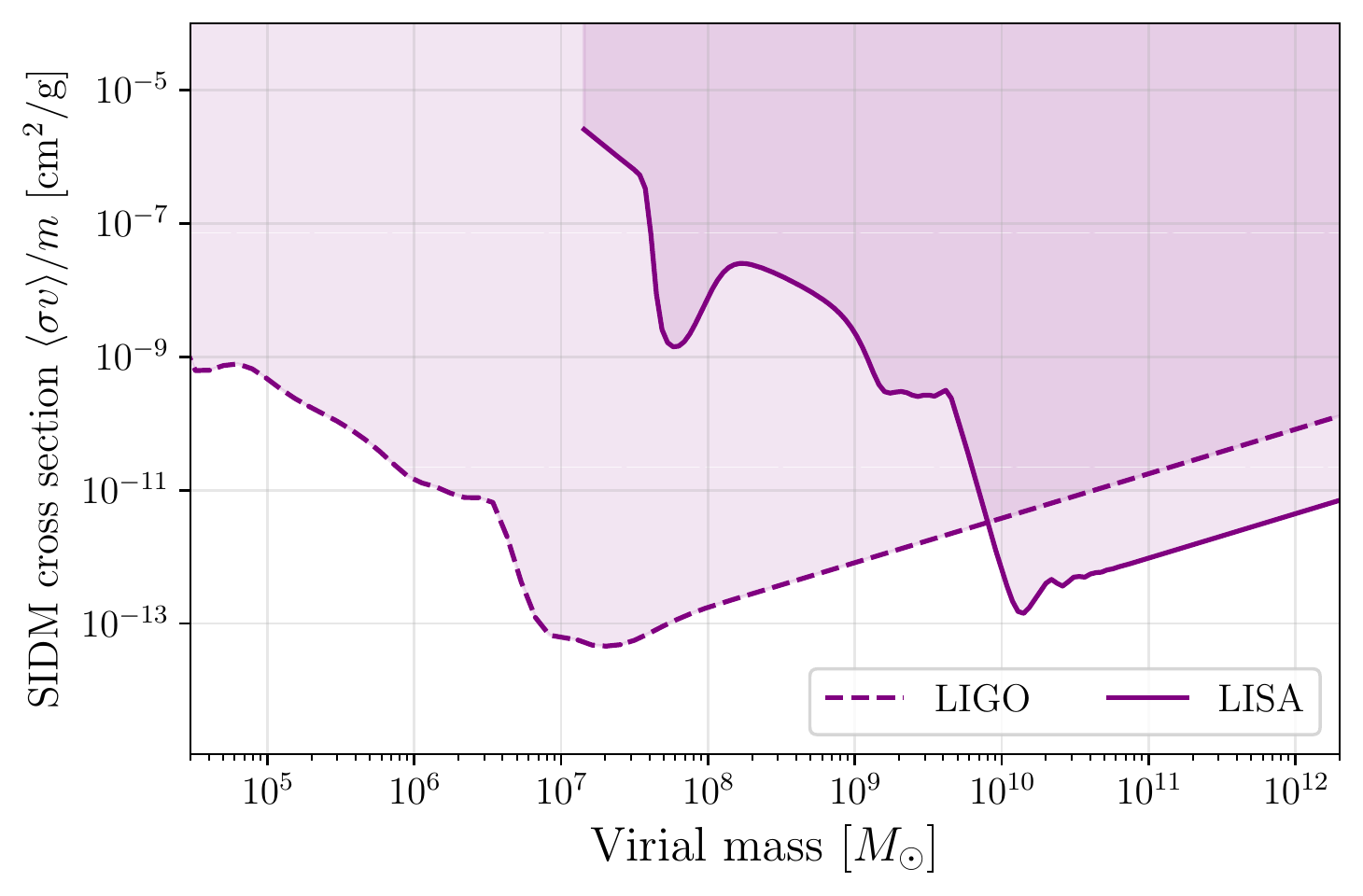}
    \caption{GW lensing as a probe of SIDM for the CIS profile. The shaded regions can be probed by strongly lensed GWs. Here $y=0.6$, $z_S=3$, $M_{\rm BBH}=10^6\, M_{\odot}$, ${\rm SNR}=10^3$ for LISA and $z_S=0.2$, $M_{\rm BBH}=30\, M_{\odot}$, ${\rm SNR}=10^2$ for LIGO. 
    The lens redshift is chosen to minimize the projected core size (see text).
    Constraints on the cross section scale as $\propto \rho(r_c)$, Eq.~\eqref{eq:sidm_core_condition}. Therefore, results for an NFW profile are expected to scale as the square root of the CIS ones, compare Eq.~\eqref{eq:nfw_xc_sidm} and \eqref{eq:sidm_cis_xc}. 
    The forecasted CIS results are orders of magnitude tighter than existing constraints \cite{Tulin:2017ara}.
    Ultimately, probes based on very small cores (low $\langle\sigma v\rangle/m$) will be limited by astrophysical uncertainties.
    }
    \label{fig:sidm_prospective_constraints}
\end{figure}

Figure \ref{fig:sidm_prospective_constraints} shows the values of the SIDM cross-section accessible to a typical lensed LIGO and LISA source ($y=0.6$), as a function of the halo mass for the CIS profile. The curves are derived by interpreting the 95\% marginalised posterior ($2\Delta x_c$, cf.~Sec.~\ref{sec:forcast_CIS}) as the minimum core size that can be probed, and using Eq.~\eqref{eq:sidm_cis_xc} to relate it to $\langle\sigma v\rangle/m$. This interpretation is supported by the convergence of the posteriors as $x_c\to 0$, Fig.~\ref{fig:forecast_LISA_core_size_dependence}. 
We limit the constraints to the region where $x_c \leq 1$, as no multiple images occur otherwise, Eq.~\eqref{eq:cis_multiple_images}. Self-Interacting DM may still be probed in that regime, but a separate analysis needs to be performed.

The capacity of lensed GWs to probe SIDM is optimal at the onset of magnification $w_{\rm ISCO}\sim 10$. While $\Delta x_c$ is smaller at higher $M_{Lz}$, lighter lenses have smaller $\xi_0$, making small cores appear larger. At masses below the onset of magnification, the decrease in sensitivity offsets the projection effect. 
Interestingly, under our conservative assumptions for $d_{\rm eff}$, the optimal LIGO sensitivity corresponds to halos dominated by DM, with very low baryonic mass. For instance, for ultra-faint dwarfs, $M_{\rm vir}\sim 10^9\, M_{\odot}$, the stellar-component mass is $M_\star \sim 10^{-5} \,M_{\rm vir}$ (see Fig.~6 in Ref.~\cite{Bullock:2017xww}).
While the uncertainty on $z_L$ implies that $M_{\rm vir}$ can not be precisely determined, the probabilistic treatment discussed in Sec.~\ref{sec:lens_mass} can be used to place plausible limits.

The above estimates rely on several simplifications that need to be revised to obtain accurate constraints on the properties of DM. We have used the CIS lens model, which might fail to capture further details of the lens, such as the density profile outside the core (different from NFW).
The fact that CIS grows as $1 / r^2$ at small $r$ (but outside of the core) gives very strong constraints while shallower inner profiles, such as NFW below the scaling radius, will plausibly lead to milder constraints, as seen by the estimate in Eq.~\eqref{eq:nfw_xc_sidm}.
Another necessary refinement is the inclusion of stellar/gas components on top of the core. This will be most important for large galactic halos, where cored profiles can arise due to baryonic effects (cf.~Fig.~13 of Ref.~\cite{Bullock:2017xww}).%
\footnote{An additional complication stems from the fact that baryonic feedback can mimic or hide the effects of DM scattering, e.g.~forming a core in a cold DM distribution \cite{Madau:2014ija,Onorbe:2015ija,Chan:2015tna,Tollet:2015gqa}. Lensed GWs can mitigate these uncertainties by constraining the halo mass, at least if low-mass, dark-matter dominated halos (e.g. $M_{\rm vir}\lesssim 10^9 \, M_\odot$) are found among the lenses.} 
This contribution may be important even for low $M_{\rm vir}$ when probing low cross sections, and hence very small cores.
Eventually, realistic analyses should extend the probabilistic treatment of Sec.~\ref{sec:lens_mass} and consider non-axisymmetric lens models. 
We briefly discuss these extensions in Sec.~\ref{sec:non_symmetric_lenses}.

\subsection{Ultra-light dark matter} \label{sec:forecast_uldm}

ULDM refers to models where DM is a bosonic particle with a very low mass $m_\phi$, typically in the range $10^{-22} \, {\rm eV} \lesssim m_\phi \lesssim 1 \, {\rm eV}$ \cite{Ferreira:2020fam,Hui:2021tkt}. These models exhibit interesting phenomenology on small scales, relative to the field's de Broglie wavelength
$\lambda_\phi \simeq 
1.92 \, {\rm kpc}\left({10^{-22}\,{\rm eV}}/{m_\phi}\right)\left({(10\,{\rm km/s)}}/{v}\right)$ \cite{Hui:2016ltb}. 
These fields appear generically in extensions of the Standard Model of particle physics e.g.~to solve the strong-CP problem (QCD axion) \cite{Graham:2015ouw,DiLuzio:2020wdo}. 
In this Section, we will consider non-interacting, scalar ULDM.

ULDM predict cored DM halos with a minimum size
\begin{equation}\label{eq:fdm_core_size}
    r_{1/2}
    \geq 
    0.33
    \,{\rm kpc}\, 
    \frac{10^9 \, M_\odot}{M_c}
    \left(
    \frac{10^{-22}\, {\rm eV }}{m_\phi}
    \right)^2
    \;,
\end{equation}
and a maximum central density.
Here $M_c$ is the mass of the solitonic core \cite{Hui:2016ltb}, and equality holds when the solitonic core forms. 
The solitonic-core mass is related to the virial mass by 
\begin{align}\label{eq:fdm_core_virial}
    M_c 
    &
    \simeq 
    6.7
    \cdot 
    10^7 \,
    M_\odot 
    (1+z_L)^{1/2}
    \times 
    \nonumber \\
    & \quad \quad
    \left(
    \frac{M_{\rm vir}}{10^{10} \, M_\odot}
    \right)^{1/3}
    \left(
    \frac{m_\phi}{10^{-22}\, {\rm eV}}
    \right)^{-1} 
    \;,
\end{align}
(see Eq.~(6) in Ref.~\cite{Schive:2014hza}, neglecting a redshift factor $\propto(\zeta(z)/\zeta(0))^{1/6}$).
\footnote{Ultra-light scalars also predict a minimum $M_{\rm vir}$ for DM halos (Eq.~(42) in Ref.~\cite{Hui:2016ltb}).
Strongly-lensed GWs constrain $M_{Lz}$, and thus can bound $M_{\rm vir}$, following Eq.~\eqref{eq:Mvir_probability}. This test is less powerful than constraining the size of the core (although less model-dependent) and we will not consider it further.}

Assuming that DM halos can be modelled by the cored isothermal sphere with $x_c\simeq r_{1/2}/\xi_0$ results in a minimum projected core size
\begin{equation}\label{eq:forecast_fdm_xc}
  x_c 
  \geq 
  \frac{50.4}{(1+z_L)^{3/2}}
  \left(\frac{10^{10} \, M_\odot}{M_{\rm vir}}\right)
  \left(\frac{10^{-22}\, \rm{eV}}{m_\phi}\right)
  \left(\frac{1\,{\rm Gpc}}{ d_{\rm eff}}\right)
  \,.
\end{equation}
This expression follows from combining Eqs.~\eqref{eq:fdm_core_size} and \eqref{eq:fdm_core_virial}, expressing $\xi_0 = \sqrt{4 GM_{Lz}d_{\rm eff}}$, Eq.~\eqref{eq:lensing_freq_dimensionless}, also in terms of the virial mass for the CIS \eqref{eq:MLz_cis}. 
We will assume the lens redshift is such that $(1+z_L)^{3/2}d_{\rm eff}$ is maximized, for fixed $z_S$ (see discussion after Eq.~\eqref{eq:sidm_cis_xc} and in Sec.~\ref{sec:lens_mass}). This gives a lower bound for the inequality in Eq.~\eqref{eq:forecast_fdm_xc}, and hence pessimistic results.

\begin{figure}
 \includegraphics[width=\columnwidth]{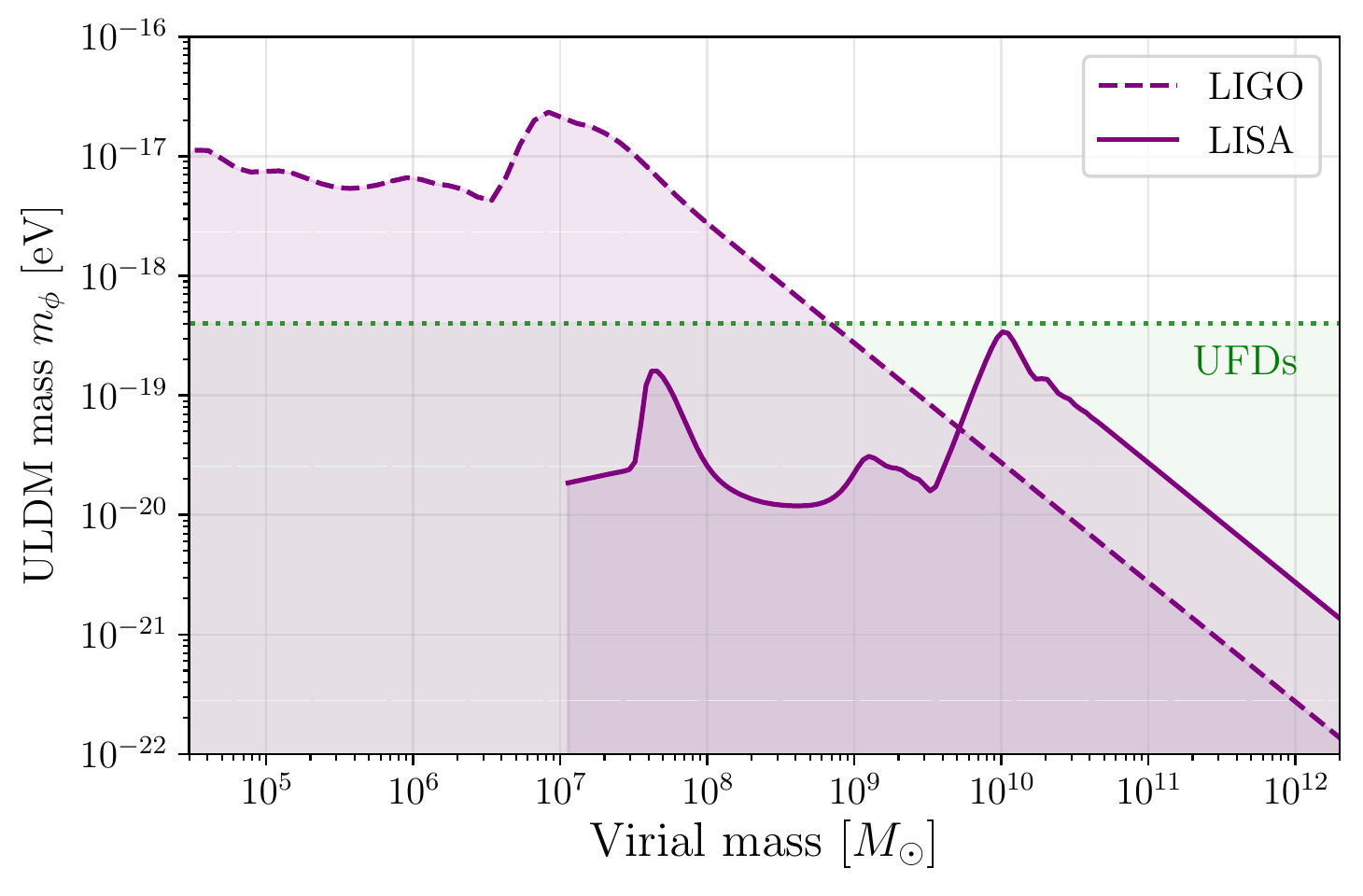}
 \caption{GW lensing as a probe of ultra-light dark matter. 
 The shaded regions show ULDM masses accessible to lensed GW in which a core can be excluded, Eq.~\eqref{eq:forecast_fdm_xc}. We set  $y=0.6$ and $z_S=3$, $M_{\rm BBH}=10^6\, M_{\odot}$, ${\rm SNR}=10^3$ for LISA, $z_S=0.2$, $M_{\rm BBH}=30\, M_{\odot}$, ${\rm SNR}=10^2$ for LIGO. The lens redshift is chosen to minimize the projected core size (see text). The green region represents comparable limits from ultra-compact dwarf galaxies \cite{Dalal:2022rmp}.
 }\label{fig:forecast_fdm}
\end{figure}

Figure \ref{fig:forecast_fdm} shows the range of DM masses that produce cores detectable by LIGO and LISA. Masses in the shaded regions and below can potentially be excluded by observations of strongly-lensed GWs (we have taken 2$\sigma$ exclusion in $\Delta x_c$, see discussion on SIDM, Sec.~\ref{sec:forecast_sidm}).
Again, we terminate the plots at low masses, where $\Delta {x_c} > 0.5$, i.e.~when no multiple images can form.
Note that the sensitivity is well above the $m_\phi \sim 10^{-22}\, {\rm eV}$ value for a broad range of halo masses, potentially probing fields with $\lambda_\phi\gtrsim \mathcal{O}(1\, {\rm kpc})$.

We expect our modelling of the core to be accurate for halos below a certain mass: cosmological simulations of ULDM show how low mass halos ($M_{\rm vir}\lesssim 10^8\, M_\odot$) flatten below a certain radius (Fig.~13 of Ref.~\cite{May:2021wwp}), consistently with our CIS modelling. 
In contrast, heavier halos develop dense and steep cores $\rho \propto (1+(r/r_c)^2)^{-8}$ \cite{Schive:2014hza}, not captured by the CIS parametrisation. More massive halos will also have a larger baryonic component contributing to the profile (Fig.~4 of Ref.~\cite{Mocz:2019uyd}). 
Outside the core, simulations predict an NFW profile with a different slope. This difference will be most relevant for large impact parameters and less so for the central image. Other aspects of the analysis need to be refined, as already discussed in the last paragraph of Sec.~\ref{sec:forecast_sidm}.

\subsection{Non-axisymmetric lens profiles}\label{sec:non_symmetric_lenses}
Let us now comment on how our results extend to non-axisymmetric lenses. 
To keep our discussion simple, we will focus on elliptical lenses, i.e.~replacing the dependence on $x \to \sqrt{x_1^2 + x_2^2/q^2}$ \cite{kormann1994isothermal}.
This introduces two parameters, the ellipticity $q$ and the angle between the lens' axis and source $\delta = \arctan{y_2/y_1}$. Ellipticity can be quite substantial (i.e.~the axis ratio of the ellipse can be few tens of percent) and degenerate with other deviations from spherical symmetry \cite{Shajib:2022con,Keeton:1996tq}. 
Elliptical lenses can form additional GO images in certain regions of the source plane \cite{kormann1994isothermal}. 

Breaking axial symmetry will introduce additional degeneracies. For example, an elliptic generalization of the SIS \cite{kormann1994isothermal} at moderate impact parameter will form only two images. Then, in the GO limit only two combinations of $M_{Lz}$, $y$, $q$ and $\delta$ can be constrained, cf.~Eq.~\eqref{eq:parameter_reconstruction_GO}. A CIS (or broad gSIS) in the same regime forms 3 images, allowing to constrain 4 combinations of $M_{Lz}$, $y$, $q$, $\delta$ and $x_c$ but leaving one combination unconstrained. 

While full reconstruction of ellipsoidal lenses is compromised in the 2-3 image GO regime, WO effects greatly improve the prospects.
In the bGO approximation, an event with 3 images allows the inference of up to  $7$ parameters, Eq.~\eqref{eq:parameter_reconstruction_bGO}. This is enough to fully constrain the ellipsoidal generalization of the CIS or broad gSIS (5 parameters). We expect the parameter reconstruction to improve in the full WO regime, in which the entire extent of the lens contributes to the amplification factor \cite{Savastano:2023spl}.

Information based on central images is robust to ellipticity and can lead to constraints even in the GO limit. 
The existence of a central image is due to the lens' profile and is not affected by ellipticity or lens orientation. Detection of a central image (i.e.~a system with 3 GO images) thus sets limits that are independent of $q$, $\delta$, such as a lower bound on the lens' core size (CIS), or an upper bound on the density slope (broad gSIS), see Fig.~\ref{fig:detectability_third_img}. Conversely, non-detection of a central image (after modelling selection effects) would favour a singular isothermal ellipsoid and limit deviations from it. 

At low impact parameter the number of images increases by two: singular lenses (e.g.~singular isothermal ellipsoid) form up to 4 images, while regular non-symmetric lenses form up to 5, with the central image being typically the faintest. These offer enough information to reconstruct all parameters within the model, even in the GO limit. Moreover, as argued above, simply counting the images should lead to an upper bound on deviations from the SIS. 
Selection effects need to be modeled, accounting for the chance of images not being detected (e.g.~due to demagnification or detector duty cycle). 
This possibility complicates the identification of the central image and requires a more sophisticated analysis. 

\section{Conclusions}\label{sec:conclusions}

We investigated the phenomenology of GW lensing and its potential to probe cosmological structures utilizing WO effects. In Sec.~\ref{sec:lensing} we reviewed the equations governing GW lensing and the methods used to obtain predictions (see Ref.~\cite{Tambalo:2022plm} for details).

We then addressed the phenomenology of two symmetric lens models that generalise the singular isothermal sphere (SIS): the power-law generalised SIS (gSIS, Sec.~\ref{sec:slope_lens}), in which the slope $k$ of the lensing potential is arbitrary and the cored isothermal sphere (CIS, Sec.~\ref{sec:cored_lens}), with a finite-density core of size $x_c$. For each lens we presented the physical scales, GO structure (images, caustics, magnifications), bGO corrections, and WO predictions, including the convergence to the SIS limit.
We mostly focused on the strong-lensing regime with mild magnification, although we showed examples of WO signatures in the single-image regime (weak lensing) and for sources near the caustics (Figs.~\ref{fig:cis_vary_xc}, \ref{fig:cis_vary_y}).

A distinctive feature of these lenses is the existence of a central image associated with the maximum of the Fermat potential. Central images produce a secondary modulation in the amplification factor, encoding additional information about the lens. For sources near a caustic, the additional modulation of the amplification becomes very broad in frequency space and pushes convergence to GO towards very high dimensionless frequency $w\gg 1/\text{min}\,|\phi_{I}-\phi_{J}| $. 
The central image vanishes for SIS and narrow gSIS profiles ($k\leq 1$): its detection is thus a direct probe of the inner structure of halos.
We discussed qualitative aspects and observational prospects (Sec.~\ref{sec:lens_detecability}), including parameter estimation in the $w\gg1$ limit, the prospect of detecting central images and mismatches between GWs lensed by gSIS and CIS, relative to SIS.

We then addressed the potential for GW detectors to probe lens features (Sec.~\ref{sec:forecasts}). We used a Fisher-matrix analysis to forecast the ability of LISA and LIGO to reconstruct the parameters of the lens (mass, impact parameter, slope/core size) and of the source (signal amplitude and initial phase), as well as their correlations. We focused on equal-mass, non-spinning sources and kept their parameters fixed (except the amplitude and overall phase). While our analysis is simplified, it produces reasonable agreement with more detailed treatments of the source and the detector \cite{Caliskan:2022hbu}.
The large degeneracy between lens parameters makes the Fisher matrix very sensitive to numerical errors in the computation. To circumvent this issue, we used approximations to the Fisher matrix at low/high frequencies and focused on typical strong-lensing situations (away from caustic) where convergence to GO occurs at relatively low dimensionless frequencies. We also addressed the validity of the linear signal approximation (LSA) in different cases.

LISA and LIGO can identify the CIS lens parameters with precision $\sim \mathcal{O}(1/\text{SNR})$ for large $w_{\rm chirp}\propto M_{Lz}/M_{\rm BBH}$ (GO limit). The precision is somewhat higher for the core size and lower for the lens mass. The marginalised constraints saturate in the GO limit because of the large degeneracies between lens parameters, although the 2D posteriors become thinner as $M_{Lz}$ increases. This behaviour is generic in strongly-lensed systems with a central image (three images total), allowing to recover up to four parameters in GO (two time delays with high accuracy $\sim 1/M_{Lz}$ and two magnification ratios $\sim 1/\text{SNR}$).
The constraints degrade at lower lens masses (such that the merger occurs below the onset of magnification), due to both parameter degeneracies and diffraction suppressing the amplification factor. 

Wave optics can substantially improve the constraints in regimes where GO does not provide enough information. We saw this explicitly for the narrow gSIS lens, which only produces two GO images. In this case, the most stringent constraints are found when the onset of magnification occurs close to the merger, $w_{\rm chirp}\propto M_{Lz}/M_{\rm BBH}\sim 1-10$. The constraints are driven by WO and degrade for increasing $M_{Lz}$, as the bGO corrections become negligible. The study of the gSIS revealed that the LSA overestimates the sensitivity at high $M_{Lz}$ due to the contribution from the lens' cusp. For this reason, in addition to full WO we quoted bGO-only results, which neglect the cusp contribution and are thus conservative for $w_{\rm chirp}\gg 1$. Our Fisher matrix results are supported by the mismatch analysis (Fig.~\ref{fig:mismatches}), which does not rely on the LSA.
Given the limitations we uncovered for the Fisher analysis on lensed signals, it would be interesting to perform more sophisticated forecasts by sampling the full likelihood and better asses past results obtained in the literature (e.g.~\cite{Takahashi:2003ix}).

Identifying lens features at percent and sub-percent precision (for LIGO and LISA respectively) opens up novel applications of GW lensing. While GW lensing measures projected quantities, limits on the lens mass and physical scales can be obtained under reasonable assumptions regarding the cosmology and the halo mass function (Sec.~\ref{sec:lens_mass} and Fig.~\ref{fig:lens_projection}). In particular $M_{\rm vir}$, $(\xi_0$) has a hard lower (upper) limit, and their 95\% credible regions can be obtained within a factor of a few, except for very high redshifts. These limitations may be lifted if the lens redshift can be identified, e.g.~via EM follow-up \cite{Hannuksela:2020xor,2022arXiv220408732W}.

The possibility of simultaneously constraining the lens mass and core size enables tests of Dark Matter scenarios. We addressed the prospects of probing self-interacting and ultra-light DM models (Sec.~\ref{sec:forecast_sidm} and \ref{sec:forecast_uldm}, respectively), which predict cored halos whose size depends on the properties of DM and the halo mass. Deriving simple expressions for the projected core size in these models, we translated the sensitivity from our forecasts into the parameter space accessible to strongly lensed GWs. 
GWs can probe SIDM cross sections as low as $10^{-4}-10^{-5} {\rm cm}^2/{\rm g}$, with the maximum sensitivity achieved around the onset of magnification (Fig.~\ref{fig:sidm_prospective_constraints}). For ULDM, the core sizes of DM halos can be tested if $m_\phi\lesssim 10^{-17}\, {\rm eV}$ (cf.~Fig.~\ref{fig:forecast_fdm}).
These bounds assume the largest possible projected core size for a given source and are thus conservative within our model. Nonetheless, the assumption of a $\rho\propto r^{-2}$ profile might give overly optimistic constraints for SIDM, compared to a shallower central profile such as NFW.

The capacity to detect very small cores (recall $x_c$ is normalized by the Einstein radius) leads to high sensitivity, particularly for lighter halos, and in some cases well beyond current limits. The capacity of GWs to constrain $M_{Lz}$ and $x_c$ independently may allow observations to discern DM-dominated halos ($M_{\rm vir}\lesssim 10^{9} \, M_\odot$), and thus whether the core is likely to represent a signature of DM properties, rather than baryonic effects.
We note that turning these prospects into actual constraints requires further modelling of lens features, astrophysical effects and lensing cross sections, as well as robust detection of lensed GW signals. To illustrate the need for more realistic modelling, we discuss how our analyses, based on axisymmetric lenses, can be generalized to elliptical lenses (Sec. \ref{sec:non_symmetric_lenses}). Despite the challenges ahead, our results warrant further investigations into the potential of DM tests with strongly-lensed GWs.

Our analysis highlights several ways in which GW lensing is highly complementary to electromagnetic observations:
\begin{enumerate}
 \item The low frequency of GW sources enables observation of WO effects in lenses with mass $M_{Lz}\gtrsim 10^2$, $10^5 \, M_\odot$ (point-lens) and $M_{\rm vir}\gtrsim 10^4$, $10^7 \, M_\odot$ (extended) for LVK and LISA, respectively. WO provides information about a region of the lens plane of size $\sim w^{-1/2}$.
 \item WO effects provide additional information through the onset of magnification and mild frequency-dependence at low/high dimensionless frequencies (beyond GO), as well as the contributions from cusps. These effects persist in the single-image (weak-lensing) regime at moderate $w$ (Fig.~\ref{fig:cis_vary_y}).
 \item GW lensing is a unique probe of the inner structure of halos. Central, demagnified images are brighter for GWs than for EM sources ($\sqrt{|\mu_H|} \gg |\mu_H|$), cf.~Fig.~\ref{fig:detectability_third_img}. Unlike EM signals, central images of GW are not blocked or outshined by matter in the inner regions of halos.
 \item Central images cause additional frequency-dependent modulation of the amplification pattern and carry the strongest frequency dependence via beyond GO corrections (Figs.~\ref{fig:slope_IS_GO}, \ref{fig:CIS_GO}).
 In the vicinity of a caustic, both the onset of this modulation and convergence to GO is pushed to very high frequencies (Fig.~\ref{fig:cis_vary_xc}).
 \item GW lensing effects enable measurements of individual lens parameters with precision $\sim \mathcal{O}(1/{\rm SNR})$, despite strongly correlated posteriors, Figs.~\ref{fig:forecast_LISA_posteriors}. Constraints on strongly-lensed signals ($y<y_{rc}$) improve for large impact parameters, due to larger time delays, Figs.~\ref{fig:forecast_LISA_impact_parameter_dependence}, \ref{fig:gsis_forecast_LISA_impact_parameter_dependence}. This feature favours observational prospects, since differential probabilities scale as $\propto y^2$.
 \item While lensed waveforms are sensitive to projected parameters ($M_{Lz}$, $y$, $x_c$), limits on the physical scales and virial mass can be placed under reasonable assumptions, Fig.~\ref{fig:lens_projection}. In particular, $M_{\rm vir}$ can be obtained within a factor of a few.
 \item Central images may enable novel tests of astrophysics and fundamental physics, such as probing DM through lens cores (Figs.~\ref{fig:sidm_prospective_constraints}, \ref{fig:forecast_fdm}). Variations of these tests may rely on measuring the slope of the inner halo cusp or the effect of microlenses and supermassive black holes close to the central image.
\end{enumerate}
Based on these prospects, we envision further extensions of our work regarding lens models, source properties, statistical analysis and towards other lensing regimes.

While we focused on simple symmetric lenses, we expect many of our conclusions to hold in more realistic situations. For instance, the central GO image probes the inner region of the lens, which may differ from the outer, averaged lens distribution. While this dependence is exact only for GO and symmetric lenses (the lens equation depends on the integrated projected mass), it should approximate more general settings. 
Further characterization may enable analyses based on central images complementary to those proposed for EM sources, e.g.~\cite{Keeton:2001kg,Evans:2002ut,Li:2011js,Hezaveh:2015oya,Quinn:2016jte}.
Still, more complex lenses may mimic or obscure some of the signatures of cores and cusps. Studying other lenses \cite{Keeton:2001ss}, adding large-scale lens features (variable slope, ellipticity, external convergence \& shear) and small-scale substructure (DM subhalos, stellar population) is thus relevant.
Future work needs to address these modelling challenges and opportunities.

Another interesting direction is a more detailed exploration of DM scenarios. By modelling DM halos as CIS, our work has shown the promise of GW lensing under rather simplistic assumptions. As discussed in Sec.~\ref{sec:dark_matter}, it will be necessary to model lenses more accurately, along the lines mentioned above.
Our work can, in principle, be extended to other DM models. Warm dark matter halos have been shown to have small cores, $r_c/r_{200}\lesssim 10^{-3}$ \cite{Villaescusa-Navarro:2010lsj}. Another promising signature is the inner halo slope $\rho \propto r^{-3/2}$ \cite{Ishiyama:2014uoa,Delos:2022yhn}. This feature is also present in other models such as compact DM structures \cite{Arvanitaki:2019rax}, with a sharp slope $\rho \propto r^{-9/4}$ (although see Ref.~\cite{Berezinsky:2010kq}) and constant density core, whose size is set by the DM velocity dispersion or annihilation cross-section \cite{Li:2012qha}.
Ultimately, deriving constraints will require a detailed assessment of lensing probabilities, including false positives due to baryonic or complex structures.

A key advantage of GWs is the existence of accurate source models.
While we considered the source parameters as fixed (except for distance and phase) and simple sources (equal mass, non-spinning binaries), further understanding the interplay between source models and lensing effects is necessary. 
Besides varying the source parameters, other directions include the study of different sources (e.g.~Extreme-mass-ratio inspirals), and more realistic waveform models (higher harmonics, spin-precession, eccentricity, see Ref.~\cite{Caliskan:2022hbu}).
Ultimately, full Bayesian parameter estimation of lensed signal injections will be necessary to derive accurate posteriors and address the shortcomings of the Fisher matrix.
These studies will allow an understanding of both parameter degeneracies and potential systematics, e.g.~if a lensed event is analyzed with an unlensed waveform.
Eventually, WO amplification factors need to be included in GW searches and parameter estimations, perhaps taking advantage of machine learning algorithms, e.g.~\cite{Dax:2021tsq}.

Our results also warrant investigating other regimes of GW lensing.
Here we have focused on the strong-lensing regime, at mild magnifications. An interesting extension is highly magnified configurations in which the source is close to a caustic \cite{Diego:2019lcd,Diego:2018fzr}, and the maximum magnification is limited by wavelength rather than source size \cite{Ohanian:1974ys,Matsunaga:2006uc}. This requires pushing our numerical methods to new limits, especially in computing forecasts. 
Another future direction is the study of line-of-sight halos \cite{CaganSengul:2020nat,Sengul:2021lxe,Fleury:2021tke}, and explore the interplay between multi-plane lensing and wave optics \cite{Feldbrugge:2020tti} to enable ray-tracing in cosmological simulations.
Finally, wave optics in the single-image regime is also promising, as the frequency dependence may be detectable up to large impact parameters \cite{Takahashi:2003ix,Gao:2021sxw,Caliskan:2022hbu,Savastano:2023spl,Caliskan:2023zqm}. Since the lensing probabilities scale as the square of the impact parameter, weak-lensing effects might be the first detectable signature, or even contribute as a systematic in the waveform reconstruction.

As the detection rate increases, GW lensing is bound to become a reality. 
It is a matter of chance whether LVK searches return confident identification of lensed signals in the coming years. However, the capacity of next-generation detectors to observe binary mergers in the entire Universe \cite{Kalogera:2021bya} will make GW lensing into a reality: even the least likely lens configurations will stand a chance to be observed. Other proposed missions will offer new opportunities, such as WO effects for galactic scale lenses from $\mu{\rm Hz}$ space-borne detectors \cite{Sesana:2019vho}.
As GW astronomy matures, wave optics effects may serve both to identify lensed sources and to study the matter distribution in the Universe.
The complementarity between lensing of GW and of EM sources may enable novel probes of the structure of halos, perhaps helping us identify the nature of DM and opening new insights into astrophysics, fundamental physics, and the history and evolution of the Universe.

\acknowledgments{
It is a pleasure to thank G.~Brando, JM.~Ezquiaga, A.~Kumar Mehta, J.~Poon, S.~Savastano and H.~Villarrubia-Rojo for useful discussions and comments. 
L.D.~acknowledges the research grant support from the Alfred P.~Sloan Foundation (Award Number FG-2021-16495). M.H.-Y.C.~is a Croucher Scholar supported by the Croucher Foundation.
M.H.-Y.C.~is also supported by NSF Grants No.~AST-2006538, PHY-2207502, PHY-090003 and PHY-20043, and NASA Grants No.~19-ATP19-0051, 20-LPS20- 0011 and 21-ATP21-0010.
}

\appendix

\section{Lensing Recap}

Here we introduce briefly wave-optics computations in the low-frequency limit (\ref{subsec:wo_low}) and via numerical methods (\ref{subsec:wo_numerical}), as well as the singular-isothermal sphere lens (\ref{sec:lens_sis}), whose generalizations we used throughout the paper.

\subsection{Wave-Optics: Low-frequency expansion} \label{subsec:wo_low}

Let us summarise the results obtained in \cite{Tambalo:2022plm} regarding the low-$w$ expansion of $F(w)$.
At low frequencies, the GO expansion fails and one needs a different approach to obtain approximations for the amplification factor. In the limit $w \ll 1$, the wavelength of the wave becomes larger than the characteristic scale of the lens, and the signal remains unperturbed: $F(w) \sim 1$. Corrections to this result, at small impact parameters and for axially-symmetric lensing potentials, are computed as a series expansion in powers of $\psi(x)$ \cite{Tambalo:2022plm}:
\begin{align} \label{eq:low_w_expansion}
  F(w) 
  &
  \simeq
  e^{-i w \phi_m}
  \bigg[
    1
    -
    \int_0^{\infty}\, \de z\, 
    z \, e^{-z^2 / 2}
    \times
  \nonumber
  \\
  & 
    \left(
      i w \psi(q)
      + 
      \frac{w^{2}}{2} \psi(q)^2
      + \mathcal{O}(w^3 \psi(q)^3)
    \right)
  \bigg]\;,
\end{align}
where $q \equiv e^{i \pi / 4}z / \sqrt{w}$. 
This expansion is well defined provided that $\psi(q)$ grows less rapidly than the quadratic part of the Fermat potential as $z\to \infty$.

For the lenses we are going to discuss (see Tab.~\ref{tab:lenses_summary}), the integrals in Eq.~\eqref{eq:low_w_expansion} can be evaluated explicitly. 
By suppressing the overall factor $e^{-i w \phi_m}$, for the point lens $\psi(x) = \log x$ we have
\begin{equation}\label{eq:low_w_pt}
  F^{\rm pl}(w) 
  \simeq 
  1 
  + 
  \frac{w}{4}
  \left(
  \pi
  +2 i \gamma_{\mathrm{E}}
  +2 i \log \frac{w}{2}
  \right)
  \;,
\end{equation}
where $\gamma_{\mathrm{E}}$ is the Euler's constant.
Similarly, for SIS
\begin{equation}
  F^{\rm SIS}(w) \simeq 1-(-1)^{3 / 4} \sqrt{\frac{\pi w}{2}}-i w\;.
\end{equation}
While for gSIS we have
\begin{equation}\label{eq:low_w_gsis}
  F^{\rm gSIS}(w) \simeq 1+(-w / 2)^{k / 2} \Gamma\left((2-k)/2\right)\;.
\end{equation}
Finally, for the CIS:
\begin{align}
  F^{\mathrm{CIS}}(w) 
  & \simeq
  1
  -\left[
  (-1)^{3 / 4} 
  \sqrt{\frac{\pi w}{2}}
  +\frac{w x_c}{4}
  \left(
  2 i \gamma_{\mathrm{E}}
  \right.\right.
  \nonumber
  \\
  &
  \left.\left.
  +\pi
  +2 i \log \frac{w}{2}
  +4 i \log \left(2 x_c\right)
  +\frac{4 i}{x_c}
  \right)
  \right]\;.
\end{align}
We will use these expressions to compute $F(w)$ and its derivatives at low frequencies.

\subsection{Wave Optics: Numerical Methods}\label{subsec:wo_numerical}

Numerical methods are necessary to compute the amplification factor at intermediate frequencies, bridging the gap between the low- and high-frequency expansions described above. In Ref.~\cite{Tambalo:2022plm} we developed and validated two numerical methods, regularized contour flow \cite{Ulmer:1994ij} and complex deformation \cite{Feldbrugge:2019fjs}. Both yield the same results at high accuracy, but the regularized contour flow is substantially faster and will be our method of choice. We will give an overview now; further details can be found in Ref.~\cite{Tambalo:2022plm}.

The starting point is the Fourier transform of the integral in \eqref{eq:lensing_wave optics}, $\tilde I(\tau) \equiv \int \de^2 \vect x \int \frac{\de w}{2\pi} \exp\left(iw\left(\phi(\vect x,\vect y)-\tau\right)\right)$, or 
\begin{equation}
 \tilde I(\tau) 
 =
 \int 
 \de^2 \vect x \,
 \delta_D\left(\phi(\vect x,\vect y)-\tau\right)
 \,,
\end{equation}
where $\delta_{D}(x)$ is the Dirac-delta function.
The above integral is computed on contours of equal Fermat potential $\phi(\vect x, \vect y) = \tau$. As contours end on critical points, the first step is to solve the lens equation to find the initial/final conditions for the contours. The integral is then sampled on the nodes that define each contour. Then the nodes are evolved to the next value of $\tau$, adding or removing more nodes depending on the curvature of the contour. The next value of $\tau$ is chosen adaptively and depends on the derivatives of $\tilde I(\tau)$.

The amplification factor \eqref{eq:lensing_wave optics} in the frequency domain is then obtained via a fast-Fourier transform (FFT). This is computed by combining all contours, interpolated on a homogeneous grid, as defined by the frequency range of the FFT. In order to improve the computation, we split $\tilde I(\tau)$ into a regular and a singular part. The singular part is the time-domain counterpart to the GO amplification factor \eqref{eq:lensing_geometric_optics}. It can be computed analytically and consists of discontinuities (for maxima/minima of $\phi$) and logarithmic divergences (for saddle points of $\phi$), which if not dealt with separately contribute spurious high-frequency noise to the FFT.

This algorithm is efficient and accurate at intermediate frequencies. At high frequency, we use the bGO approximation described in Eq.~\eqref{eq:bGO}. The transition between the two regimes, $w_{\rm high}$, is chosen such that the relative deviation $|F_{\rm bGO}/F_{\rm WO}-1|$ is sub-percent. At the same time, $w_{\rm high}$ is chosen relatively low to avoid numerical aliasing from the FFT, usually $w_{\rm high} \sim \mathcal{O}(100)$. The choice for $w_{\rm high}$ is more delicate when computing derivatives of $F(w)$ with respect to the lens parameters since numerical noise is higher in these cases: Appendix \ref{sec:forecast_implementation} discusses this issue in the context of Fisher-matrix forecasts.

\subsection{Singular Isothermal Sphere} \label{sec:lens_sis}

The SIS is characterised by the following density distribution
\begin{equation}\label{eq:rho_SIS}
 \rho(r) 
 = 
 \frac{\sigma_v^2}{2\pi G r^2}
 \;,
\end{equation}
where $\sigma_v$ is the line of sight velocity dispersion. This lens is often used to model dark matter halos. Because of the falloff as $\sim 1/r^2$ of the density profile, the halo mass is formally infinite, and the model thus requires a physical cutoff for $r$, after which the density is imagined to drop to zero quickly. In the context of lensing, this is not an issue as long as the cutoff is much larger than the scale $\xi_0$ specified below. 

The projected mass density of this lens is (Eq.~\eqref{eq:proj_mass_density_def})
\begin{equation}
    \Sigma(\vect \xi) 
    = 
    \frac{\sigma_v^2}{2G |\vect \xi|}
    \,.
\end{equation}
This simple lens model falls within the class of axisymmetric lenses since the projected mass only depends on $|\vect \xi|$. 
The lensing potential is then obtained from Eq.~\eqref{eq:dimensionless_lensing_pot} as 
 $\psi(x) = \sigma_v^2 x/ (\xi_0 G \, \Sigma_{\rm cr})$. In our work we use a standard choice for the normalisation of the lens, $\xi_0 = \sigma_v^2/(G \,\Sigma_{\rm cr})$. With this choice the lensing potential becomes particularly simple
\begin{equation}
    \psi(x) = x
    \;.
\label{eq:psi_SIS}
\end{equation}

In the GO limit we can have one or two images depending on whether the impact parameter is outside or inside the caustic $y_{\rm cr} = 1$ (see Fig.~\ref{fig:lens_models}, Panel C). For $y < y_{\rm cr}$, the two images have magnifications $\mu_{\pm} = 1/y\pm 1$, time delays $\phi_{\pm} = \mp y -1/2$ and Morse phases $n_+ = 0$ (minimum), $n_- = 1/2$ (saddle). For $y > y_{\rm cr}$ only the minimum image survives. Consistently with the prescription of Eq.~\eqref{eq:time_delay_def}, the minimum time delay needs to be set to zero by adding the appropriate $\phi_m(y)  = -y - 1/2$.

In order to properly interpret our results, we now discuss the relationship between the quantities above and the physical properties of the SIS. 
To define the total mass of the SIS it is necessary to truncate the mass density at a finite radius. We will follow the standard definition of considering the virial radius $r_{\rm vir}$ as the radius at which the density reaches $200 \rho_c$
, with $\rho_c = 3H_0^2/(8 \pi G)$ being the critical density of the Universe today ($H_0 = 100\,  h \, \rm{km}\, \rm{s}^{-1}\, \rm{Mpc}^{-1}$ is the Hubble constant), and choose this as the truncation radius. Then, the virial mass $M_{\rm vir}$ is defined as the mass within the virial radius and coincides with the mass of the halo. Using Eq.~\eqref{eq:rho_SIS} and the expression for $r_{\rm vir}$, $M_{\rm vir}$ is obtained as 
\begin{equation}\label{eq:virial_mass_sis}
\begin{aligned}
    M_{\rm vir} 
    &= 4 \pi \int_0^{r_{\rm vir}}\de r\, r^2 \rho(r)  = \frac{2 \sigma_v^3}{5\sqrt{6}G H_0}
    \\
    & = 
    1.8 \cdot 10^{11} \, M_\odot\left(\frac{\sigma_v}{70 \, {\rm km/s}}\right)^3 
    \;.
\end{aligned}
\end{equation}
Alternatively, the expression above can be seen as a relation between the velocity dispersion and the halo mass.

With our normalisations, the effective lens mass is instead given by
$M_{Lz} = 4 \pi^2 (1+z_L)^2\sigma_v^4 d_{\rm eff}/G$ 
(see Eq.~\eqref{eq:lens_mass_def} and our choice of $\xi_0$). Using Eq.~\eqref{eq:virial_mass_sis} we can then relate $M_{Lz}$ with the virial mass:
\begin{equation}\label{eq:MLz_sis}
\begin{aligned}
    M_{Lz}
    &= 
    \frac{4\pi^2}{G}(1+z_L)^2 d_{\rm eff} 
    \left(\frac{5\sqrt{6}}{2} G H_0 M_{\rm vir} \right)^{4/3}
    \\
    &=
    2.3 \cdot 10^6 \, M_{\odot} 
    (1 + z_{L})^2
    \left(\frac{d_{\rm eff}}{1\, {\rm Gpc}}\right) 
    \left(\frac{M_{\rm vir}}{10^9\, M_\odot}\right)^{4/3}\;.
\end{aligned}
\end{equation}
This relation, together with Eq.~\eqref{eq:w_typical_LISA}, suggests that for LISA frequencies and virial masses around $M_{\rm vir}\sim 10^9 \, M_\odot$, lensing is best described by wave optics.

\section{Forecast implementation and tests} \label{sec:forecast_implementation}

\begin{figure*}
\centering
\includegraphics[width=1\columnwidth]{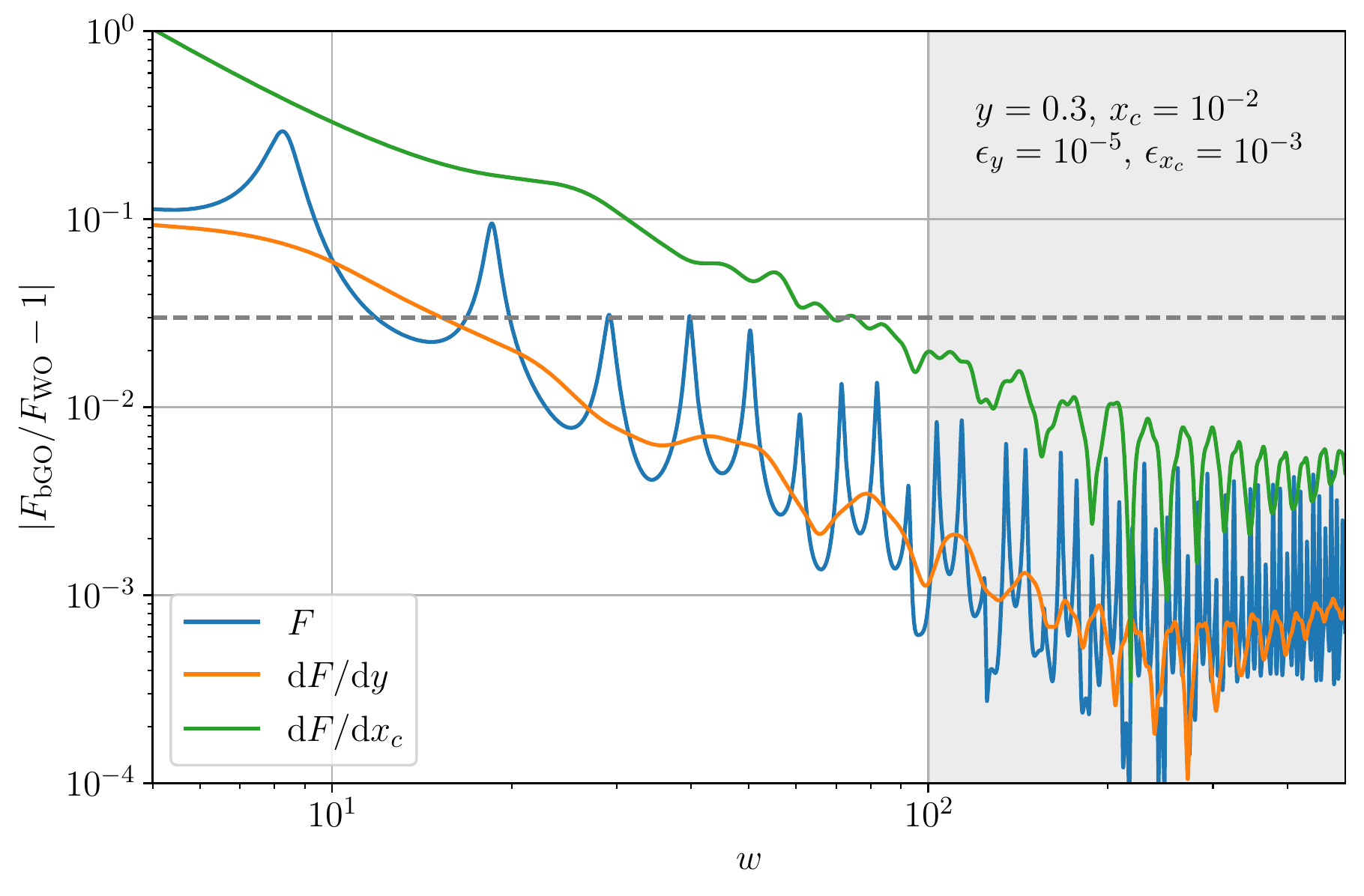}
\includegraphics[width=1\columnwidth]{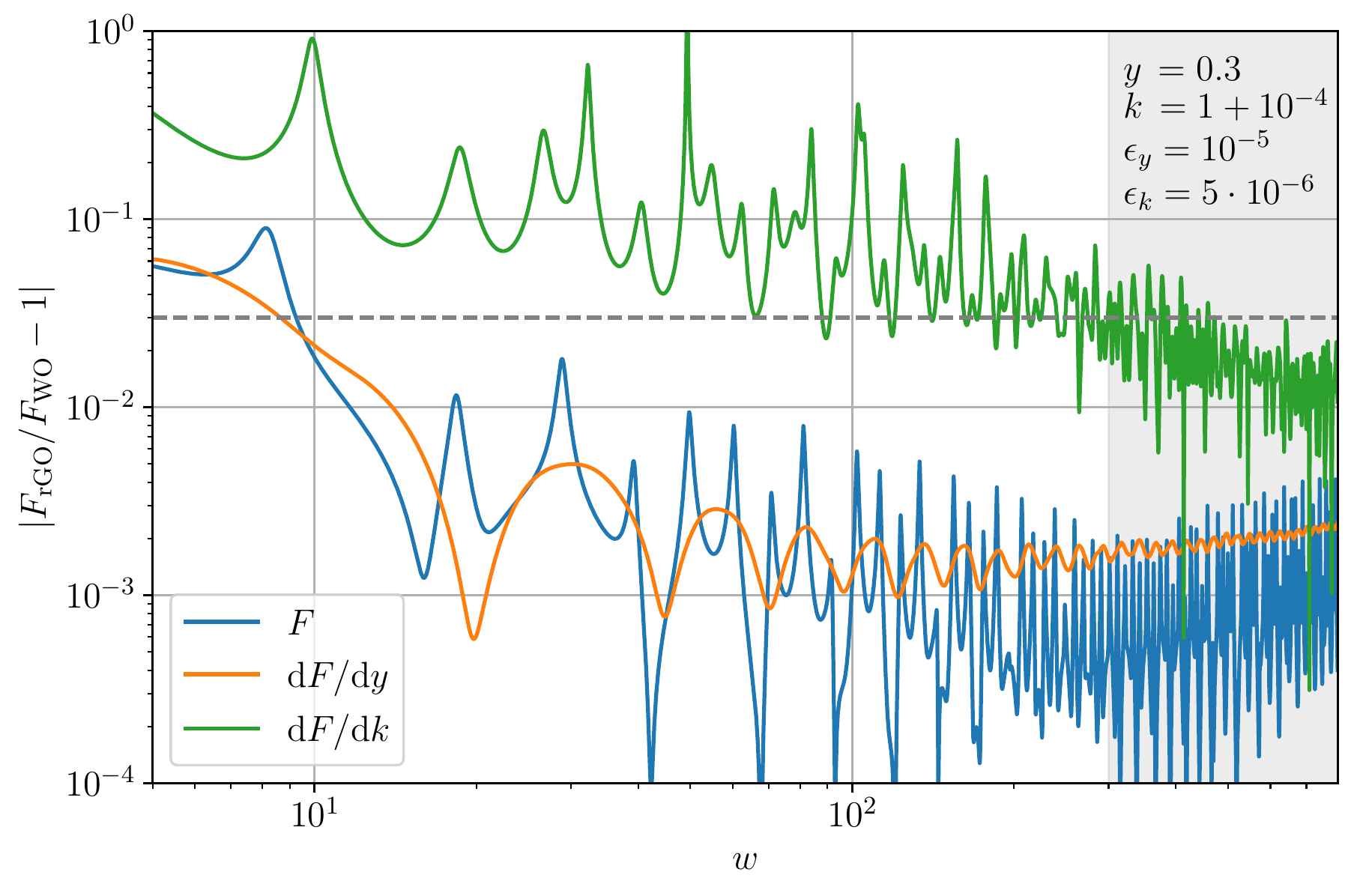}
\caption{Convergence of the amplification factor and the Fisher derivatives to bGO and rGO respectively for a CIS lens (left) and gSIS (right). The bGO/rGO values are used for high frequencies (shaded region) to avoid aliasing noise. 
\label{fig:forecast:high_freq_test}}
\end{figure*}

In this appendix, we are going to discuss the methods and validation procedures we employed in evaluating Fisher-matrix elements. The numerical derivatives of $F(w)$ with respect to the lens parameters $\theta_l$ are particularly delicate. Indeed, as we are going to mention soon, the large degeneracies in the Fisher matrix can amplify numerical errors when computing its inverse.
It is then also important to develop approximation methods at low and high frequencies in order to validate our results in these regions. 

At high frequencies we compute derivatives of the amplification factor using the beyond GO expression, Eq.~\eqref{eq:bGO}, which can be arranged as a sum over GO images $I$,
\begin{equation}\label{eq:fisher_bGO_deriv}
 \partial_l F(w) 
 = 
 \sum_I
 \left( 
 iw\psi_{I,l}+\frac{\mu_{I,l}}{2\mu_I} + \frac{i}{w}\Delta_{I,l} 
 \right)
 F_I(w)
 \,.
\end{equation}
Here uppercase Latin indices $I,J,...$ refer to GO images, lowercase indices $l,m,...$ to lens parameters and commas denote partial derivatives with respect to the lens parameters ($f_{,l} \equiv \partial f/\partial\theta_l$). Additionally, $F_I(w)$ is the GO amplification of the $I$-th image, Eq.~\eqref{eq:lensing_geometric_optics}.
Note that the Morse phases $n_I$ are assumed to be constant in the derivatives.

Parametrizing the effective lens mass by its logarithm provides a convenient rescaling of the Fisher matrix.
Since $M_{Lz}$ only enters through the dimensionless frequency $w$, Eq.~\eqref{eq:lensing_freq_dimensionless}, the associated derivative reads
\begin{equation}
    \frac{\partial F(w)}{\partial \log M_{Lz}}= \frac{\partial F(w)}{\partial \log w}
    = \sum_I\left( i w\psi_{I} - \frac{i}{w}\Delta_{I} \right)F_I(w)\,.
\end{equation}
Here the first equality is fully general and will be used in the intermediate and low-frequency regimes. The second expression is used in the bGO regime.
Similar expressions are used for the derivatives of the cusp terms appearing in the gSIS lens, Eq.~\eqref{eq:F_cusp}.

At low frequencies we compute $F(w)$ and its derivatives using the low-$w$ expansion of Sec.~\ref{subsec:wo_low} (one needs to re-introduce the minimum time delay $\phi_m$ in these expressions, as it depends on the lensing parameters and so it affects the derivatives). Moreover, we checked these results do not change when using the full WO result at low frequency (computed using Eq.~(16) of Ref.~\cite{Tambalo:2022plm}).

Derivatives of $F(w)$ in the intermediate region are computed in the time domain via finite differences and then Fourier transformed. This process significantly worsens the numerical noise at high frequencies, in a manner analogous to the introduction of terms $\propto w$ in the bGO derivatives, Eq.~\eqref{eq:fisher_bGO_deriv}. Given the sensitivity of the inverse Fisher matrix to numerical errors, it is necessary to achieve sufficient accuracy. To this end, we use a supplementary regularization procedure, described in App.~A of Ref.~\cite{Tambalo:2022plm}. In addition, we choose $w_{\rm high}$, the transition between WO and bGO, such that the errors between both computations are smaller than $1\%$ for $F(w)$ and $3\%$ for the derivatives with respect to the lens parameters. Figure \ref{fig:forecast:high_freq_test} shows how these tests are implemented for the CIS and gSIS. We have found that achieving our target precision is hardest for the core size and the density slope.

The large degeneracy between lens parameters makes the inversion of the Fisher matrix (i.e.~the posterior) very sensitive to numerical errors. 
This can be seen in the condition numbers, which are typically $\mathcal{O}(10^4)$ at best and much larger at very low frequencies $w \ll 1$, as the lack of magnification prevents the recovery of lens parameters (see Fig.~\ref{fig:forecast_LISA_fisher_degeneracies} in the main text and Fig.~\ref{fig:forecast_LISA_fisher_matrix} for the dependence of the eigenvalues with the lens mass in a specific example). The conditioning number also grows as $M_{Lz}^2$ when $w\gg 1$ due to the degeneracies in lens parameters in the geometric optics limit. These degeneracies are responsible for the saturation of the precision at high lens masses (for fixed SNR), e.g.~Figs.~\ref{fig:forecast_LISA_source_mass_dependence}, \ref{fig:forecast_LISA_impact_parameter_dependence} and \ref{fig:forecast_LISA_core_size_dependence}.

As a further test of the stability of the computation, we have checked the convergence between our full calculation, GODA and bGODA (introduced in App.~\ref{sec:forecast_GODA}) at sufficiently high lens masses (see Fig.~\ref{fig:GODA_tests} in App.~\ref{sec:forecast_GODA}). In the gSIS case with $k>1$ (only two images) the Fisher matrix for GODA is not invertible: with only two images one can extract at most two lens parameters. The inclusion of bGO terms allows instead to extract more information, although the precision falls with $M_{Lz}$, since GODA is recovered in this limit. Finally, the rGODA terms (containing also the cusp contribution) allow to formally obtain constant precision at high $M_{Lz}$, cf.~Fig.~\ref{fig:GODA_tests_gSIS} in App.~\ref{sec:forecast_GODA}. We explain this (paradoxical) result in Sec.~\ref{sec:forcast_gSIS}, motivating that it cannot be trusted, as the LSA becomes invalid in this limit. 

\section{Geometric Optics Diagonal Approximation}\label{sec:forecast_GODA}

\begin{figure*}
\includegraphics[width=\columnwidth]{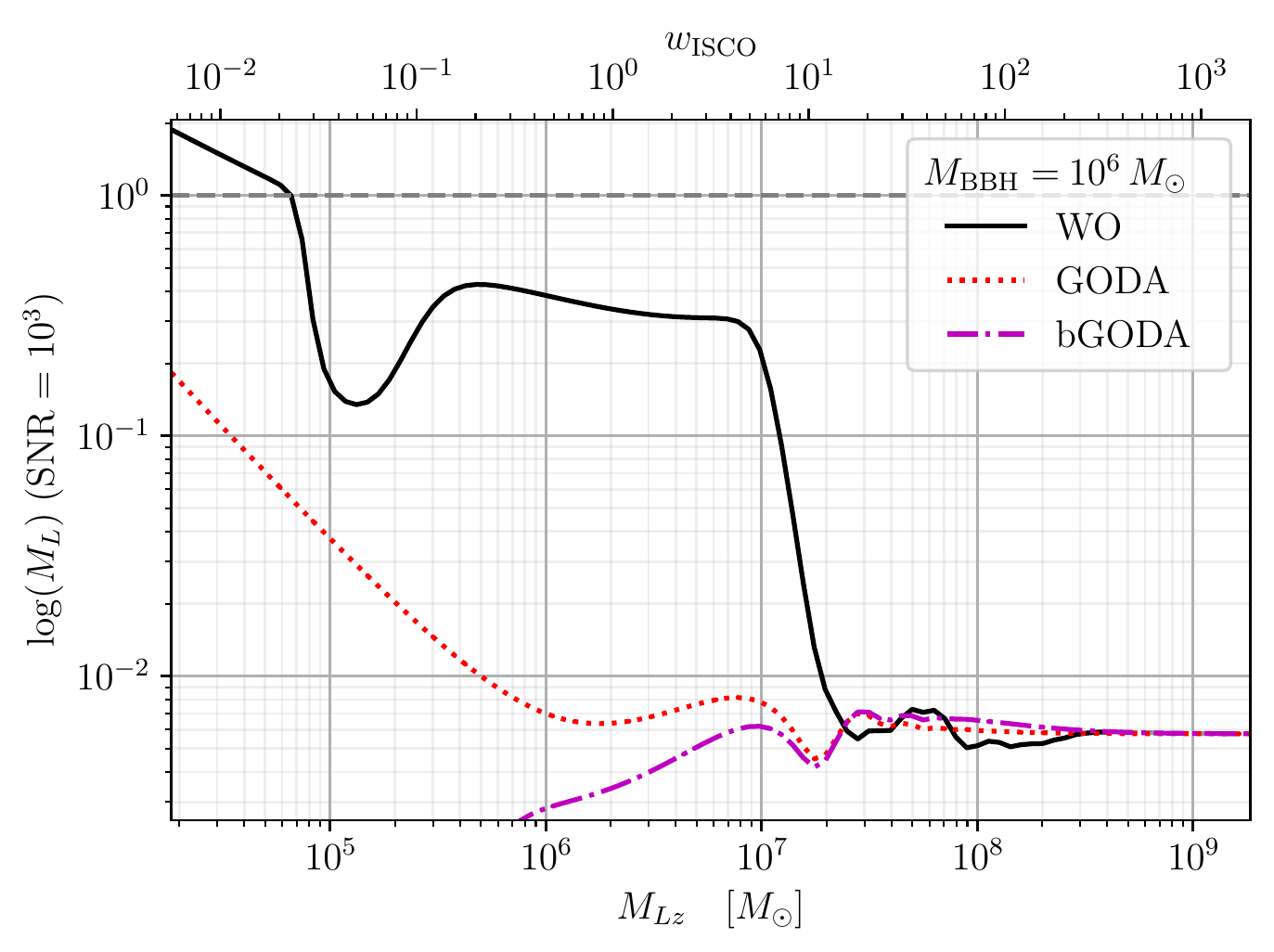}
\includegraphics[width=\columnwidth]{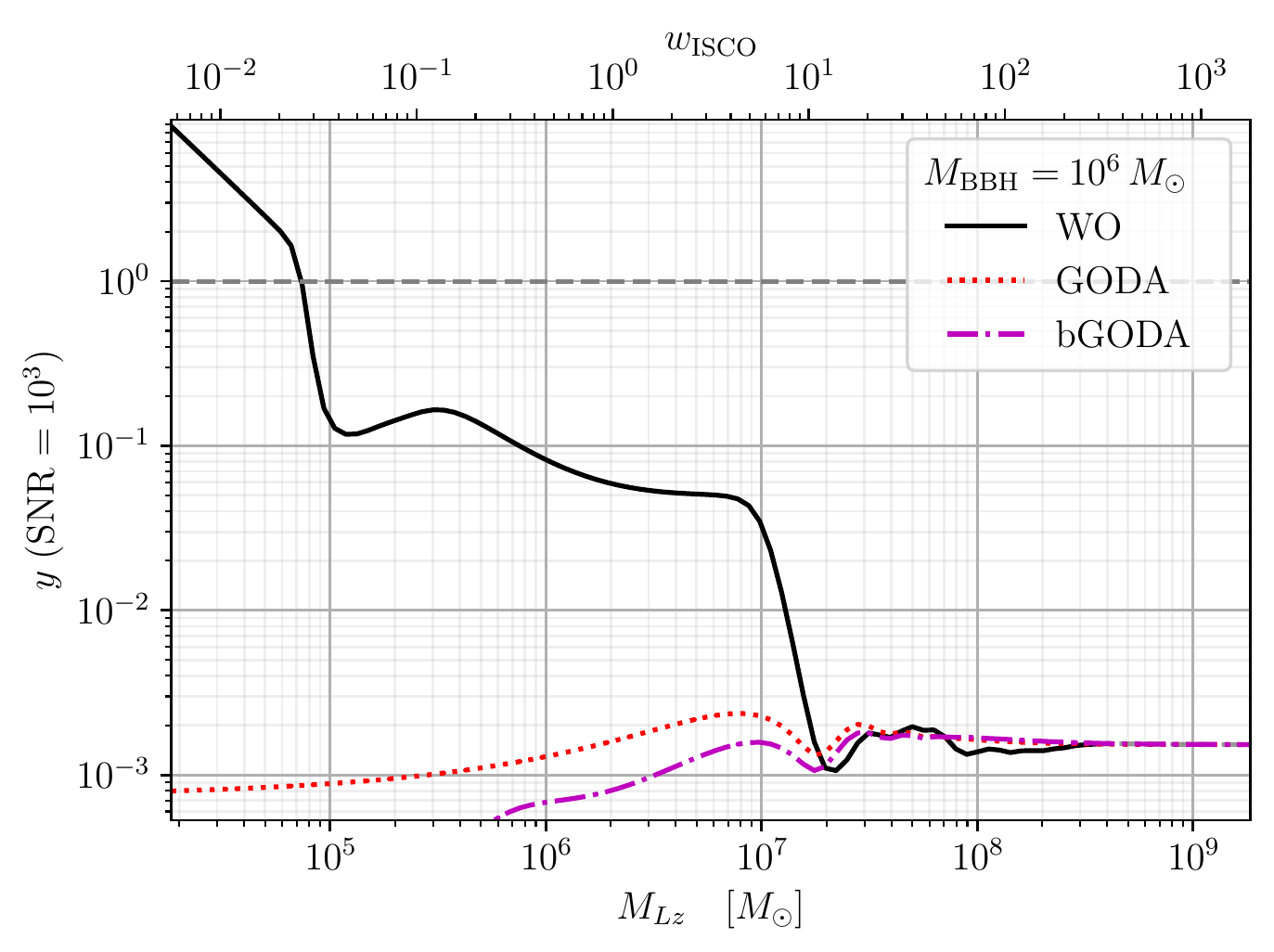}
\includegraphics[width=\columnwidth]{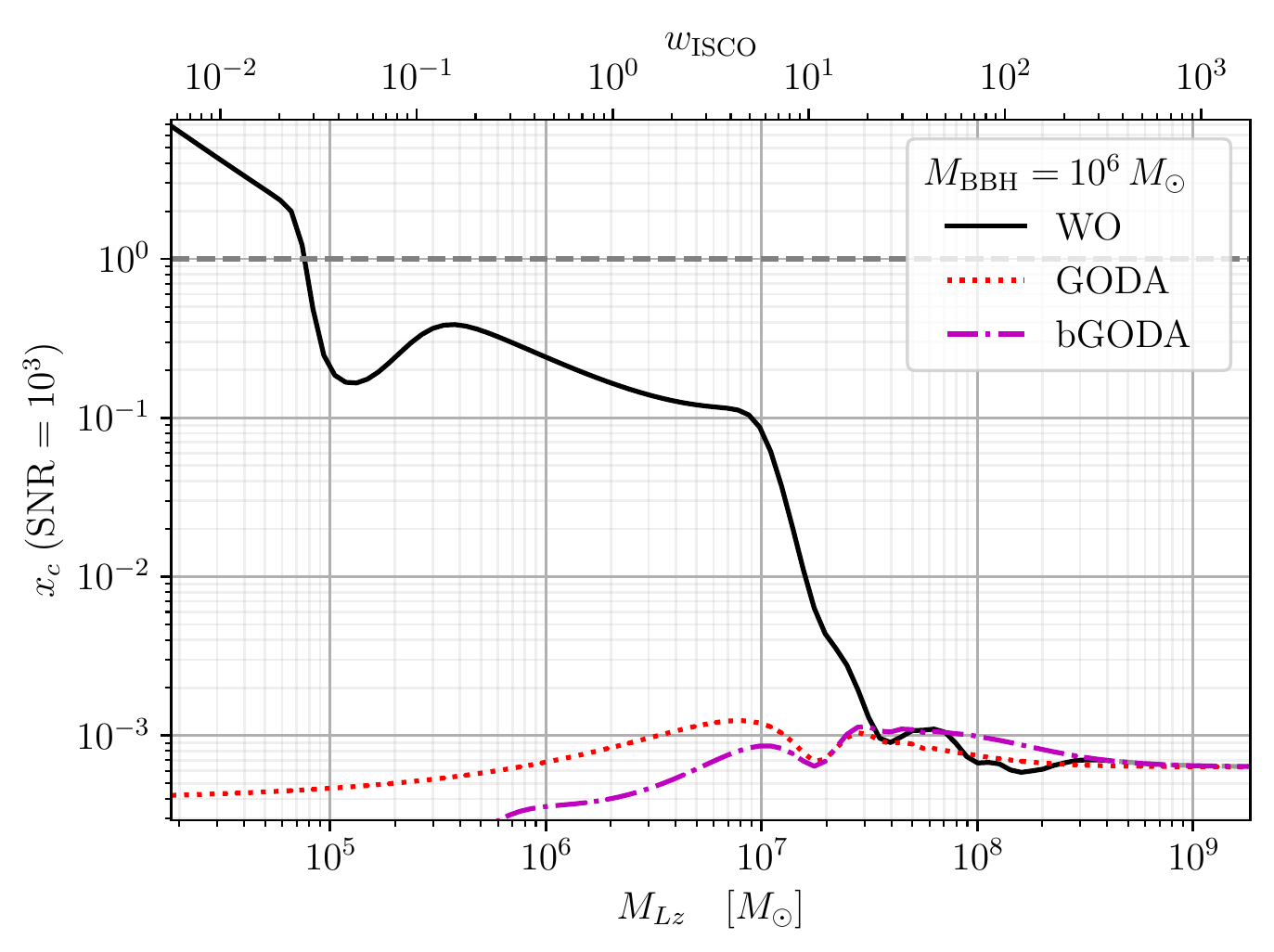}
\includegraphics[width=\columnwidth]{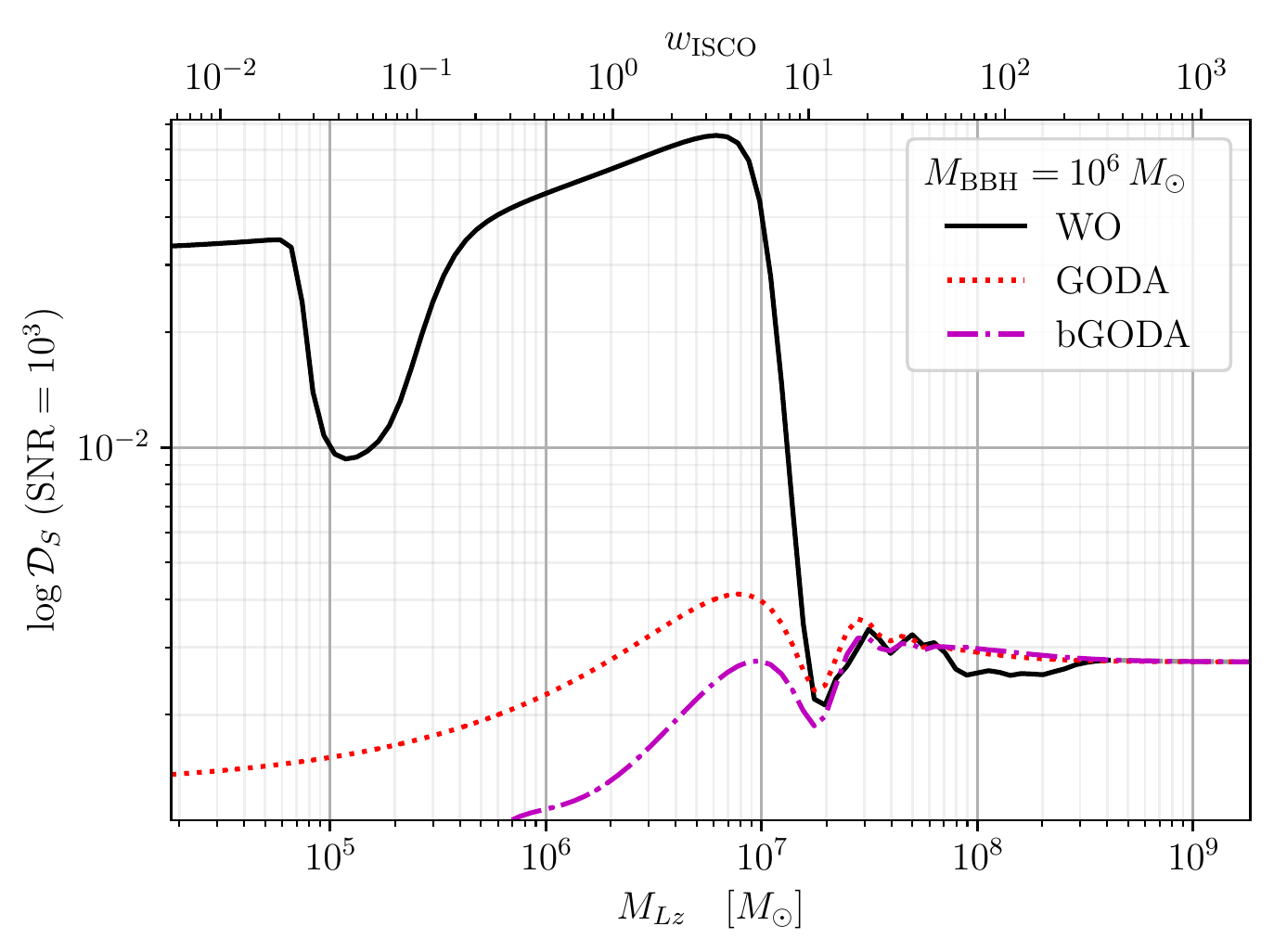}
\caption{68\% c.l.~marginalised constraints on the lens parameters as a function of the lens mass. $y=0.3$, $x_c=10^{-2}$ and the source $ 10^6\, M_\odot$ at $z=3$. Results are normalized to a lensed ${\rm SNR} = 10^3$. The solid black line shows the marginalised errors using the full calculation, including wave-optics effects. The red-dotted and magenta dash-dotted correspond to Geometric-optics diagonal approximation and its beyond GO extension. The upper horizontal axes show the dimensionless frequency corresponding to the innermost stable circular orbit (ISCO) given the source and lens mass. 
} \label{fig:GODA_tests}
\end{figure*}

\begin{figure*}
\includegraphics[width=\columnwidth]{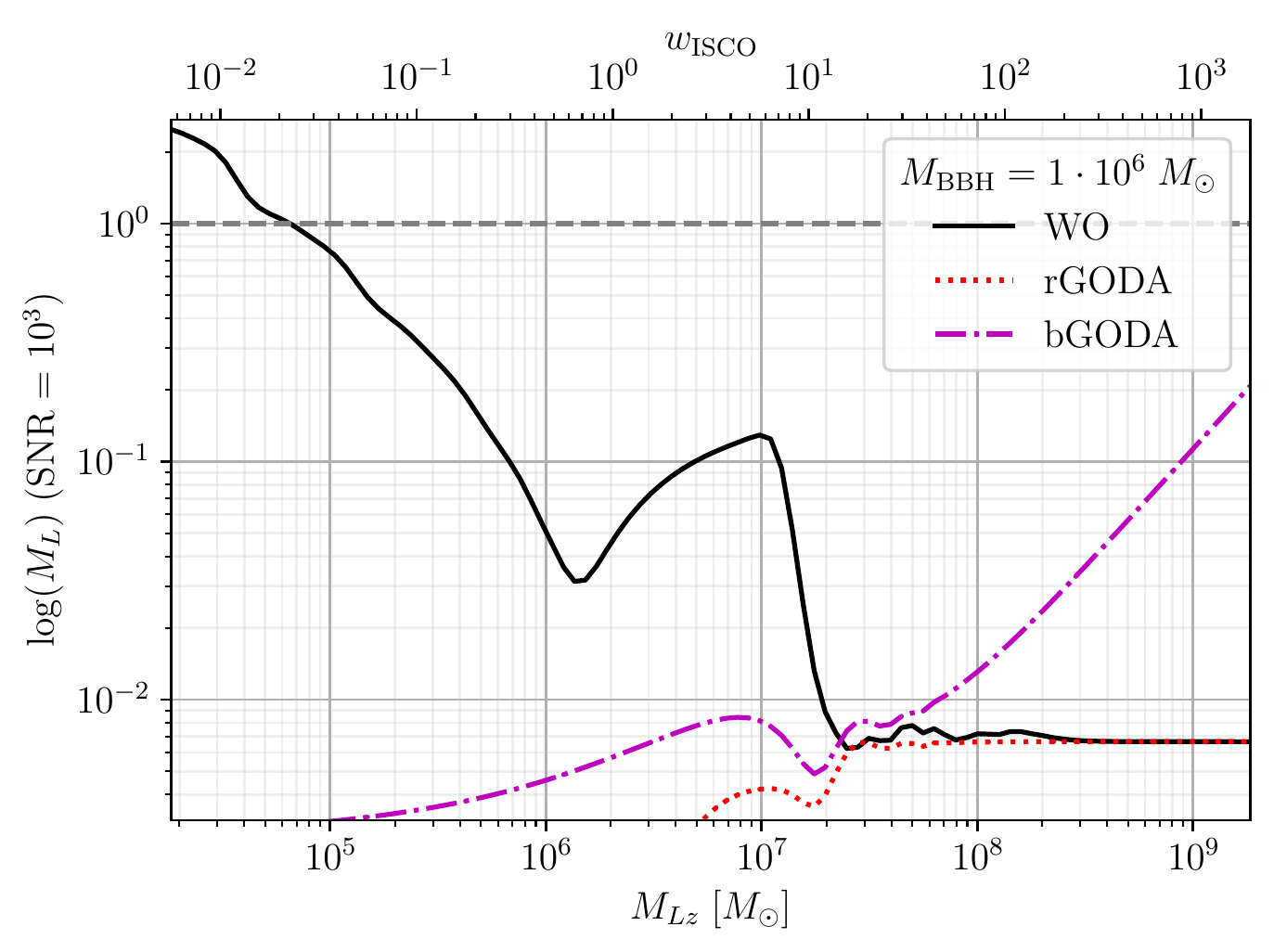}
\includegraphics[width=\columnwidth]{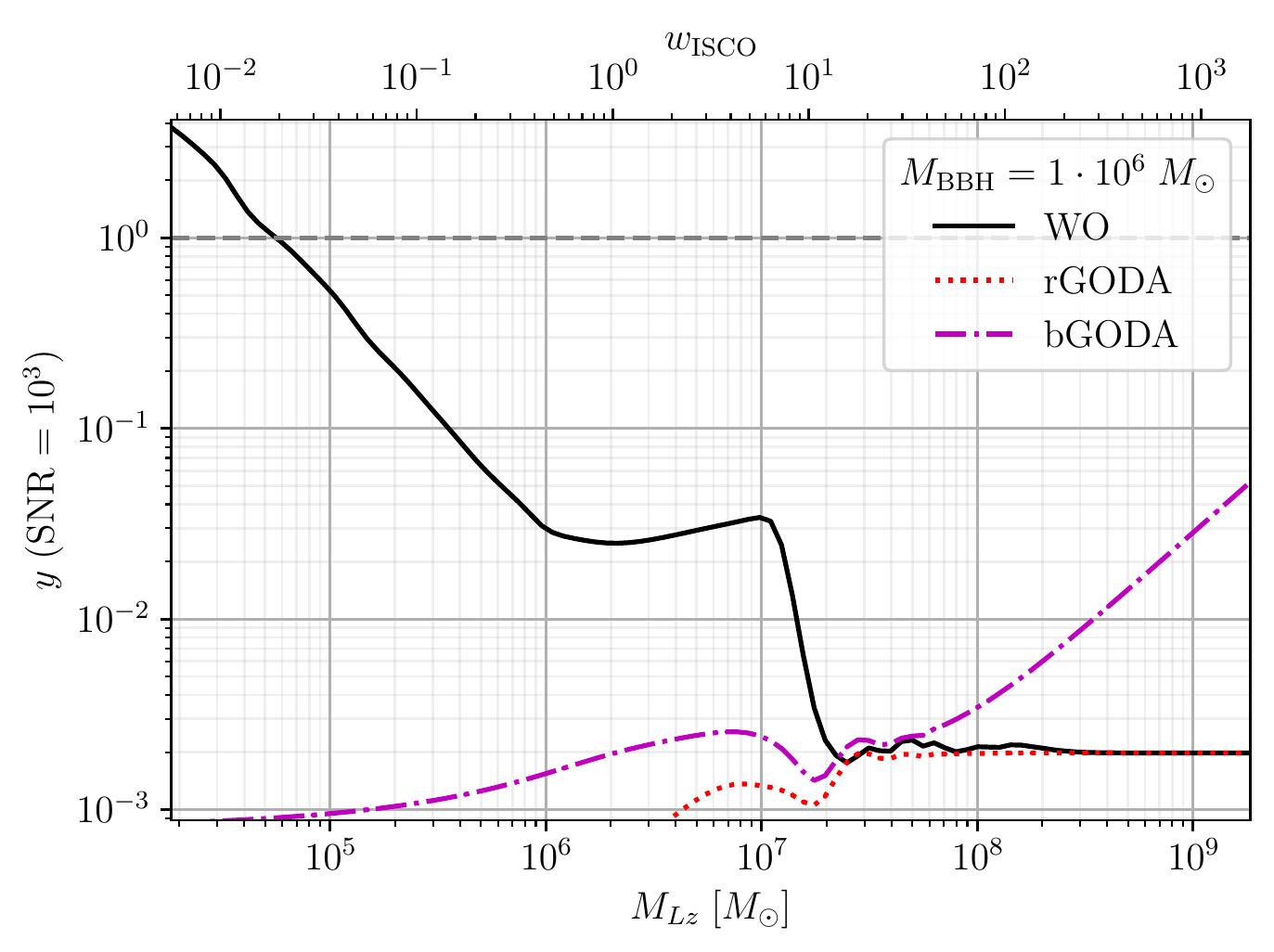}
\includegraphics[width=\columnwidth]{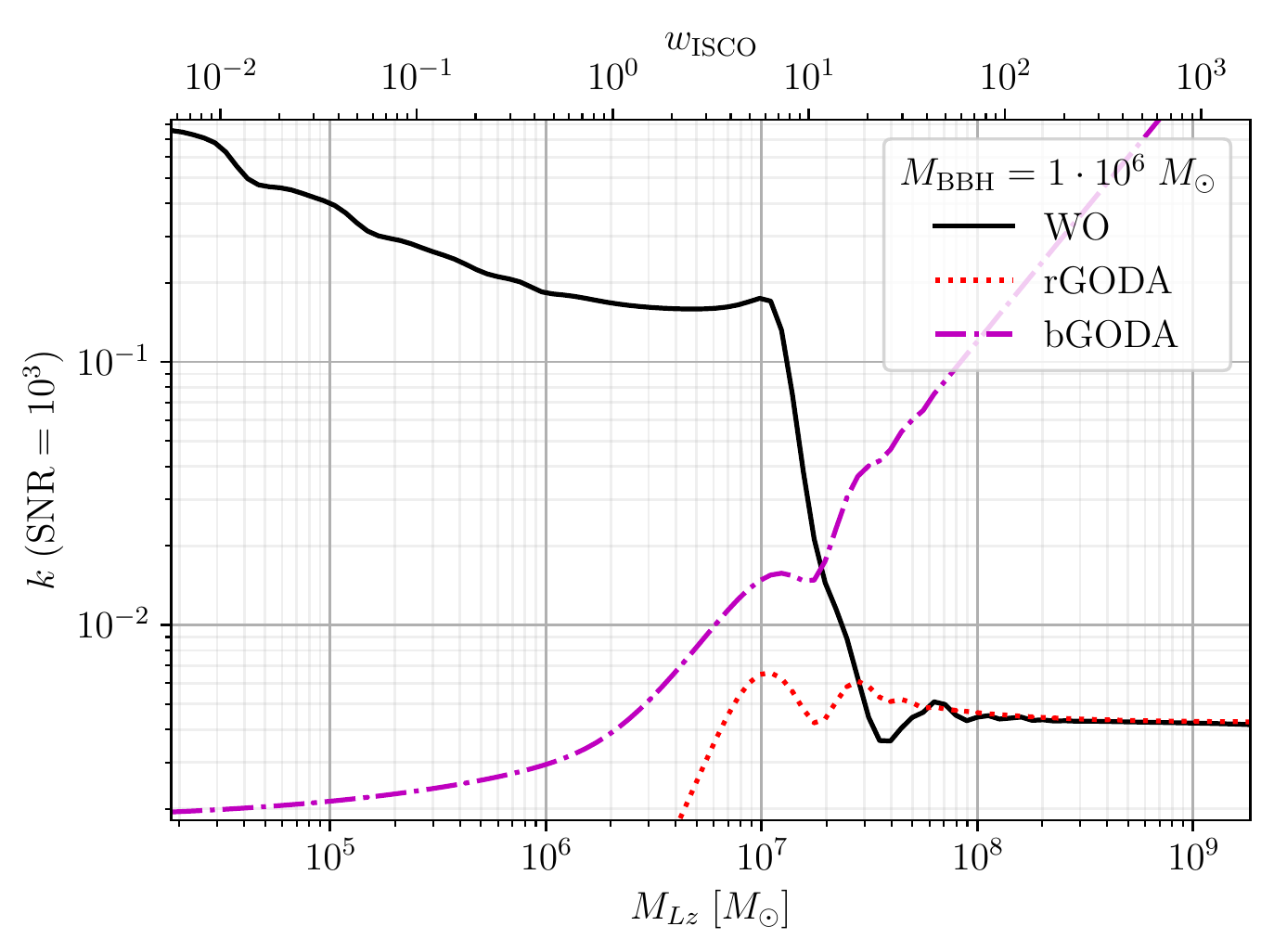}
\includegraphics[width=\columnwidth]{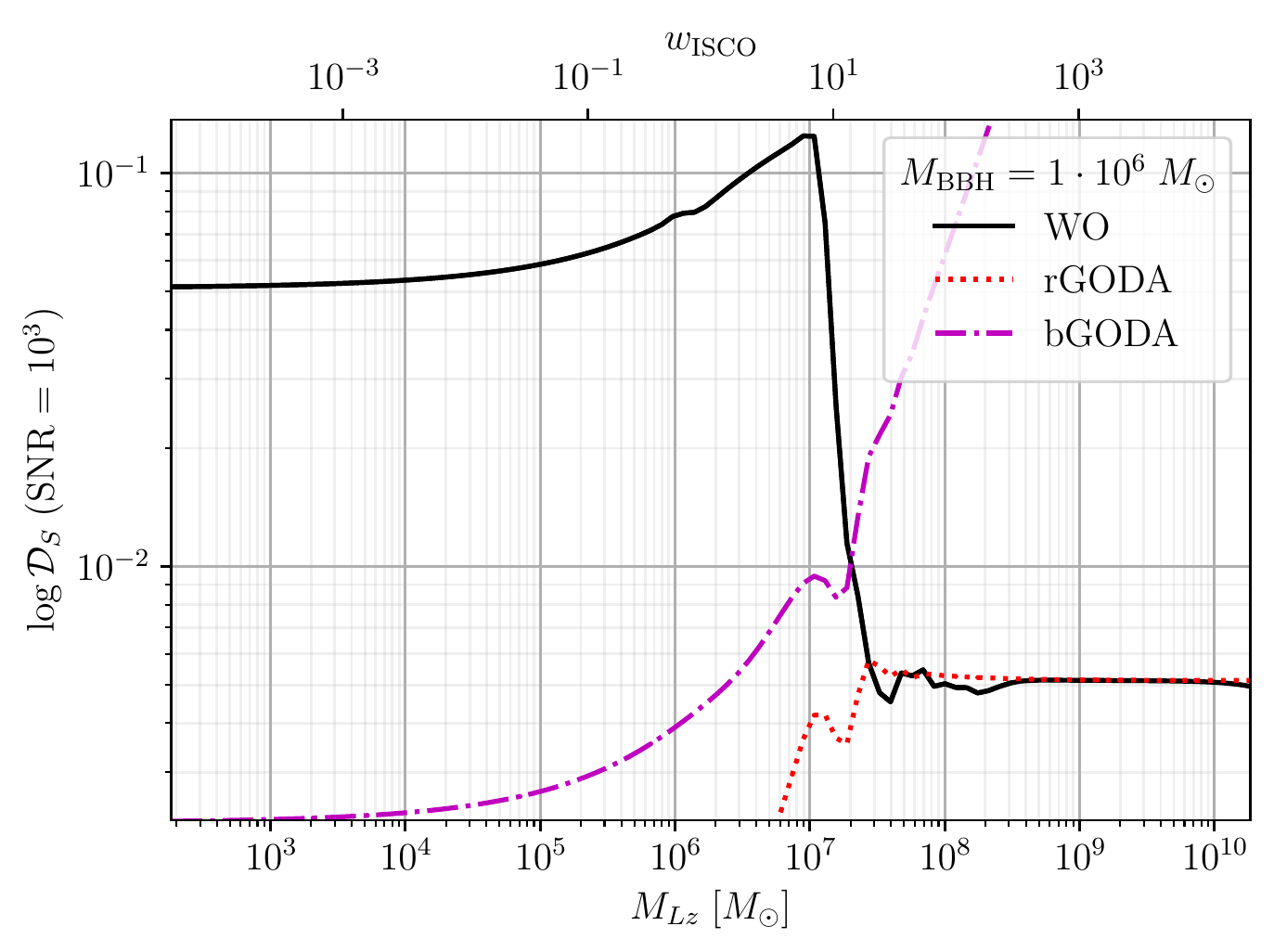}
\caption{68\% c.l.~marginalised constraints on the lens parameters for gSIS as a function of the lens mass. The fiducial parameters are $y=0.3$, $k=1+10^{-4}$ and the source mass $10^6 M_\odot$ at $z=3$. Results are normalized to a lensed ${\rm SNR} = 10^3$. The solid black line shows the marginalised errors using the full calculation, including wave-optics effects. The red-dotted and magenta dash-dotted correspond to beyond Geometric-optics diagonal and the rGO diagonal approximations. The upper horizontal axes show the dimensionless frequency corresponding to the innermost stable circular orbit (ISCO) given the source and lens mass. See discussions in Sec.~\ref{sec:forecast_implementation} and \ref{sec:forcast_gSIS}.}
\label{fig:GODA_tests_gSIS}
\end{figure*}

We will introduce an approximation valid at high frequencies, based on the bGO Fisher derivatives \eqref{eq:fisher_bGO_deriv}. Here we will give expressions for GO, but bGO (and rGO) generalizations can be obtained straightforwardly.
In the GO limit the Fisher-matrix elements involving lens parameters $l,m$ read
\begin{equation}
\mathcal{F}_{lm}^{\rm GO} 
= 
\sum_{IJ} 
\sqrt{|\mu_I\mu_J|}
\int \de f
\frac{|\tilde h_0(f)|^2}{S_n(f)}
\mathcal{I}_{Il, Jm}
\;.
\end{equation}
The integrand associated with each term is
\begin{align}\label{eq:integrand_fisher_GO}
 \mathcal{I}_{Il, Jm}
 &=
 \left(
 \frac{\mu_{I,l}\mu_{J,m}}{\mu_I\mu_J} 
 + 4 w^2\phi_{I,l}\phi_{J,m} 
 \right)
 \cos(\Delta_{IJ})
 \nonumber 
 \\
 &
 \quad
 + 2 w
 \left(
 \phi_{I,l}\frac{\mu_{J,m}}{\mu_J} 
 -\frac{\mu_{I,l}}{\mu_I}\phi_{J,m} 
 \right)
 \sin(\Delta_{IJ})
 \;.
\end{align}
The oscillatory pieces are given by the phase difference between images
\begin{equation}
 \Delta_{IJ} = w(\phi_I-\phi_J)+n_I-n_J \;.
\end{equation}

At high frequencies, the GO Fisher matrix will in general be dominated by the diagonal terms $I=J$, for which the oscillatory contribution cancels. 
We, therefore, introduce the \textit{Geometric Optics Diagonal Approximation} (GODA) for the Fisher matrix, defined by neglecting the cross-image terms as follows:
\begin{align}
 \mathcal{F}_{lm}^{\rm GD} 
 &
 \equiv 
 \sum_I 
 \bigg[
 \frac{\mu_{I,l} \mu_{I,m}}{4|\mu_I|}(h_0|h_0) 
 \nonumber \\
 &
 \quad
 \quad
 \quad
 \quad
 +|\mu_I|\phi_{I,l}\phi_{I,m}(w h_0| w h_0) 
 \bigg]
 \;.
\end{align}
The first term is proportional to the unlensed signal's $\text{SNR}^2$. The second term involves the signal weighted by the dimensionless frequency $(w h_0| w h_0)\propto M_{Lz}^2 (f h_0| f h_0)$ and typically dominates in the high-mass limit.
Note that GODA is equivalent to GO in the weak-lensing regime, as there is a single image.

Figure \ref{fig:GODA_tests} shows the convergence of the full calculation to GO and bGO for an example lensed LISA source.
Good convergence is achieved above lens masses such that the dimensionless frequency corresponding to the innermost stable circular orbit (ISCO) is large, $w_{\rm ISCO}>10$.
For lower lens masses the GODA overestimates the sensitivity: it traces the shape of the error in the mass, but completely overestimates other lens parameters. This is because information about the mass is only contained in the phase/time delays, and hence lost in the limit $w(\phi_I-\phi_J)\lesssim 1$. Other lens parameters affect the GO magnification factors, and hence appear to be measurable in the GODA limit even when $w \ll 1$, as GODA fails to account for the lack of magnification in the low-$w$ regime. 
WO has an additional source of information on the lens mass, namely the onset of magnification around $w\sim 1$.

The neglected cross-image terms ($I\neq J$) lead to oscillatory integrals and are in general subdominant for large $w$. This can be seen as follows, assuming that the coefficient $|h_0|^2/S_n$ depends weakly on frequency. The integral of a term proportional to $w^\alpha$ in Eq.~\eqref{eq:integrand_fisher_GO} would be proportional to $w^{\alpha+1}$ for $I = J$, while to $\Delta_{IJ}^{-1}w^\alpha\cos(\Delta_{IJ} w)\leq w^{\alpha+1}$ for $I\neq J$. 
Cross-image terms may be important in some cases, e.g.~if the source is observed over a narrow range of frequencies. In addition, if an image is very faint compared to other images (e.g.~$\mu_I^2\ll |\mu_I\mu_J|$), its GODA contribution can be smaller than the $I
\neq J$ cross terms. Finally, non-diagonal terms might be important if the unlensed waveform $\tilde h_0$ has a modulation with period $w \sim 1/\Delta_{IJ}$.

\section{LSA validity for CIS}
\label{sec:fisher_validity}

At this point we would like to check the validity of the Fisher-matrix formalism for our analysis, starting from the CIS lens.
Let us write the full likelihood $\mathcal L$ as
\begin{equation}
    \mathcal L = e^{-\Lambda} = e^{-\frac{1}{2}(s-h(\vect \theta)|s-h(\vect \theta))}\;,
\label{eq:def_likelihood}
\end{equation}
We then require that the $1\sigma$ posteriors of the full likelihood 
\eqref{eq:def_likelihood} fall within the 2$\sigma$ contours obtained through the Fisher analysis.\footnote{This criterion is less conservative than the one proposed in \cite{Vallisneri:2007ev} to check the validity of the LSA: ours can be satisfied even in cases where the full likelihood and the one from the Fisher analysis differ substantially. Nonetheless, we believe it to be informative since it bounds the size of the correlations of the full likelihood.}
If this condition is satisfied, it means we are not missing important degeneracies that could hinder our results.
In practice, we check a simplified version of this condition, namely that the full log-likelihood $\Lambda$ evaluated at the $2\sigma$ contours (obtained from the Fisher) is larger than the Fisher log-likelihood evaluated at $1\sigma$ (from our definition of $\Lambda$ in Eq.~\eqref{eq:def_likelihood} this means the probability density is smaller).

In order to evaluate this likelihood for a large number of points, we work in the regime of high $w$ (high $M_{Lz}$), where we can obtain lensed signals quickly using the GO/bGO amplification factor. This is also the region where the forecast gives the best sensitivity and, as we have motivated in the main text (see Sec.~\ref{sec:forecast_framework}), the breakdown of the LSA condition is expected to appear.
A point on the $2\sigma$ contours is parameterized as $\Delta \theta^i = \sum_{k = 1}^5 u_{(k)}^i c_{k} (\lambda_{(k)}/2)^{-1/2}$, where $\lambda_{(k)}$ and $ u_{(k)}^i$ are the eigenvalues and eigenvectors defined in Eq.~\eqref{eq:fisher_eigen} and $c_{k}$ is a unit-norm vector on the 5D parameter-space sphere. To generate more points, we draw the values of $c_{k}$ from a normal distribution \cite{Vallisneri:2007ev}.
We then evaluate the full log-likelihood $\Lambda = (h(\vect{\bar \theta}) - h(\vect{\bar \theta} + \Delta \vect \theta)|h(\vect{\bar \theta}) - h(\vect{\bar \theta} + \Delta \vect{\theta}))/2$. Again, $h(\vect \theta)$ is computed for simplicity using bGO. Our criterion is satisfied if $\Lambda$ is larger than the Fisher log-likelihood for most of the points $\Delta \vect \theta$.

This criterion is found to be satisfied $99.7\, \%$ of the times in our typical fiducial values $y = 0.3$, $x_c = 10^{-2}$, $M_{Lz} = 10^{10}\, M_{\odot}$, $M_{\rm BBH} = 10^{6}\, M_{\odot}$ and $z = 3$ (for $10^3$ points).

Even when the condition is satisfied, we find that the log-likelihood $\Lambda$ is not matching the Fisher one. This does not necessarily call for concern, in the sense we just explained, but in any case points towards a breakdown of the LSA in the high-$M_{Lz}$ limit. 
These conclusions motivate the explanation given in the main text, Sec.~\ref{sec:forecast_framework} and \ref{sec:forcast_CIS}.

Notice that this issue is common to all lens models in the strong-lensing regime, not only to the CIS. To our knowledge, this issue was not investigated previously in the literature. It would be interesting to revisit the Fisher results in the literature and characterize the sensitivities with a more sophisticated analysis. 

As we are soon going to mention, this issue becomes more problematic in the gSIS case, where the Fisher analysis yields very optimistic results that contradict the number of parameters that can be reconstructed in the GO limit, Eq.~\eqref{eq:parameter_reconstruction_GO}. We give a concrete analytical example of this issue in App.~\ref{sec:SPA}. As we comment at the end, the analysis of this example also supports the conclusion that a violation of the LSA condition for the CIS should not call for concern.
\begin{figure*}
    \centering
    \includegraphics[width=0.8\textwidth]{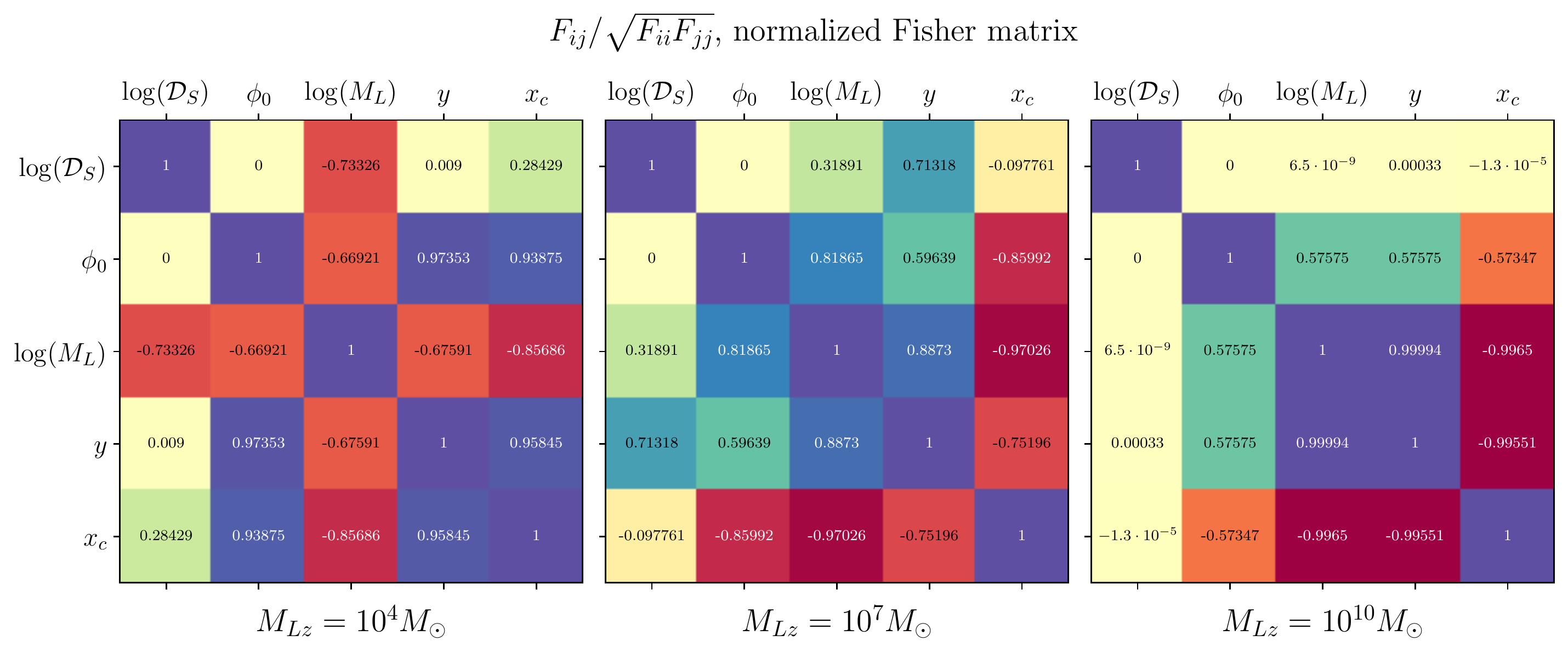}
    \includegraphics[width=0.8\textwidth]{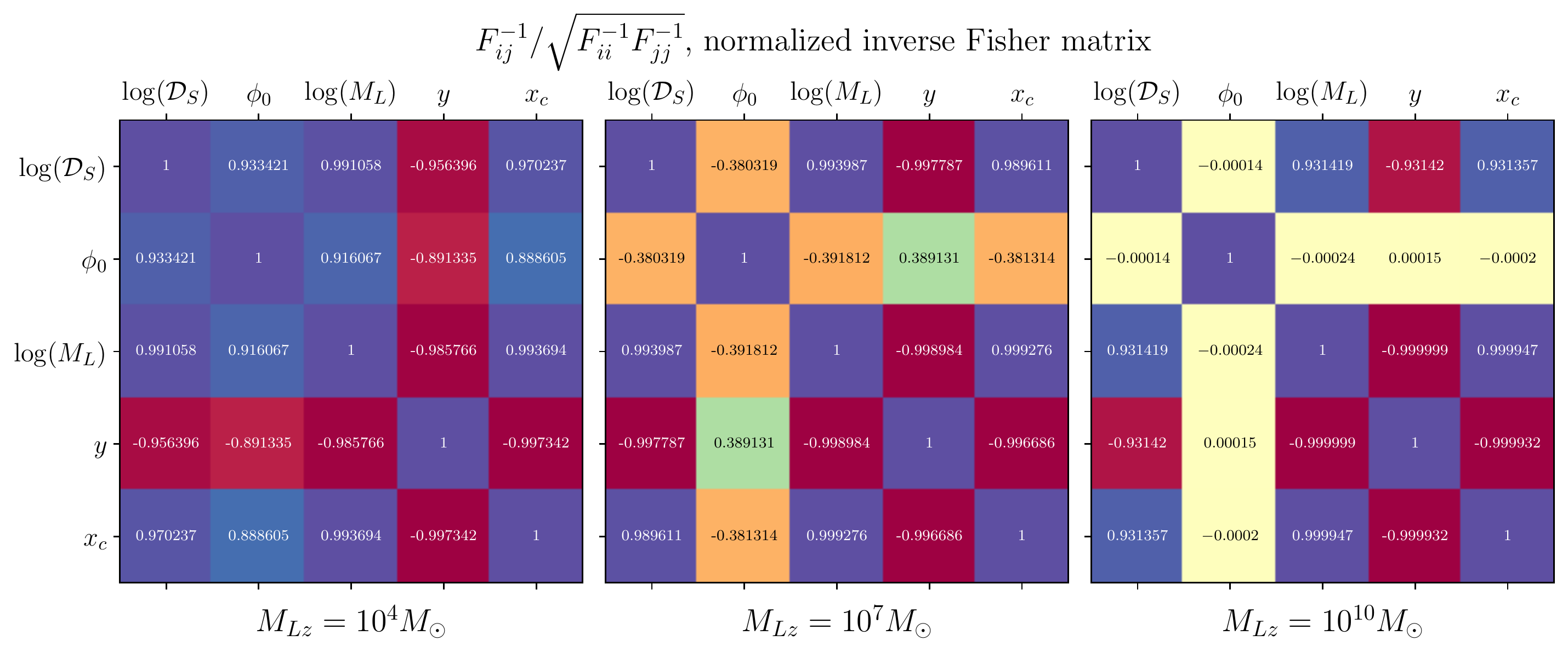}
    \includegraphics[width=0.8\textwidth]{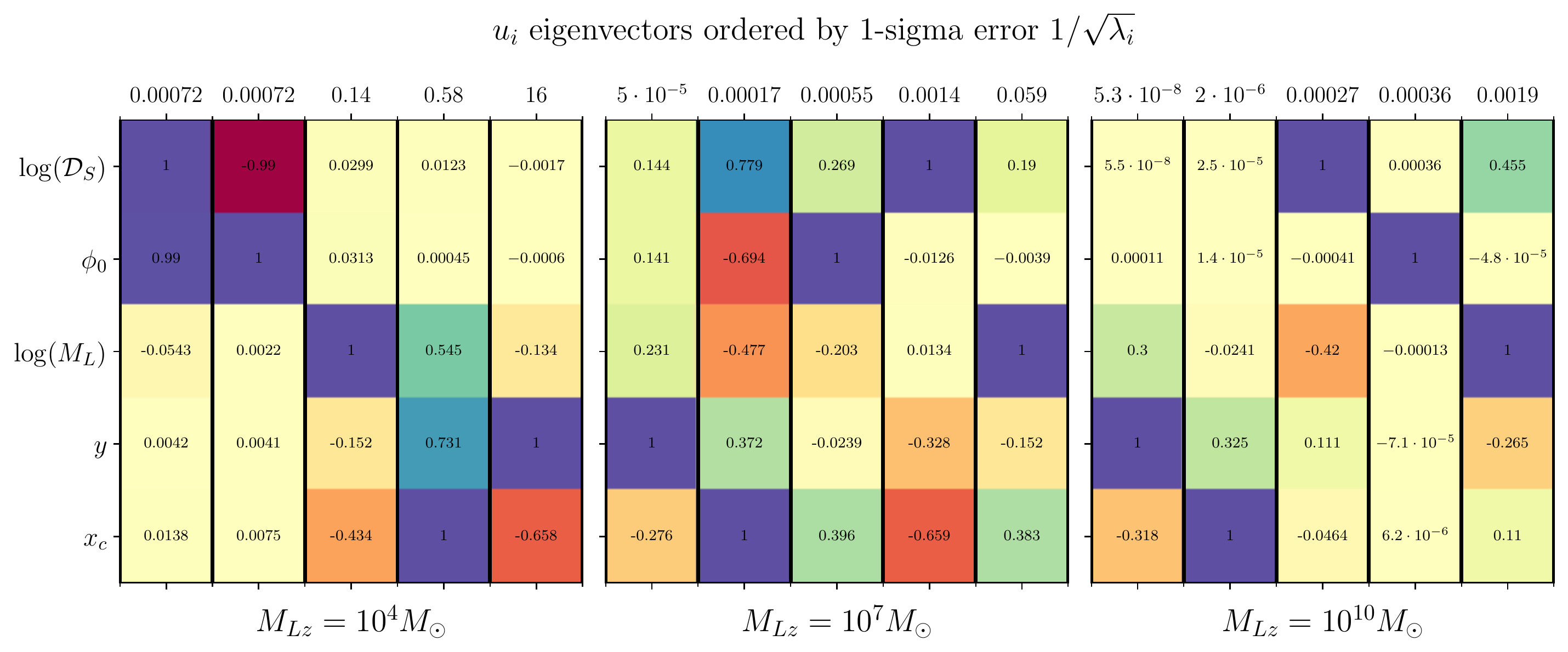}
\caption{\textbf{Top}: Normalized Fisher matrix for a LISA source with $M_{\rm BBH}=10^6\, M_{\odot}$ and CIS with $M_{Lz}=10^4\, M_\odot$, $10^7\, M_\odot$ and $10^{10}\, M_\odot$, $y=0.3$, $x_c=10^{-2}$.
\textbf{Middle}: Normalized inverse Fisher matrix. 
\textbf{Bottom}: Principal components of the Fisher matrix (left to right, normalized to the largest component). The precision is given on top of each eigenvector and it is rescaled to a lensed ${\rm SNR}=10^3$.
} \label{fig:forecast_LISA_fisher_matrix}
\end{figure*}

\section{LSA approximation for cusps}
\label{sec:SPA}

In the main text we argued that the linearisation of the likelihood used in the Fisher matrix formalism might not be always appropriate, even for moderately high $\rm SNR$. 
In turn, the linearisation procedure can overestimate the sensitivity on parameters, yielding unrealistic, or even paradoxical, results.  
Specifically, this seems to be the case for the parameter $k$ in the gSIS forecast. For a fiducial value $k = 1$ the Fisher matrix predicts $\Delta k \propto 1/{\rm SNR}$ in the GO limit (large lens masses $M_{Lz}$). On the other hand, in this limit the characteristic feature due to $k$ (the signal from the cusp) quickly decays. Thus, no meaningful constraint on $k$ should be present.
Here we provide a toy example highlighting this feature: by studying the full probability distribution we can see that indeed the error on $k$ grows for large $M_{Lz}$.

Let us consider a toy lensing signal with a single generic lensing parameter $\theta$ (that could represent $k$ in the case of the gSIS). The lensed GW waveform $h_L(f)$ is taken to be 
\begin{equation}
    h_L(f) = h_0(f) \left(1+\frac{e^{i w \theta}}{w}\right)\;,
\label{eq:lens_toy}
\end{equation}
where $h_0(f)$ is the unlensed waveform, $w = 8 \pi G M_{Lz}$ and $M_{Lz}$ is taken as fixed. This signal is indeed analogous to the cusp obtained for the gSIS in the limit of $k \to 1$, see Eq.~\eqref{eq:F_cusp}. One can easily generalise what follows to cases with different $w$ dependencies.

Let us write $\theta = \bar \theta + \Delta \theta$, where $\bar \theta$ is the fiducial value. We want to compute the probability distribution for $\Delta \theta$, assuming some analytic behaviour for $h_0(f)$ and the noise.

Let us call $\Delta h \equiv h(\bar \theta)-h(\theta)$. Then, the exponent $\Lambda$ of the likelihood is defined as
\begin{equation}
  \Lambda = \frac{1}{2}\left(\Delta h|\Delta h\right)  \;,
\end{equation}  
where we are neglecting the noise term. In this way the likelihood is written as $\mathcal L \propto e^{-\Lambda}$.
We can write the full $\Lambda$ without expanding in $\Delta \theta$. It is easy to get
\begin{equation}
\Lambda 
= 
8 \int \frac{\de f}{S_n(f)} \frac{|h_0(f)|^2}{w^{2}}\left[1-\cos(w \Delta \theta)\right]
\;.
\end{equation}
If we were to expand the square bracket for small $\Delta \theta$, we would recover the Fisher formalism. Notice that if we expand and trust the result for $w \Delta \theta >1$, then we are \emph{overestimating} $\Lambda$. In the full result, we see indeed that the square bracket remains bounded for large $\Delta \theta$.

At this point, we can understand the behaviour of the integral by taking a simplified form for $h_0(f)$ and $S_n(f)$.
The lower band of the LISA noise is characterised by $S_n(f) \sim s_n f^{-4}$, where $s_n$ is a dimensionful constant that depends on the noise properties of LISA. This is a good approximation for events with large masses $M_{\rm BBH} \gtrsim 10^7\, M_{\odot}$ since the merger happens before the noise curve reaches its minimum. 
For the waveform, we take simply $h_0(f) = A f^{-7/6} e^{i \Psi(f)}$, where $A$ is the amplitude and $\Psi(f)$ the phase evolution \cite{maggiore2007gravitational}. We perform the integration over $f$ from some low $f_0$, where the signal enters the LISA band, and we cut the signal at the ISCO frequency, which we indicate as $f_1$. 

The SNR for this signal is then
\begin{equation}
{\rm SNR}^2 = (h_0|h_0) \simeq \frac{3 A^2}{2 s_n}f_1^{8/3}\;,
\end{equation}
where we assumed $f_1\gg f_0$.

We can use the expression above to simplify $\Lambda$, and get
\begin{equation}
\Lambda \simeq \frac{8\, {\rm SNR}^2}{3 f_1^{8/3}}\int_{f_0}^{f_1} \de f \, \frac{f^{5/3}}{w^2}\left[1-\cos(w \Delta \theta)\right] \;.
\label{eq:exp_likelihood_toy}
\end{equation} 
This integral can be solved explicitly in terms of exponential integrals. However, for our purposes, we can focus on the two limits of $w \Delta \theta \ll 1$ and $w \Delta \theta \gg 1$, where $w$ represents the dimensionless frequency contributing mainly to the integral (i.e.~$w$ at the ISCO).

For small $w \Delta \theta$ we can expand the square bracket, recovering a quadratic function of $\Delta \theta$: $\Lambda \simeq  {\rm SNR}^2 \Delta \theta^2$. This corresponds with the Fisher-matrix result. Naively, it would predict $\Delta \theta \sim 1 /{\rm SNR}$, even though the correction due to $\theta$ in Eq.~\eqref{eq:lens_toy} becomes negligible at high $M_{Lz}$. From this discussion it is now clear that this result is only applicable for $w_{\rm ISCO} \Delta \theta < 1$ and therefore breaks down at high $M_{Lz}$.
This is analogous to the LSA condition in Eq.~\eqref{eq:LSA_cond2} for this simplified case.

Instead, in the opposite limit, we can estimate Eq.~\eqref{eq:exp_likelihood_toy} by noting that the cosine function becomes rapidly oscillating. This means its contribution to the integral is negligible: the log-likelihood then becomes independent of $\Delta \theta$:
\begin{equation}
\Lambda \simeq \frac{4\, {\rm SNR}^2}{w_{\rm ISCO}^2}\;.
\end{equation}
This result shows $\theta$ cannot be efficiently constrained when $w_{\rm ISCO} \sim {\rm SNR}$, since its probability distribution stops being localized around $\Delta \theta = 0$ and becomes flat, with $\mathcal L \propto e^{-\Lambda} \sim 1$. 
In the more realistic situation where, on top of the cusp contribution, there are more GO images the situation is of course more involved. Indeed since the cusp is very faint, its contribution to the likelihood might be dominated by the product with other images. The general logic is expected to apply nonetheless.  

In conclusion, we expect the Fisher approximation to stop being accurate around $\Delta \theta \sim w_{\rm ISCO}^{-1}$, with the constraint on $\theta$ becoming weaker and weaker as we increase $M_{Lz}$. Eventually $\theta$ becomes unconstrained when we reach $w_{\rm ISCO} \sim {\rm SNR}$.

We can also comment on the case in which the lensing effects are not suppressed by $1/w$, as in Eq.~\eqref{eq:lens_toy}, but are due to a GO image. (Notice also that bGO terms do not have the form of Eq~\eqref{eq:lens_toy}, since the phase factor would be factorised outside of the bracket.) It is straightforward to obtain the analogous of Eq.~\eqref{eq:exp_likelihood_toy} in this case (the factor $w^{-2}$ disappears). Linearisation becomes unreliable when $\Delta \theta \sim w_{\rm ISCO}^{-1}$, as in the previous case. However, the likelihood is still highly suppressed for large $w_{\rm ISCO}$, since $\Lambda \sim {\rm SNR}^2$ (the tails of the distribution differ from a Gaussian, but the associated probability remains small). 
Applied to the CIS case, this suggests the Fisher analysis still reliably captures the size of the sensitivities even after the LSA condition is violated.

\bibliography{gw_lensing}

\end{document}